\documentclass[final]{ws-procs9x6}
\usepackage{graphics}
\usepackage{axodraw}
\usepackage{natbib}

\usepackage{amsmath}
\usepackage{amssymb}
\usepackage{latexsym}
\usepackage{epsf}
\usepackage{psfrag}
\usepackage{feynarts}
\usepackage{graphicx}

\newcommand{ \slashchar }[1]{\setbox0=\hbox{$#1$}   
   \dimen0=\wd0                                     
   \setbox1=\hbox{/} \dimen1=\wd1                   
   \ifdim\dimen0>\dimen1                            
      \rlap{\hbox to \dimen0{\hfil/\hfil}}          
      #1                                            
   \else                                            
      \rlap{\hbox to \dimen1{\hfil$#1$\hfil}}       
      /                                             
   \fi}                                             %

\newcommand{\gsim}{\gtrsim}
\newcommand{\lsim}{\lesssim}

\bibpunct{[}{]}{,}{n}{,}{,}

\begin{document}

\def\draftnote{}
\def\trimmarks{}
\setlength{\topmargin}{-0.01cm}

\pagestyle{plain}

\title{TASI 2008 Lectures on Dark Matter}

\author{Dan Hooper}

\address{Theoretical Astrophysics Group, Fermi National Accelerator Laboratory,\\
Department of Astronomy and Astrophysics, University of Chicago \\[3mm]
\textnormal{\texttt{dhooper@fnal.gov}}}

\maketitle

\abstracts{Based on lectures given at the 2008 Theoretical Advanced Study Institute (TASI), I review here some aspects of the phenomenology of particle dark matter, including the process of thermal freeze-out in the early universe, and the direct and indirect detection of WIMPs. I also describe some of the most popular particle candidates for dark matter and summarize the current status of the quest to discover dark matter's particle identity.}

\section{Evidence For Dark Matter}

A wide variety of evidence has accumulated in support of dark matter's existence. At galactic and sub-galactic scales, this evidence includes galactic rotation curves~\cite{rotationcurves}, the weak gravitational lensing of distant galaxies by foreground structure~\cite{weaklensing}, and the weak modulation of strong lensing around individual massive elliptical galaxies~\cite{strong}. Furthermore, velocity dispersions of stars in some dwarf galaxies imply that they contain as much as $\sim 10^3$ times more mass than can be attributed to their luminosity. On the scale of galaxy clusters, observations (of radial velocities, weak lensing, and X-ray emission) indicate a total cosmological matter density of $\Omega_M\approx 0.2-0.3$~\cite{clusters}, which is much larger than the corresponding density in baryons. In fact, it was measurements of velocity dispersions in the Coma cluster which led Fritz Zwicky to claim for the first time in 1933 that large quantities of non-luminous matter are required to be present~\cite{zwicky}. On cosmological scales, observations of the anisotropies in the cosmic microwave background have lead to a determination of the total matter density of $\Omega_M h^2 = 0.1326 \pm 0.0063$, where $h$ is the Hubble parameter in units of 100 km/sec per Mpc (this improves to $\Omega_M h^2 = 0.1358^{+0.0037}_{-0.0036}$ if distance measurements from baryon acoustic oscillations and type Ia supernovae are included)~\cite{wmap}. In contrast, this information combined with measurements of the light chemical element abundances leads to an estimate of the baryonic density given by $\Omega_B h^2 = 0.02273\pm 0.00062$ ($\Omega_B h^2 = 0.02267^{+0.00058}_{-0.00059}$ if BAO and SN are included)~\cite{bbn,wmap}. Taken together, these observations strongly lead us to the conclusion that 80-85\% of the matter in the universe (by mass) consists of non-luminous and non-baryonic material.

The process of the formation of large scale structure through the gravitational clustering of collisionless dark matter particles can be studied using N-body simulations. When the observed structure in our universe~\cite{lss} is compared to the results of cold (non-relativistic at time of structure formation) dark matter simulations, good agreement has been found. The large scale structure predicted for hot dark matter, in contrast, is in strong disagreement with observations.

Although there are many pieces of evidence in favor of dark matter, it is worth noting that they each infer dark matter's presence uniquely through its gravitational influence. In other words, we currently have no conclusive evidence for dark matter's electroweak or other non-gravitational interactions. Given this, it is natural to contemplate whether, rather than being indications of dark matter's existence, these observations might instead be revealing departures from the laws of gravity as described by general relativity.

Since first proposed by Milgrom in 1983~\cite{mond}, efforts have been made to explain the observed galactic rotation curves without dark matter within the context of a phenomenological model known as modified Newtonian dynamics, or MOND. The basic idea of MOND is that Newton's second law, $F=ma$, is modified to $F=ma \times \mu(a)$, where $\mu$ is very closely approximated by unity except in the case of very small accelerations, for which $\mu$ behaves as $\mu=a/a_0$. Applying the modified form of Newton's second law to the gravitational force acting on a star outside of a galaxy of mass $M$ leads us to
\begin{equation}
F=\frac{GMm}{r^2} = ma \mu,
\end{equation}
which in the low acceleration limit (large $r$, $a \ll a_0$) yields
\begin{equation}
a=\frac{\sqrt{GMa_0}}{r}.
\end{equation}
Equating this with the centrifugal acceleration associated with a circular orbit, we arrive at
\begin{equation}
\frac{\sqrt{GMa_0}}{r}=\frac{v^2}{r} \,\,\,\,\,\,\,\,\Longrightarrow\,\,\,\,\,\,\,\, v=(GMa_0)^{1/4}.
\end{equation}

In other words, MOND yields the prediction that galactic rotation curves should become flat (independent of $r$) for sufficiently large orbits. This result is in good agreement with galaxy-scale observations for a value of $a_0 \sim 1.2 \times 10^{-10}$ m/s$^{2}$, even without the introduction of dark matter. For this value of $a_0$, the effects of MOND are imperceptible in laboratory or Solar System scale experiments.

MOND is not as successful in explaining the other evidence for dark matter, however. In particular, MOND fails to successfully describe the observed features of galaxy clusters. Other evidence, such as the cosmic microwave background anisotropies and large scale structure, are not generally able to be addressed by MOND, as MOND represents a phenomenological modification of Newtonian dynamics and thus is not applicable to questions addressed by general relativity, such as the expansion history of the universe. Efforts to develop a viable, relativistically covariant theory which yields the behavior of MOND in the non-relativistic, weak-field limit have mostly been unsuccessful. A notable exception to this is Tensor-Vector-Scalar gravity, or TeVeS~\cite{Bekenstein:2004ne}. TeVeS, however, fails to explain cluster-scale observations without the introduction of dark matter~\cite{Skordis:2005xk}. This problem has been further exacerbated by recent observations of two merging clusters, known collectively as the bullet cluster. In the bullet cluster, the locations of the baryonic material and gravitational potential (as determined using X-ray observations and weak lensing, respectively) are clearly spatially separated, strongly favoring the dark matter hypothesis over modifications of general relativity~\cite{bullet}.

\section{The Production of Dark Matter in the Early Universe}
\label{relic}

The nucleons, electrons and neutrinos that inhabit our universe can each trace their origin back to the first fraction of a second following the Big Bang.  Although we do not know for certain how the dark matter came to be formed, a sizable relic abundance of weakly interacting massive particles (WIMPs) is generally expected to be produced as a byproduct of our universe's hot youth. In this section, I discuss this process and the determination of the relic abundance of a WIMP~\cite{Srednicki:1988ce,Gondolo:1990dk,kolbturner}.

Consider a stable particle, $X$, which interacts with Standard Model particles, $Y$, through some process $X\bar{X} \leftrightarrow Y \bar{Y}$ (or $XX \leftrightarrow Y \bar{Y}$ if $X$ is its own antiparticle). In the very early universe, when the temperature was much higher than $m_X$, the processes of $X\bar{X}$ creation and annihilation were equally efficient, leading $X$ to be present in large quantities alongside the various particle species of the Standard Model. As the temperature of the universe dropped below $m_X$, however, the process of $X\bar{X}$ creation became exponentially suppressed, while $X\bar{X}$ annihilation continued unabated. In thermal equilibrium, the number density of such particles is given by
\begin{equation}
n_{X, \, {\rm eq}} = g_X \bigg(\frac{m_X T}{2 \pi}\bigg)^{3/2} e^{-m_X/T},
\end{equation}
where $g_X$ is the number of internal degrees of freedom of $X$. 

If these particles were to remain in thermal equilibrium indefinitely, their number density would become increasingly suppressed as the universe cooled, quickly becoming cosmologically irrelevant. There are ways that a particle species might hope to avoid this fate, however.  For example, baryons are present in the universe today because of a small asymmetry which initially existed between the number of baryons and antibaryons; when all of the antibaryons had annihilated with baryons, a small residual of baryons remained. The baryon-antibaryon asymmetry prevented the complete annihilation of these particles from taking place.

While it is possible that a particle-antiparticle asymmetry is also behind the existence of dark matter, there is an even simpler mechanism which can lead to the survival of a sizable relic density of weakly interacting particles. In particular, the self-annihilation of weakly interacting species can be contained by the competing effect of Hubble expansion. As the expansion and corresponding dilution of WIMPs increasingly dominates over the annihilation rate, the number density of $X$ particles becomes sufficiently small that they cease to interact with each other, and thus survive to the present day. Quantitatively, the competing effects of expansion and annihilation are described by the Boltzmann equation:
\begin{equation}
\frac{dn_X}{dt} + 3 H n_X = -<\sigma_{X\bar{X}} |v|> (n^2_X - n^2_{X,\,{\rm eq}}),
\label{boltman}
\end{equation}
where $n_{X}$ is the number density of WIMPs, $H \equiv \dot{R}/R = (8\pi^3 \rho/3 M_{\rm Pl})^{1/2}$ is the expansion rate of the universe, and $<\sigma_{X\bar{X}} |v|>$ is the thermally averaged $X\bar{X}$ annihilation cross section (multiplied by their relative velocity). 

From Eq.~\ref{boltman}, we can identify two clear limits.  As I said before, at very high temperatures ($T \gg m_X$) the density of WIMPs is given by the equilibrium value, $n_{X,\, {\rm eq}}$. In the opposite limit ($T \ll m_X$), the equilibrium density is very small, leaving the terms $3Hn_X$ and $<\sigma_{X\bar{X}}|v|> n_X^2$ to each further deplete the number density. For sufficiently small values of $n_X$, the annihilation term becomes insignificant compared to the dilution due to Hubble expansion. When this takes place, the comoving number density of WIMPs becomes fixed --- thermal freeze-out has occurred. 

\begin{figure}[t]
\centerline{\includegraphics[width=0.7\hsize]{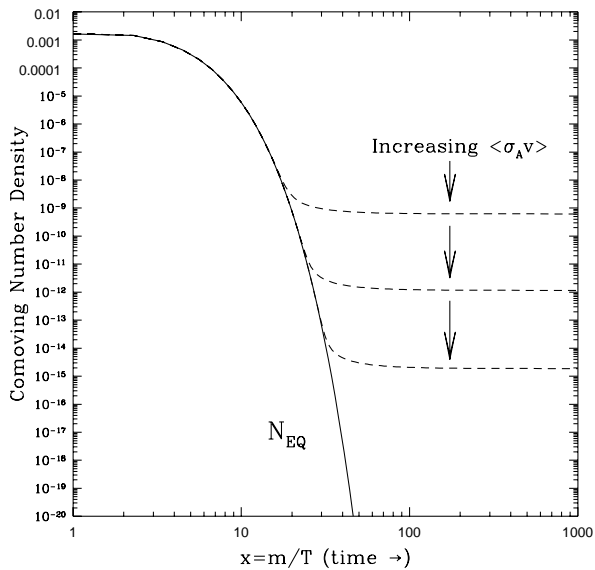}}
\caption{A schematic of the comoving number density of a stable species as it evolves through the process of thermal freeze-out. }
\label{freezeout}
\end{figure}

The temperature at which the number density of the species $X$ departs from equilibrium and freezes out is found by numerically solving the Boltzmann equation.  Introducing the variable $x \equiv m_X/T$, the temperature at which freeze-out occurs is approximately given by
\begin{equation}
x_{\rm FO} \equiv \frac{m_X}{T_{\rm FO}} \approx \ln\bigg[ c(c+2) \sqrt{\frac{45}{8}} \frac{g_X}{2\pi^3} \frac{m_X M_{\rm Pl} (a+6b/x_{\rm FO})}{g^{1/2}_{\star} x^{1/2}_{\rm FO}} \bigg]. 
\label{xfo}
\end{equation}
Here, $c \sim 0.5$ is a numerically determined quantity, $g_{\star}$ is the number of external degrees of freedom available (in the Standard Model, $g_{\star}\sim 120$ at $T\sim 1$ TeV and $g_{\star}\sim 65$ at $T\sim 1$ GeV), and $a$ and $b$ are terms in the non-relativistic expansion, $<\sigma_{X\bar{X}} |v|> = a + b <v^2> + \mathcal{O}(v^4)$. The resulting density of WIMPs remaining in the universe today is approximately given by
\begin{equation}
\Omega_X h^2 \approx \frac{1.04 \times 10^9 \, {\rm GeV}^{-1}}{M_{\rm Pl}}\frac{x_{\rm FO}}{g^{1/2}_{\star}(a + 3b/x_{\rm FO})}.
\label{omegah2}
\end{equation}
If $X$ has a GeV-TeV scale mass and a roughly weak-scale annihilation cross section, freeze-out occurs at $x_{\rm FO} \approx  20-30$, resulting in a relic abundance of
\begin{equation}
\Omega_X h^2 \approx 0.1 \, \bigg(\frac{x_{\rm FO}}{20}\bigg) \bigg(\frac{g_{\star}}{80}\bigg)^{-1/2} \bigg(\frac{a+3b/x_{\rm FO}}{3\times 10^{-26} {\rm cm}^3/{\rm s}}\bigg)^{-1}.
\end{equation}
In other words, if a GeV-TeV scale particle is to be thermally produced with an abundance similar to the measured density of dark matter, it must have a thermally averaged annihilation cross section on the order of $3 \times 10^{-26}$ cm$^3$/s.  Remarkably, this is very similar to the numerical value arrived at for a generic weak-scale interaction. In particular, $\alpha^2/(100\,{\rm GeV})^2 \sim  {\rm pb}$, which in our choice of units (and including a factor of velocity) is $\sim 3 \times 10^{-26}$ cm$^3$/s. The similarly between this result and the value required to generate the observed quantity of dark matter has been dubbed the ``WIMP miracle''. While far from constituting a proof, this argument has lead many to conclude that dark matter is likely to consist of particles with weak-scale masses and interactions and certainly provides us with motivation for exploring an electroweak origin of our universe's missing matter.

\subsection{Case Example -- The Thermal Abundance of a Light or Heavy neutrino}
\label{neutrinocase}

At first glance, the Standard Model itself appears to contain a plausible candidate for dark matter in the form of neutrinos. Being stable and weakly interacting, neutrinos are a natural place to start in our hunt for dark matter's identity. 
 
In the case of a Standard Model neutrino species, the relatively small annihilation cross section ($<\sigma |v|> \sim 10^{-32}$ cm$^3$/s) and light mass leads to an overall freeze-out temperature on the order of $T_{\rm FO} \sim {\rm MeV}$, and a relic density of 
\begin{equation}
\Omega_{\nu + \bar{\nu}} h^2 \approx 0.1 \,\bigg(\frac{m_{\nu}}{9 \, {\rm eV}}\bigg).
\end{equation}
As constraints on Standard Model neutrino masses require $m_{\nu}$ to be well below 9 eV, we are forced to conclude that only a small fraction of the dark matter could possibly consist of Standard Model neutrinos. Furthermore, even if these constraints did not exist, such light neutrinos would be highly relativistic at the time of freeze-out ($T_{\rm FO}/m_{\nu} \sim {\rm MeV}/m_{\nu} \gg 1$) and thus would constitute hot dark matter, in conflict with observations of large scale structure. 

Moving beyond the Standard Model, we could instead consider a heavy 4th generation Dirac neutrino. In this case, the annihilation cross section can be much larger, growing with the square of the neutrino's mass up to the $Z$ pole, $m_{\nu} \sim m_Z/2$, and declining  with $m^{-2}_{\nu}$ above $m_{\nu} \sim m_Z/2$. For a GeV-TeV mass neutrino, the process of freeze-out yields a cold relic ($T_{\rm FO}/m_{\nu} \sim \mathcal{O}(0.1)$), with an abundance approximately given by
\begin{eqnarray}
\Omega_{\nu + \bar{\nu}} h^2 &\approx& 0.1 \,\bigg(\frac{5.5 \, {\rm GeV}}{m_{\nu}}\bigg)^2, \,\,\,\,\,\,\,\,\,\, {\rm MeV} \ll m_{\nu} \ll m_Z/2 \\
\Omega_{\nu + \bar{\nu}} h^2 &\sim& 0.1 \,\bigg(\frac{m_{\nu}}{400 \, {\rm GeV}}\bigg)^2, \,\,\,\,\,\,\,\,\,\,\,\,\,\,\,\,\,\, m_{Z}/2 \ll m_{\nu}.
\end{eqnarray}
Thus, if the relic density of a heavy neutrino species is to constitute the bulk of the observed dark matter abundance, we find that it must have a mass of approximately 5 GeV~\cite{leeweinberg}, or several hundred GeV. The former case is excluded by LEP's measurement of the invisible width of the $Z$, however, which rules out the possibility of a fourth active neutrino species lighter than half of the $Z$ mass~\cite{pdg}. The later case, although consistent with the bounds of LEP and other accelerator experiments, is excluded by the limits placed by direct dark matter detection experiments (which I will discuss in Sec.~\ref{direct}).


\subsection{A Detour Into Coannihilations}

In some cases, particles other than the WIMP itself can play an important role in the freeze-out process~\cite{Griest:1990kh}. Before such a particle can significantly impact the relic density of a WIMP, however, it must first manage to be present at the temperature of freeze-out. 

The relative abundances of two species at freeze-out can be very roughly estimated by
\begin{equation}
\frac{n_Y}{n_X} \sim \frac{e^{-m_Y/T_{\rm FO}}}{e^{-m_X/T_{\rm FO}}}.
\end{equation}
Considering, for example, a particle with a mass twice that of the WIMP and a typical freeze-out temperature of $m_X/T_{\rm FO} \approx 20$, there will be only $\sim e^{-40}/e^{-20} \sim 10^{-9}$ $Y$ particles for every $X$ at freeze-out, thus making $Y$ completely irrelevant. If $m_Y$ were only 10\% larger than $m_X$, however, we estimate $n_Y/n_X \sim e^{-22}/e^{-20} \sim 10^{-1}$. In this quasi-degenerate case, the additional particle species can potentially have a significant impact on the dark matter relic abundance.

To quantitatively account for other species in the calculation of the relic abundance of a WIMP, we make the following substitution (for both $a$ and $b$) into Eqs.~\ref{xfo} and~\ref{omegah2}:
\begin{equation}
\sigma_{\rm Ann} \rightarrow \sigma_{\rm Eff}(x)=\sum_{i,j} \sigma_{i,j}\frac{g_i g_j}{g^2_{\rm Eff}(x)} (1+\Delta_i)^{3/2}  (1+\Delta_j)^{3/2} e^{-x (\Delta_i+\Delta_j)}, 
\end{equation}
where the double sum is over all particle species ($i,j$=1 denoting the WIMP itself) and $\sigma_{i,j}$ is the cross section for the coannihilation of species $i$ and $j$ (or self-annihilation in the case of $i=j$) into Standard Model particles. As the effective annihilation cross section has a strong dependence on $x$, we must integrate Eqs.~\ref{xfo} and~\ref{omegah2} over $x$ (or $T$). The quantities $\Delta_i=(m_i-m_1)/m_1$ denote the fractional mass splittings between the species $i$ and the WIMP. The effective number of degrees of freedom, $g_{\rm Eff}(x)$, is given by:
\begin{equation}
g_{\rm Eff}(x) = \sum_i g_i (1+\Delta_i)^{3/2} e^{-x \Delta_i}.
\end{equation}

To better understand how the introduction of particles other than the WIMP can effect the process of freeze-out, lets consider a few simple cases.  First, consider one additional particle with a mass only slightly above that of the WIMPs ($\Delta_2 \ll 1$), and with a comparatively large coannihilation cross section, such that $\sigma_{1,2} \gg \sigma_{1,1}$.  In this case, $g_{\rm Eff} \approx g_1 + g_2$, and $\sigma_{\rm Eff} \approx \sigma_{1,2}\, g_1 g_2/(g_1+g_2)^2$. Since $\sigma_{\rm Eff}$ is much larger than the WIMP's self-annihilation cross section, the relic density of WIMPs will be sharply suppressed. This is the case that is usually meant by the term ``coannihilation''.

Alternatively, consider the opposite case in which the WIMP and the additional quasi-degenerate particle do not coannihilate efficiently ($\sigma_{1,2} \ll \sigma_{1,1}, \sigma_{2,2}$). Here, $\sigma_{\rm Eff} \approx \sigma_{1,1} \, g^2_1/(g_1+g_2)^2 + \sigma_{2,2} \, g^2_2/(g_1+g_2)^2$, which in some cases can actually be {\it smaller} than that for the process of self-annihilation alone, leading to an {\it enhanced} relic abundance. Physically speaking, what is going on here is that the two species are each freezing out independently of each other, after which the heavier species decays, producing additional WIMPs as a byproduct.

As an extreme version of this second case, consider a scenario in which the lightest state is not a WIMP, but is instead a purely gravitationally interacting particle. A slightly heavier particle with weak interactions will self-annihilate much more efficiently than it will coannihilate with the lightest particle ($\sigma_{1,2}$ is negligible), leading the two states to freeze-out independently. The gravitationally interacting particle, however, never reaches thermal equilibrium, so could potentially have not been produced in any significant quantities up until this point.  Well after freezing out, the heavier particles will eventually decay, producing the stable gravitationally interacting lightest state. Although the resulting particles are not WIMPs (they do not have weak interactions), they are naturally produced with approximately the measured dark matter abundance because of the WIMP-like properties of the heavier state. In other words, this case -- known as the  ``superWIMP'' scenario~\cite{Feng:2003xh} -- makes use of the coincidence between the electroweak scale and the measured dark matter abundance without the dark matter actually consisting of WIMPs. Because gravitationally interacting particles and other much less than weakly interacting particles are almost impossible to detect astrophysically, superWIMPs are among the dark matter hunter's worst nightmares.

\section{Beyond The Standard Model Candidates For Dark Matter}
\label{candidates}

There has been no shortage of dark matter candidates proposed over the years. A huge variety of beyond the Standard Model physics models have been constructed which include a stable, electrically neutral, and colorless particle, many of which could serve as a phenomenologically viable candidate for dark matter. I could not possibly list, must less review, all of the proposed candidates here. Finding the ``WIMP miracle'' (as discussed in Sec.~\ref{relic}) to be fairly compelling (along with the hierarchy problem, which strongly suggests the existence of new particles at or around the electroweak scale), I chose to focus my attention on dark matter in the form of weak-scale particles. So although the dark matter of our universe could plausibly consist of particles ranging from $10^{-6}$ eV axions to $10^{16}$ GeV WIMPzillas, I will ignore everything but those particle physics frameworks which predict the existence of a stable, weakly interacting particle with a mass in the few GeV to few TeV range.

\subsection{Supersymmetry}

For a number of reasons, supersymmetry is considered by many to be among the most
attractive extensions of the Standard Model. In particular, weak-scale
supersymmetry provides us with an elegant solution to the hierarchy
problem~\cite{susyreview}, and enables grand unification by causing the gauge
couplings of the Standard Model to evolve to a common scale~\cite{gut}. From
the standpoint of dark matter, the lightest superpartner is naturally stable in models that conserve $R$-parity. $R$-parity is defined as $R=(-1)^{3B+L+2S}$ ($B$, $L$ and $S$ denoting baryon number, lepton number and spin), and thus is assigned as $R=+1$ for all Standard Model particles and $R=-1$ for all superpartners. $R$-parity conservation, therefore, requires superpartners to be created or destroyed in pairs, leading the lightest supersymmetric particle (LSP) to be stable, even over cosmological timescales.

The identity of the LSP depends on the hierarchy of the supersymmetric spectrum, which in turn is determined by the details of how supersymmetry is broken. The list of potential LSPs which could constitute a plausible dark matter candidate is somewhat short, however. The only electrically neutral and colorless superparnters in the minimal supersymmetric standard model (MSSM) are the four neutralinos (superpartners of the neutral gauge and Higgs bosons), three sneutrinos, and the gravitino. The lightest neutralino, in particular, is a very attractive and throughly studied candidate for dark matter~\cite{neutralinodm}.

Before discussing the phenomenology of neutralino dark matter, lets briefly contemplate the possibility that sneutrinos might make up the dark matter of our universe. In many respects, sneutrino dark matter would behave very similarly to a heavy 4th generation neutrino, as discussed in Sec.~\ref{neutrinocase}. In particular, like neutrinos, sneutrinos are predicted to annihilate to Standard Model fermions efficiently through the $s$-channel exchange of a $Z$ boson (as well as through other diagrams). As a result, sneutrinos lighter than about 500-1000 GeV would be under produced in the early universe (a $\sim$$10 \,{\rm GeV}$ sneutrino would also be produced with approximately the measured dark matter abundance, but is ruled out by LEP's invisible $Z$ measurement).

The Feynman diagram corresponding to sneutrino annihilation into quarks through a $s$-channel $Z$ exchange can be turned on its side to produce an elastic scattering diagram with quarks in nuclei. When the elastic scattering cross section of a $\sim$100-1000 GeV sneutrino is calculated, we find that it is several orders of magnitude larger than current experimental constraints~\cite{falksneutrino}. We are thus forced to abandon MSSM sneutrinos as candidates for dark matter.

In the MSSM, the superpartners of the four Standard Model neutral bosons (the bino, wino and two neutral higgsinos) mix into four physical states known as neutralinos. Often times, the lightest of these four states is simply referred to as ``the neutralino''. The neutralino mass matrix can be used to determine the masses and mixings of these four states. In the $\widetilde{B}$-$\widetilde{W}^3$-$\widetilde{H}_1$-$\widetilde{H}_2$ basis, this matrix is given by
\begin{equation}
{\mathcal M}_{\chi^0}= \,\,\,\,\,\,\,\,\,\,\,\,\,\,\, \,\,\,\,\,\,\,\,\,\,\,\,\,\,\, \,\,\,\,\,\,\,\,\,\,\,\,\,\,\, \,\,\,\,\,\,\,\,\,\,\,\,\,\,\, \,\,\,\,\,\,\,\,\,\,\,\,\,\,\, \,\,\,\,\,\,\,\,\,\,\,\,\,\,\, \,\,\,\,\,\,\,\,\,\,\,\,\,\,\, \,\,\,\,\,\,\,\,\,\,\,\,\,\,\, \,\,\,\,\,\,\,\,\,\,\,\,\,\,\, \,\,\,\,\,\,\,\,\,\,\,\,\,\,\,
\end{equation}
\begin{eqnarray}
\arraycolsep=0.01in
\left( \begin{array}{cccc}
M_1 & 0 & -m_Z\cos \beta \sin \theta_W^{} & m_Z\sin \beta \sin \theta_W^{}
\\
0 & M_2 & m_Z\cos \beta \cos \theta_W^{} & -m_Z\sin \beta \cos \theta_W^{}
\\
-m_Z\cos \beta \sin \theta_W^{} & m_Z\cos \beta \cos \theta_W^{} & 0 & -\mu
\\
m_Z\sin \beta \sin \theta_W^{} & -m_Z\sin \beta \cos \theta_W^{} & -\mu & 0
\end{array} \right)\nonumber \;, 
\end{eqnarray}
where $M_1$ and $M_2$ are the bino and wino masses, $\mu$ is the higgsino mass parameter, $\theta_W$ is the Weinberg angle, and $\tan \beta \equiv \upsilon_2/\upsilon_1$ is the ratio of the vacuum expectation values of the Higgs doublets. This matrix can be diagonalized into mass eigenstates by the unitary matrix $N$,
\begin{equation}
{\mathcal M}_{\chi^0}^{\rm{diag}} = N^*  M_{\chi^0} N^{-1}.
\end{equation}
In terms of the elements of the matrix, $N$, the lightest neutralino is given by the following mixture of gaugino and higgsino components:
\begin{equation}
\chi^0 = N_{11}\tilde{B}     +N_{12} \tilde{W}^3
          +N_{13}\tilde{H}_1 +N_{14} \tilde{H}_2.
\label{eq3}
\end{equation}
The quantities $|N_{11}|^2+|N_{12}|^2$ and  $|N_{13}|^2+|N_{14}|^2$ are often referred to as the gaugino fraction and higgsino fraction of the lightest neutralino, respectively.

\begin{figure}[t]
\begin{center}
$\begin{array}{c@{\hspace{0.5in}}c}
\includegraphics[width=0.6\textwidth,clip=true]{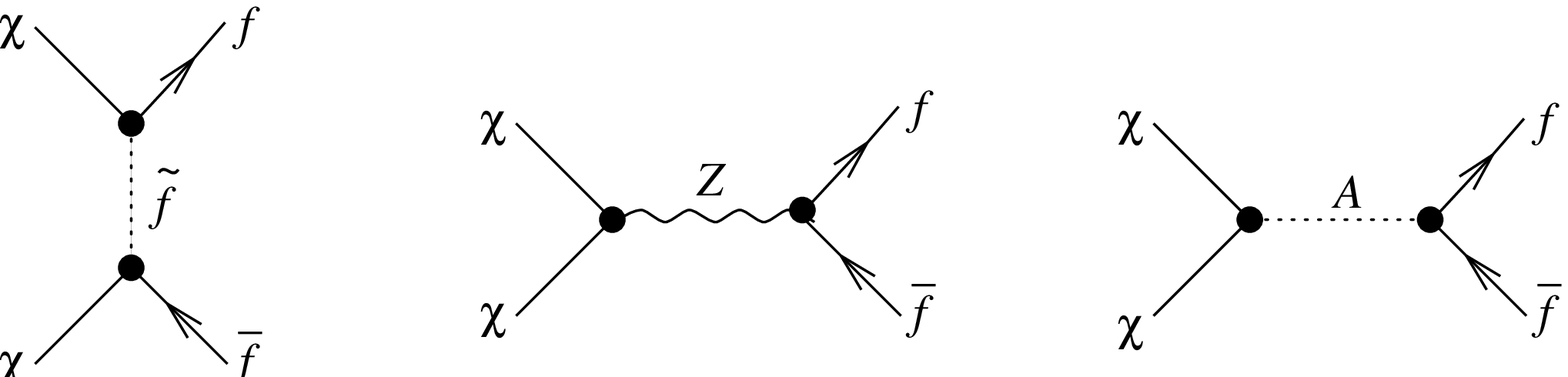}\hspace{-6.5cm} & \\ [0.4cm]
\includegraphics[width=0.4\textwidth]{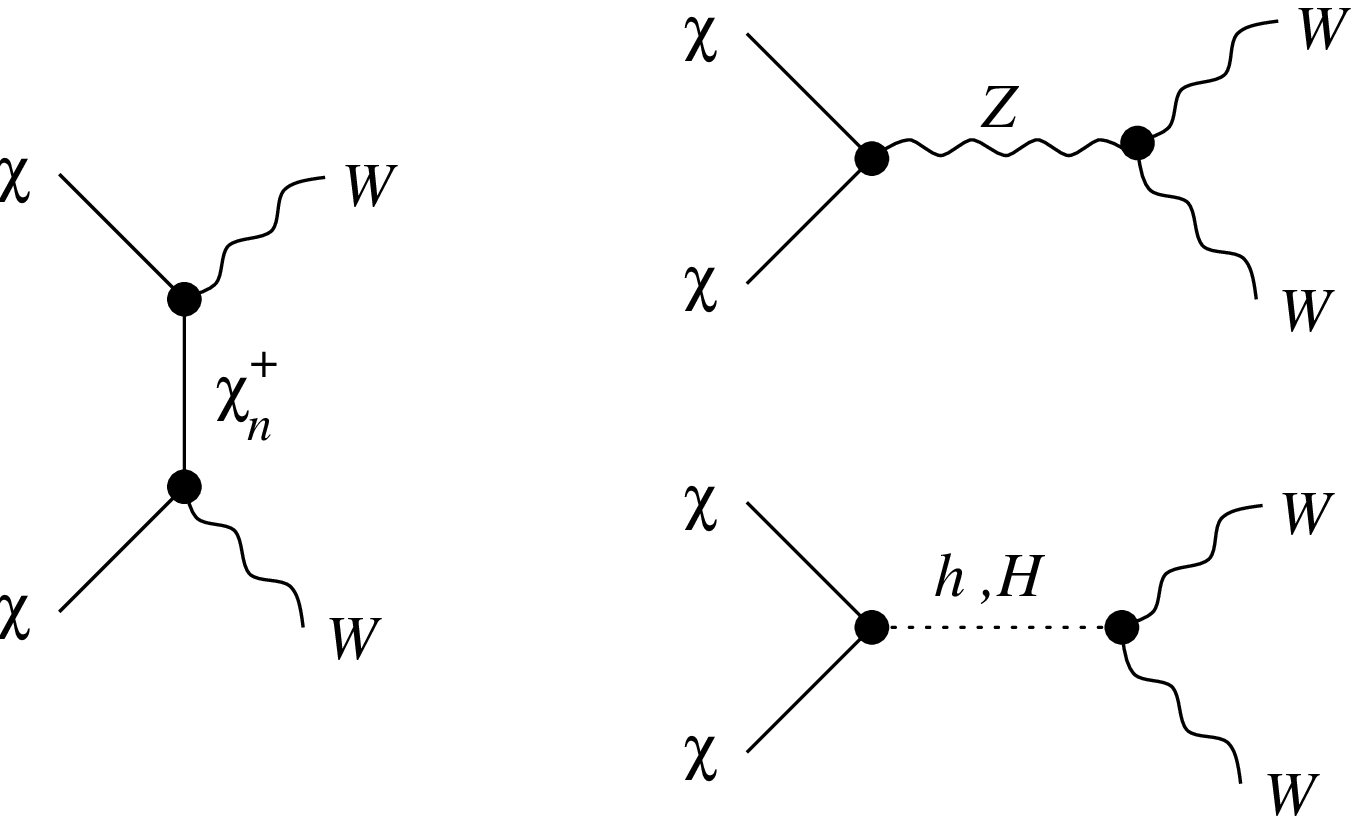} &
\includegraphics[width=0.4\textwidth]{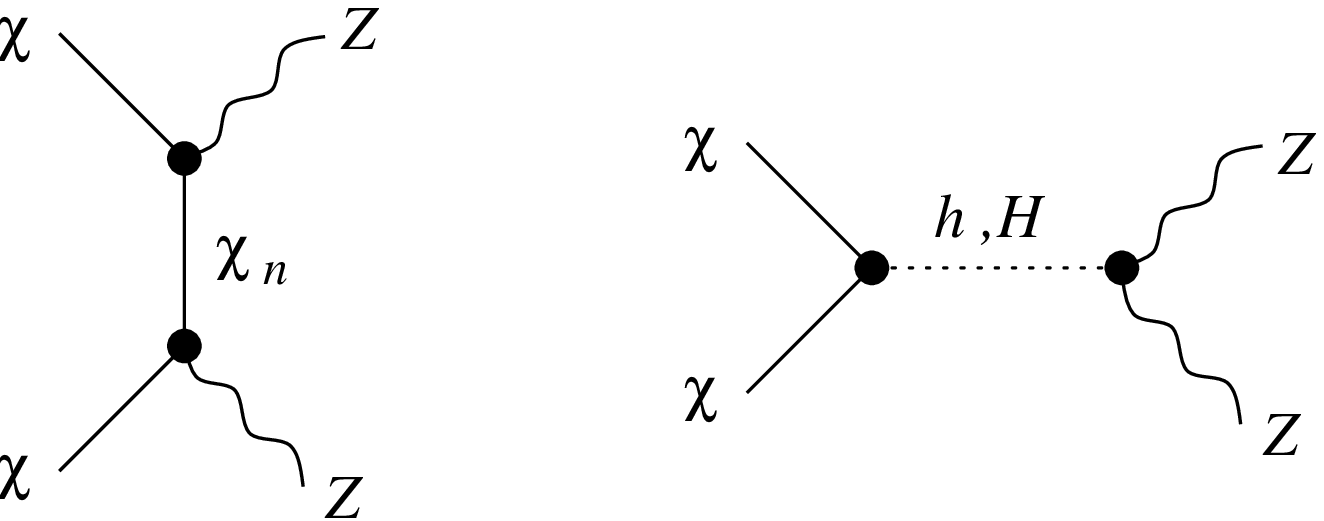} \\ [0.4cm]
\includegraphics[width=0.4\textwidth]{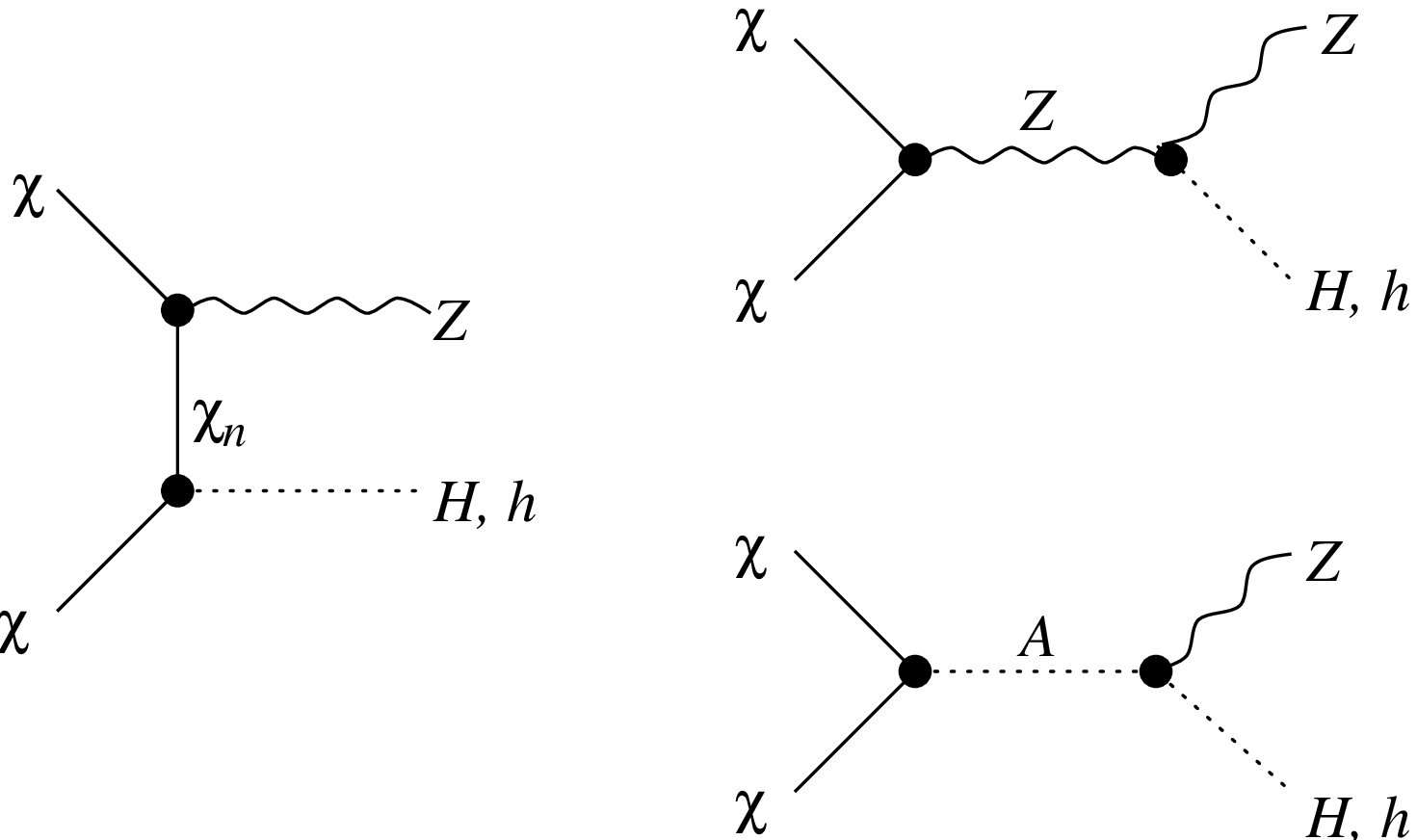} &
\includegraphics[width=0.4\textwidth]{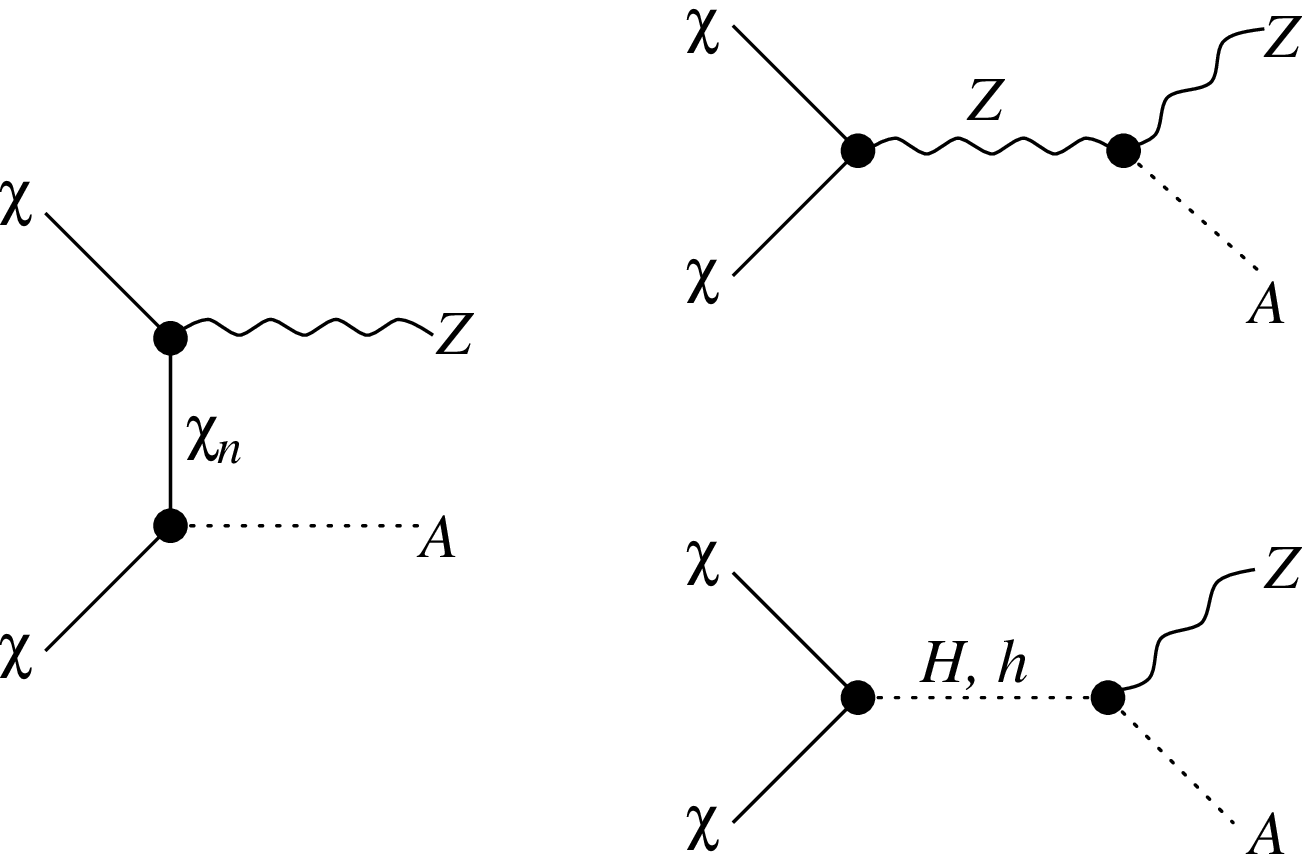} \\ [0.4cm]
\includegraphics[width=0.4\textwidth]{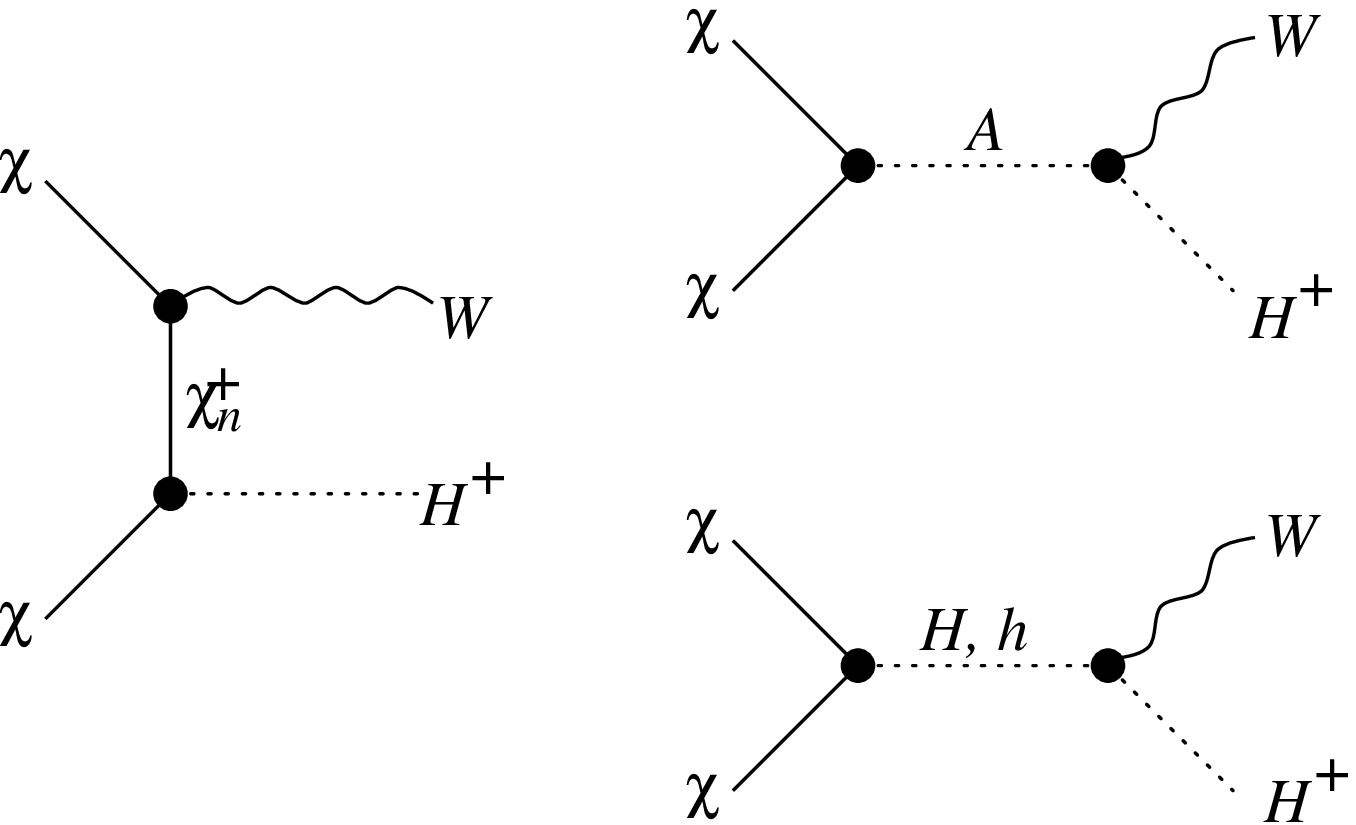} &
\includegraphics[width=0.4\textwidth]{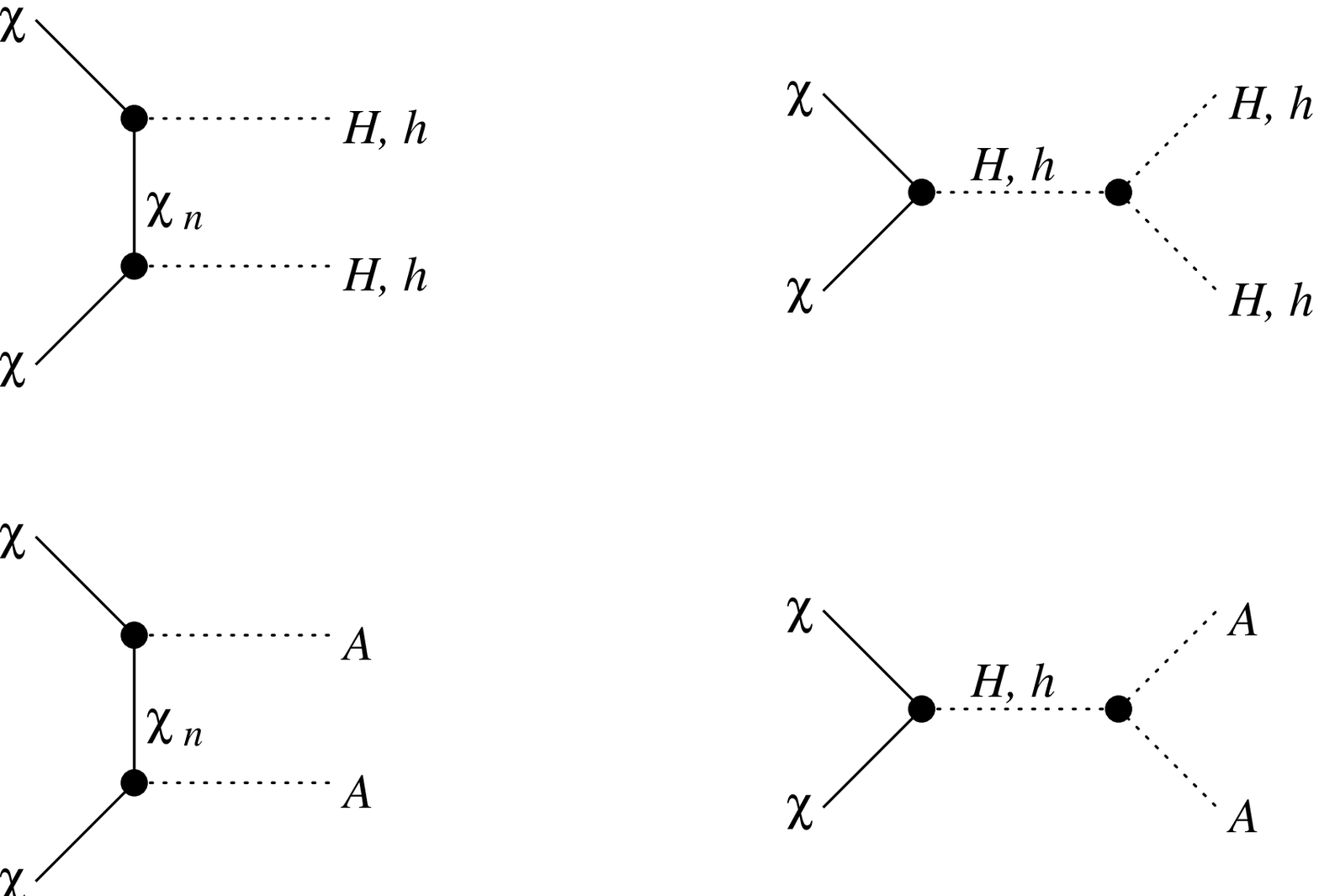} \\ [0.4cm]
\includegraphics[width=0.45\textwidth]{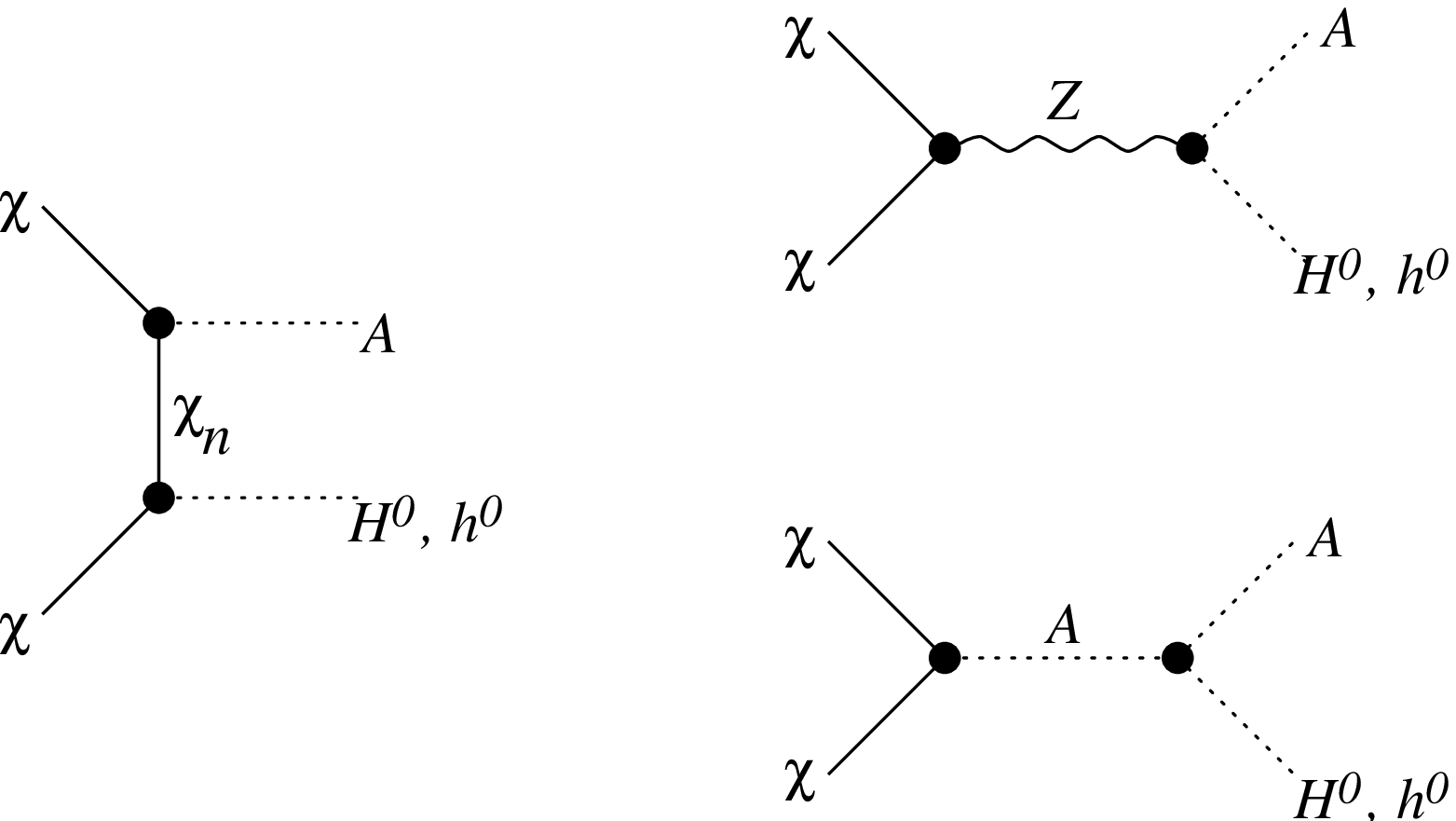} \hspace{-6.5cm} & 
\end{array}$
\end{center}
\caption[]
{Some of the most important Feynman diagrams for neutralino annihilation.}
\label{neutralinoannihilation}
\end{figure}

The lightest neutralino can annihilate through a wide variety of Feynman diagrams. In Fig.~\ref{neutralinoannihilation}, we show some of the most important of these; although it is far from an inclusive list. Which of these diagrams dominate the process of thermal freeze-out in the early universe depends on the composition of the lightest neutralino, and on the masses and mixings of the exchanged particles~\cite{jungman}. Since so many different diagrams can potentially contribute to neutralino annihilation (not to mention the many possible coannihilation processes~\cite{Edsjo:1997bg,Ellis:1999mm,Ellis:2001nx,Edsjo:2003us}), the resulting relic density depends on a large number of supersymmetric parameters and is not trivial to calculate accurately. Publicly available tools such as DarkSUSY~\cite{darksusy} and MicroOmegas~\cite{micromegas} are often used for this purpose.

The mass and composition of the lightest neutralino is a function of four supersymmetric parameters: $M_1$, $M_2$, $\mu$ and $\tan \beta$. This becomes further simplified if the gaugino masses are assumed to evolve to a single value at the GUT scale, yielding a ratio at the electroweak scale of $M_1 = \frac{5}{3} \tan^2\theta_W M_2 \approx 0.5 M_2 $. In this case, the lightest neutralino has only a small wino fraction and is largely bino-like (higgsino-like) for $M_1 \ll |\mu|$ ($M_1 \gg |\mu|$).

Over much or most of the supersymmetric parameter space, the relic abundance of neutralinos is predicted to be in excess of the observed dark matter density. To avoid this, we are forced to consider the regions of parameter space which lead to especially efficient neutralino annihilation in the early universe. In particular, the following scenarios are among those which can lead to a phenomenologically viable density of neutralino dark matter:
\begin{itemize}
\item{If the lightest neutralino has a significant higgsino or wino fraction, it can have fairly large couplings and, as a result, annihilate very efficiently.}
\item{If the mass of the lightest neutralino is near a resonance, such as the CP-odd Higgs pole, it can annihilate efficiently, even with relatively small couplings.}
\item{If the lightest neutralino is only slightly lighter than another superpartner, such as the lightest stau, coannihilations between these two states can very efficiently deplete the dark matter abundance.}
\end{itemize}

\begin{figure}[t]
\centering\leavevmode
\includegraphics[width=2.0in,angle=-90]{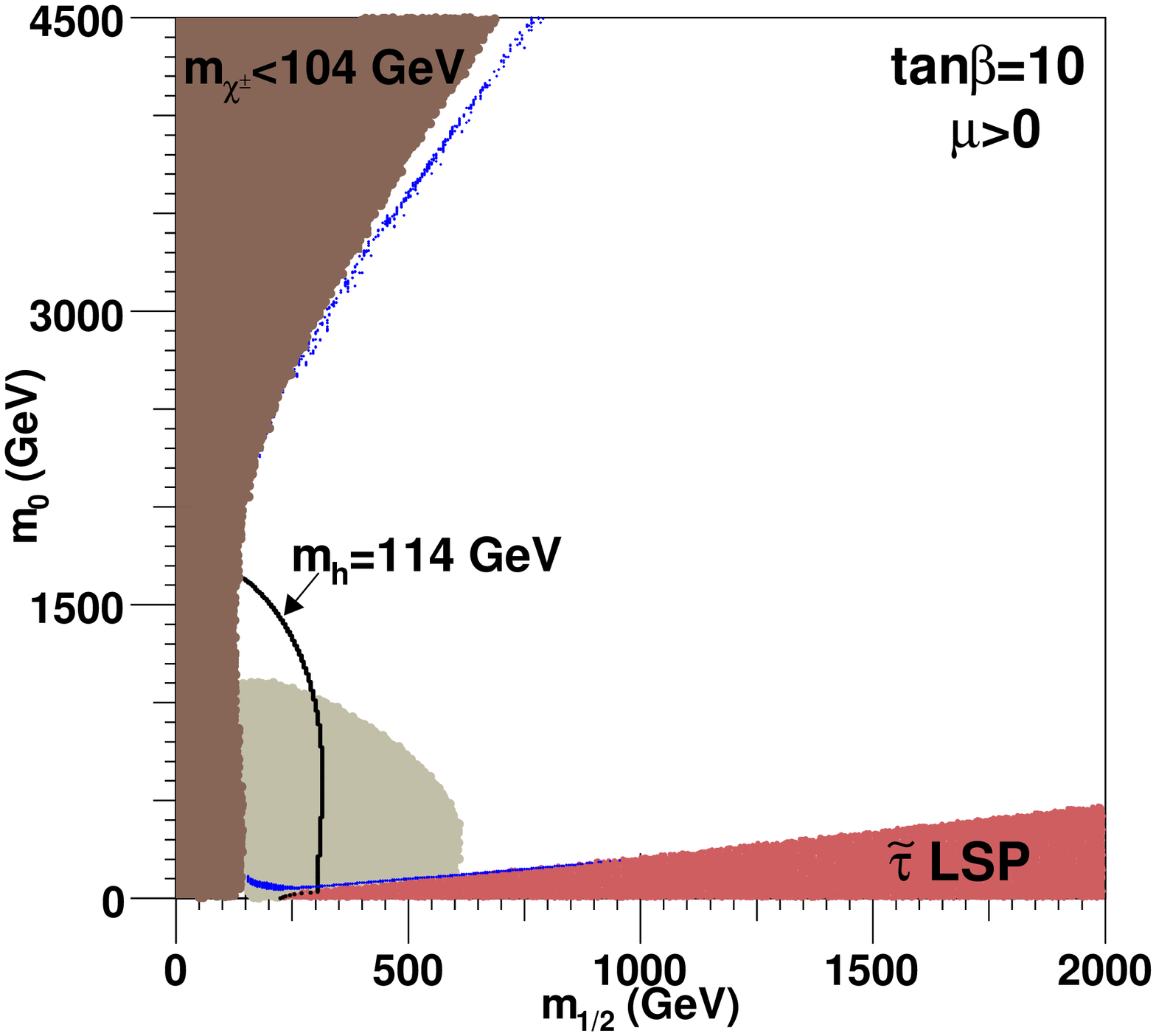}
\includegraphics[width=2.0in,angle=-90]{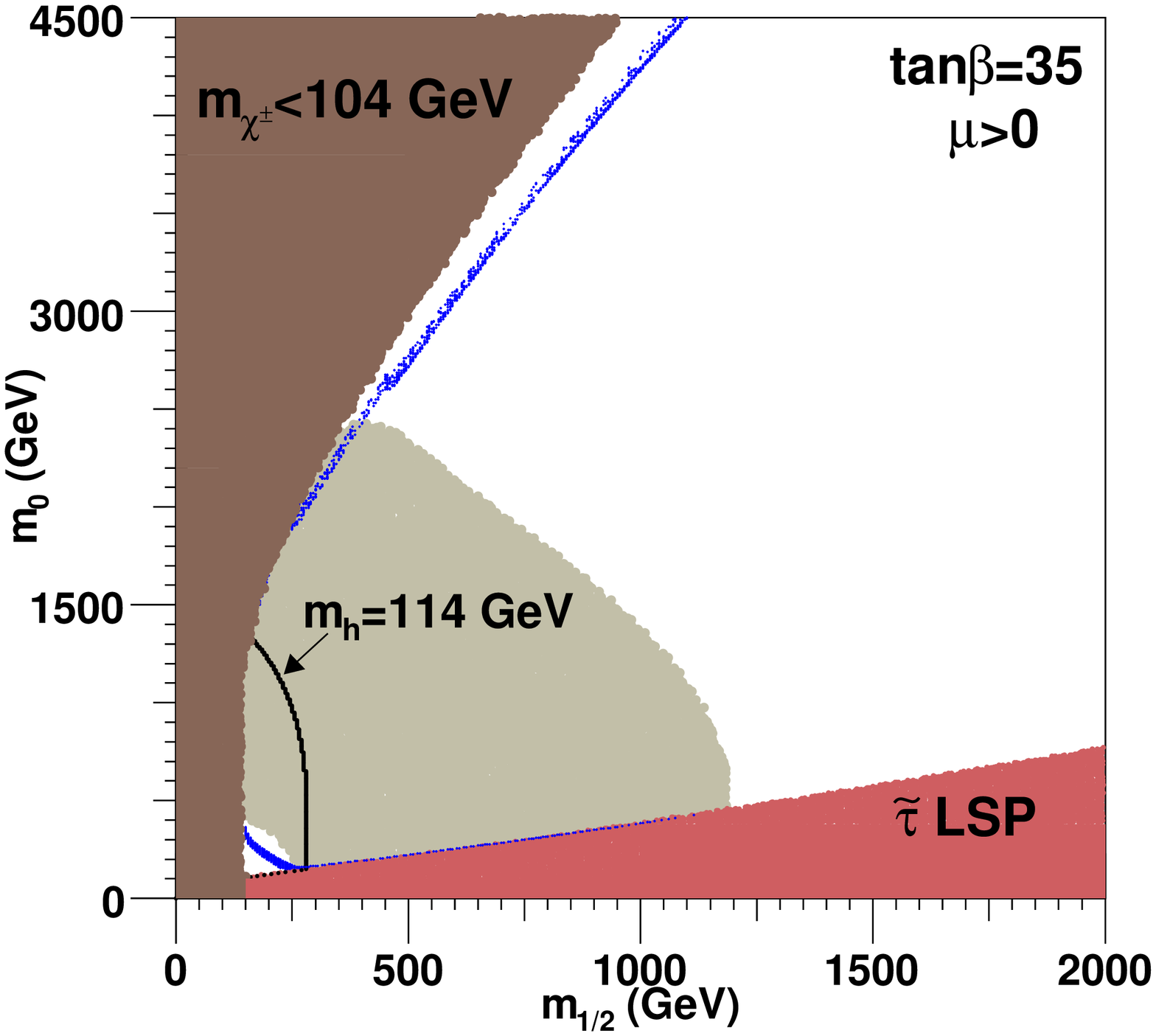}\\
\vspace{0.7cm}
\includegraphics[width=2.0in,angle=-90]{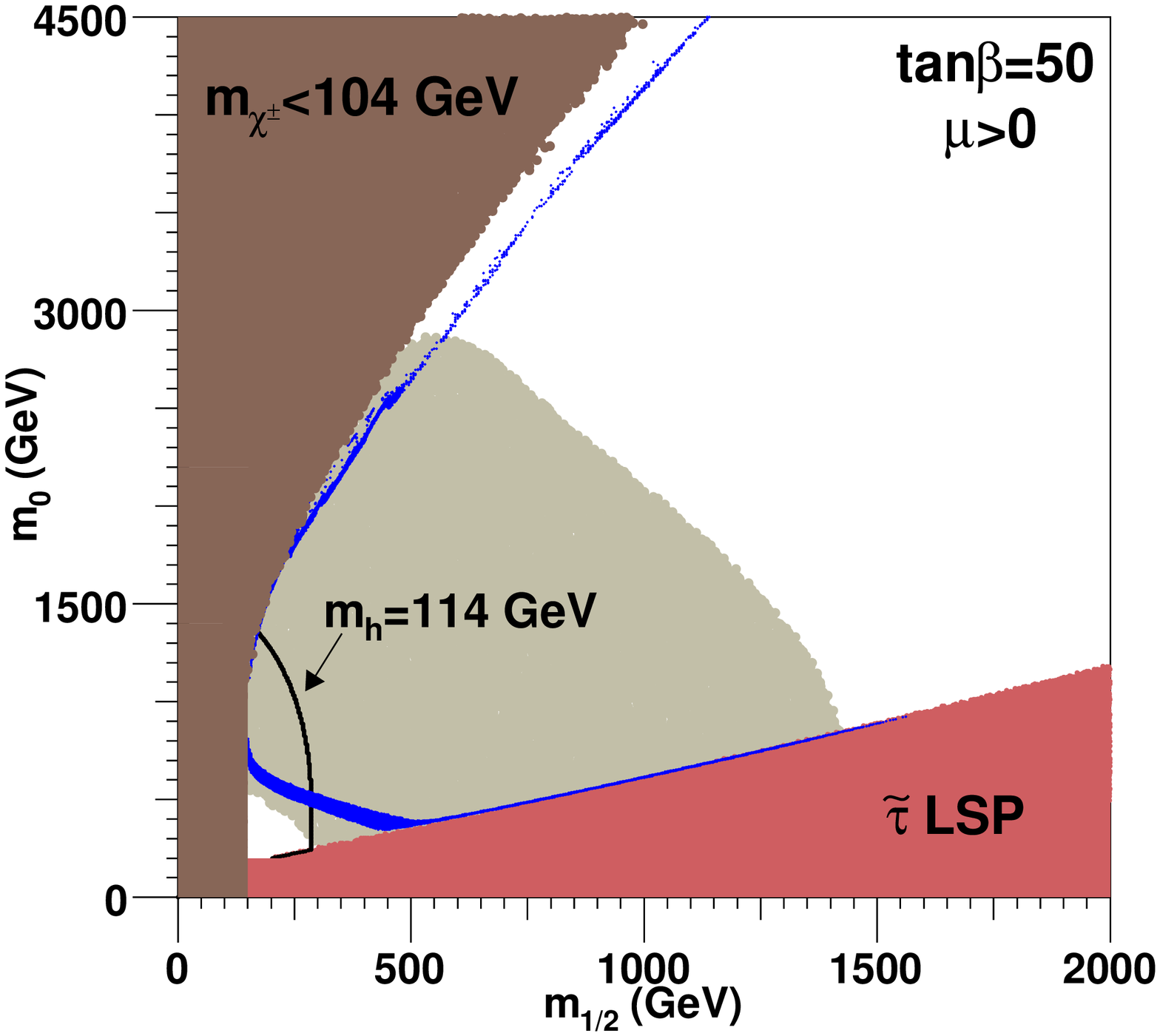}
\includegraphics[width=2.0in,angle=-90]{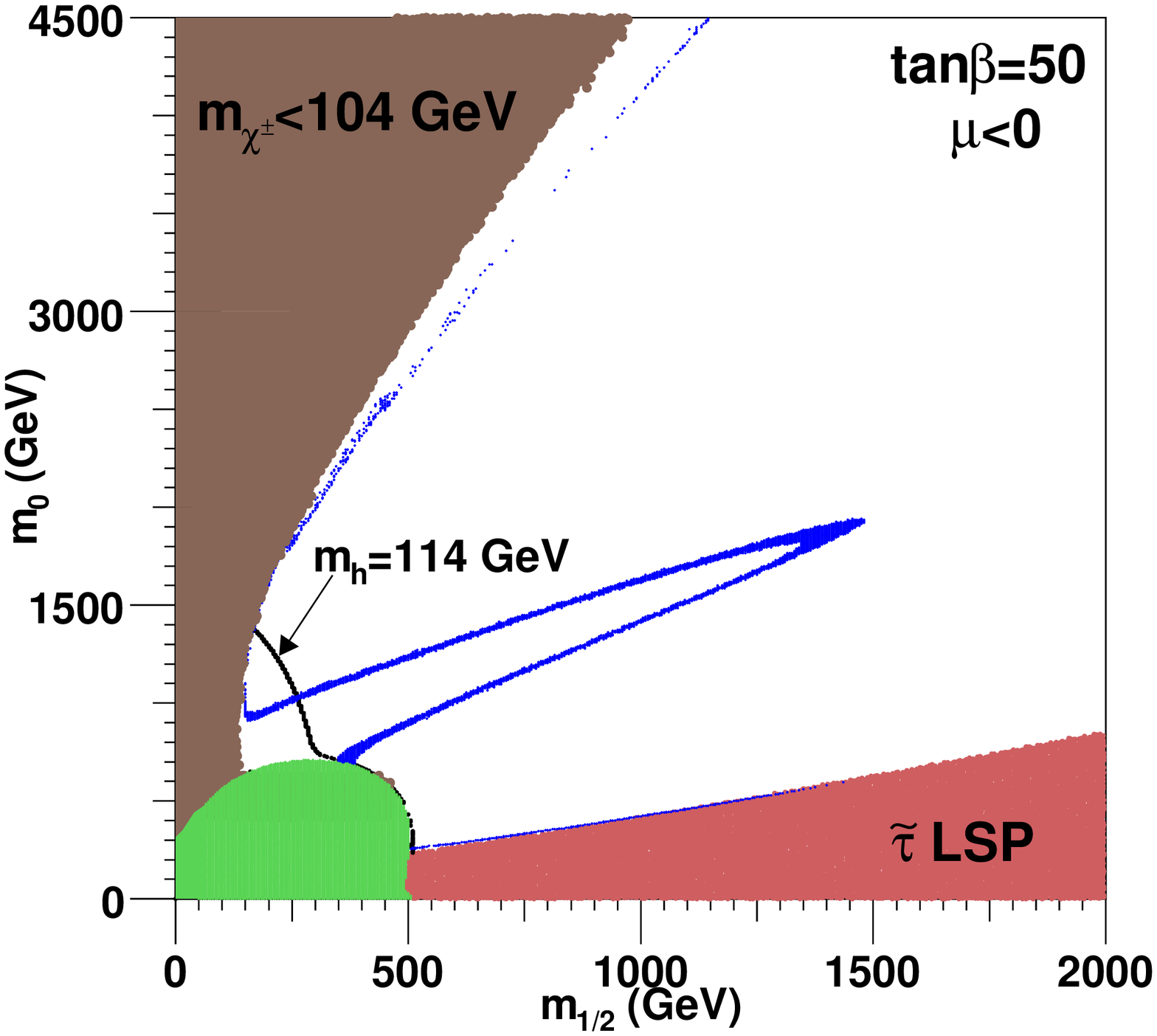}
\caption{Representative regions of the CMSSM parameter space. The blue regions predict a neutralino density consistent with the measured dark matter abundance. The shaded regions to the upper left and lower right are disfavored by the LEP chargino bound and as a result of containing a stau LSP, respectively. The LEP bound on the light Higgs mass is shown as a solid line ($m_h=114$ GeV). The region favored by measurements of the muon's magnetic moment are shown as a light shaded region (at the 3$\sigma$ confidence level)~\cite{gm2SM}. In each frame, we have used $A_0=0$ and $\mu > 0$.  (Figures generated by Gabriel Caceres, using the package DarkSUSY~\cite{darksusy}.)}
\label{mzeromhalf}
\end{figure}

In Fig.~\ref{mzeromhalf}, we illustrate these regions within the context of a specific subset of the MSSM known as the CMSSM (C stands for constrained). In this framework, all of the scalar masses are set to a common value $m_0$ at the GUT scale, from which the electroweak scale values are determined by RGE evolution. Similarly the three gaugino masses are each set to $m_{1/2}$ at the GUT scale. In each frame, the narrow blue regions denote the parameter space in which neutralino dark matter is predicted to be generated with the desired abundance ($0.0913 < \Omega_{\chi^0} h^2 < 0.1285$). In the corridor along side of the LEP chargino bound ($m_{\chi^{\pm}} > 104$ GeV), $\mu$ and $M_1$ are comparable in magnitude, leading to a mixed bino-higgsino LSP with large couplings. Within the context of the CMSSM, this is often called the ``focus point'' region. In the bottom portion of each frame, the lightest stau ($\tilde{\tau}$) is the LSP, and thus does not provide a viable dark matter candidate. Just outside of this region, however, the stau is slightly heavier than the lightest neutralino, leading to a neutralino LSP which efficiently coannihilates with the nearly degenerate stau.  In the lower right frame, a viable region also appears along the CP-odd Higgs resonance ($m_{\chi^0} \approx m_A/2$). This is often called the $A$-funnel region.

\subsection{Kaluza-Klein Dark Matter in Models With Universal Extra Dimensions}
\label{kkdm}

Supersymmetry is not by any means the only particle physics framework from which viable a dark matter candidate can arise.  As an alternative, I will discuss in this section the possibility of Kaluza-Klein dark matter in models with a universal extra dimension. 

In recent years, interest in theories with extra spatial dimensions has surged. In particular, a colossal amount of attention has been given to two classes of extra dimensional theories over the past decade: scenarios featuring one or more large (millimeter-scale), flat extra dimensions (ADD)~\cite{Arkani-Hamed:1998rs,Arkani-Hamed:1998nn}, and the Randall-Sundrum scenario, which introduces an additional small dimension with a large degree of spatial curvature~\cite{Randall:1999ee}. A somewhat less studied but interesting class of extra-dimensional models, which goes by the name of universal extra dimensions (UED), postulates the existence of a flat extra dimension (or dimensions) through which all of the Standard Model (SM) fields are free to propagate (rather than being confined to a brane as some or all of them are in the ADD and Randall-Sundrum models)~\cite{Appelquist:2000nn}.

For a number of reasons, extra dimensions of size $R \sim \rm{TeV}^{-1}$ are particularly well motivated within the context of UED~\cite{uedreview} (much smaller than those found in the ADD model). Among these reasons is the fact that a TeV-scale Kaluza-Klein (KK) state, if stable, colorless, and electrically neutral, could potentially serve as a viable candidate for dark matter~\cite{Servant:2002aq,Cheng:2002ej}.

Standard Model fields with momentum in an extra dimension appear as heavy particles, called KK states. This leads to a tower of KK states for each Standard Model field, with tree-level masses given by:
\begin{equation}
m^2_{X^{(n)}} = \frac{n^2}{R^2} + m^2_{X^{(0)}},
\label{tree}
\end{equation}
where $X^{(n)}$ is the $n$th Kaluza-Klein excitation of the Standard Model field, $X$, and $R\sim \rm{TeV}^{-1}$ is the size of the extra dimension. $X^{(0)}$ denotes the ordinary Standard Model particle (known as the zero mode).

If the extra dimensions were simply wrapped (compactified) around a circle or torus, then extra dimensional momentum conservation would ensure the conservation of KK number ($n$) and make the lightest first level KK state stable. Realistic models, however, require an orbifold to be introduced, which leads to the violation of KK number conservation. A remnant of KK number conservation called KK-parity, however, can remain and lead to the stability of the lightest KK particle (LKP) in much the same way that $R$-parity conservation prohibits the decay of the lightest supersymmetric particle.

In order for the LKP to be a viable dark matter candidate, it must be electrically neutral and colorless. Possibilities for such a state include the first KK excitation of the photon, $Z$, neutrinos, Higgs boson, or graviton. Assuming that $R^{-1}$ is considerably larger than any of the Standard Model zero mode masses, Eq.~\ref{tree} leads us to expect a highly degenerate spectrum of Kaluza-Klein states at each level (although this picture is somewhat modified when radiative corrections and boundary terms are included). Of our possible choices for the LKP, the relatively large zero-mode mass of the Higgs make its first level KK excitation an unlikely candidate. Furthermore, KK neutrinos are excluded by direct detection experiments, just as sneutrinos or heavy 4th generation Dirac neutrinos are. For these reasons, we focus on the mixtures of the KK photon and KK $Z$ as our dark matter candidate (note that, unlike higgsino and gauginos, the KK Higgs has a different spin than the KK photon and KK $Z$, and thus does not mix with these states).

The mass eigenstates of the KK photon and KK $Z$ are very nearly identical to their gauge eigenstates,  $B^{(n)}$ and $W^{3(n)}$.  The reason for this can be seen from their mass matrix:
\begin{equation}
\left( \begin{array}{cc} 
\frac{n^2}{R^2} + \delta m^2_{B^{(n)}} + \frac{1}{4} g_1^2 v^2 
  & \frac{1}{4} g_1 g_2 v^2 \\
\frac{1}{4} g_1 g_2 v^2 
  & \frac{n^2}{R^2} + \delta m^2_{W^{(n)}} + \frac{1}{4} g_2^2 v^2 
\end{array} \right).
\end{equation}
Here $v\approx 174$ GeV is the Higgs vacuum expectation value. In the well known zero-mode case ($n=0$), there is significant mixing between $B^{(0)}$ and $W^{3(0)}$ ($\sin^2\theta_W \approx 0.23$). In the absence of radiative corrections ($\delta m^2_{B^{(n)}}=\delta m^2_{W^{(n)}}=0$), the same mixing angle is found at the first KK level as well. If the difference between $\delta m^2_{B^{(n)}}$ and $\delta m^2_{W^{(n)}}$ is larger than the (rather small) off diagonal terms, however, the mixing angle between these two KK states is driven toward zero. Using typical estimates of these radiative corrections, the effective first KK level Weinberg angle is found to be approximately $\sin^2 \theta_{W,1} \sim 10^{-3}$. Thus the mass eigenstate often called the ``KK photon'' is not particularly photon-like, but instead is nearly identical to the state $B^{(1)}$.

The KK state $B^{(1)}$ annihilates largely to Standard Model (zero-mode) fermions through the $t$-channel exchange of KK fermions, with a cross section given by 
\begin{equation}
\sigma v (B^{(1)}B^{(1)} \rightarrow f \bar{f})=  \frac{95}{32,256} \sum_f \frac{N_c (Y^4_{f_L}+Y^4_{f_R}) g_1^4}{\pi \, m^2_{B^{(1)}}}.  
\end{equation}
As the $B^{(1)}$ couples to the fermions' hypercharge, the cross section scales with $Y^4_{f_L}+Y^4_{f_R}$ and most of its annihilations proceed to charged lepton pairs. 

\begin{figure}[t]
\centerline{\includegraphics[width=0.7\hsize]{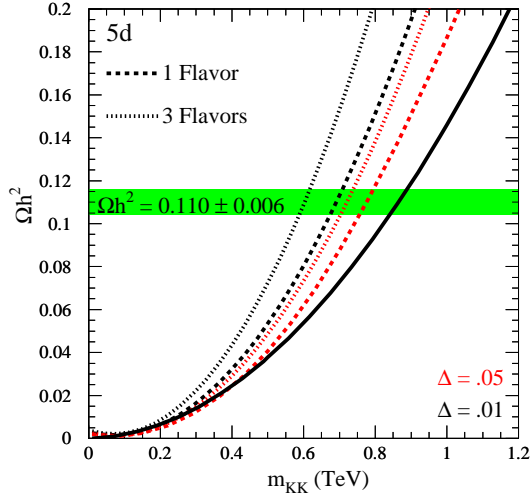}}
\caption{The thermal relic abundance of KK dark matter ($B^{(1)}$) without the effects of any other KK species (solid line) and including the effects of KK leptons 5\% and 1\% heavier than the LKP (dashed and dotted lines). Shown as a horizontal band is the measured dark matter abundance~\cite{wmap}. Adapted from Ref.~\cite{Servant:2002aq}.}
\label{kkrelic}
\end{figure}

If Fig.~\ref{kkrelic}, the thermal relic abundance of KK dark matter is shown as a function of its mass. Because the first level KK spectrum is expected to be quasi-degenerate, coannihilations are likely to play an important role. In the figure, results are shown ignoring the effects of other KK states (solid line) and including the effects of KK leptons 5\% or 1\% heavier than the LKP (dashed and dotted lines, respectively). Note that the KK leptons lead to a larger relic abundance, due to the fact that they freeze-out quasi-independently from the LKP and then increase the number of LKPs through their decays. Depending on the details of the KK spectrum, many other states could potentially effect the relic density of KK dark matter as well.  LKP masses from approximately 500 GeV to several TeV can potentially lead to a relic density consistent with the measured dark matter abundance~\cite{Burnell:2005hm,Kong:2005hn}.

\subsection{A Note On Other Possibilities For TeV-Scale Dark Matter}

At this time, I would like to make a general comment about some of the many other possibilities for the particle identity of dark matter. It is interesting to note that a wide range of solutions to the gauge hierarchy problem also introduce a candidate for dark matter. In particular, in order to be consistent with electroweak precision measurements, new physics at the TeV scale typically must possess a discrete symmetry. Supersymmetry, for example, accomplishes this through $R$-parity conservation.  Similarly, Little Higgs models can avoid problems with electroweak precision tests by introducing a discrete symmetry called $T$-parity. And just as $R$-parity stabilizes the lightest superpartner, $T$-parity can enable the possibility of dark matter by stabilizing the lightest $T$-parity odd state in the theory~\cite{tparity}. So, thinking outside of any specific particle physics framework, we can imagine a much larger class of TeV-scale physics scenarios which address the hierarchy problem without violating electroweak precision measurements by introducing a discrete symmetry, which in turn leads to a stable dark matter candidate. In this way, many of the motivations and much of the phenomenology discussed within the context of supersymmetry can actually be applied to a more general collection of particle physics models associated with the electroweak scale.

\section{Direct Detection}
\label{direct}

Turning our attention now to dark matter detection, we begin with those experiments which attempt to detect dark matter particles through their elastic scattering with nuclei, including CDMS~\cite{cdms}, XENON~\cite{xenon}, ZEPLIN~\cite{zeplin}, EDELWEISS~\cite{edelweiss}, CRESST~\cite{cresst}, CoGeNT~\cite{cogent}, DAMA/LIBRA~\cite{dama}, COUPP~\cite{coupp}, WARP~\cite{warp}, and KIMS~\cite{kims}. This class of techniques is collectively known as direct detection, in contrast to indirect detection efforts which attempt to observe the annihilation products of dark matter particles.

A WIMP striking a nucleus will induce a recoil of energy given by
\begin{equation}
E_{\rm recoil} = \frac{|\vec{q}|^2}{2 M_{\rm nucleus}}=\frac{2 \mu^2 v^2 (1-\cos\theta)}{2 M_{\rm nucleus}}=\frac{m^2_{X} M_{\rm nucleus} \, v^2 (1-\cos\theta)}{(m_X+M_{\rm nucleus})^2},
\end{equation}
where $\vec{q}$ is the WIMP's momentum, $v$ is its velocity, and $\mu$ is the reduced mass. For $m_X \gg M_{\rm nucleus}$ and a velocity of $\sim$300 km/s, we expect typical recoil energies of $E_{\rm recoil} \sim M_{\rm nucleus} \, v^2 \sim 1$-$100$ keV.

WIMPs scatter with nuclei in a target at a rate given by
\begin{equation}
R \approx \int^{E_{\rm max}}_{E_{\rm min}} \int^{v_{\rm max}}_{v_{\rm min}} \frac{2 \rho}{m_X} \, \frac{d \sigma}{d |\vec{q}|} \, v \, f(v) \, dv \, dE_{\rm recoil}, 
\label{directrate}
\end{equation}
where $\rho$ is the dark matter density, $\sigma$ is the WIMP-nuclei elastic scattering cross section, and $f(v)$ is the velocity distribution of WIMPs. The limits of integration are set by the galactic escape velocity, $v_{\rm max}\approx 650$ km/s, and kinematically by $v_{\rm min} = (E_{\rm recoil}\, M_{\rm nucleus}/2 \mu^2)^{1/2}$. The minimum energy is set by the energy threshold of the detector, which is typically in the range of several keV to several tens of keV.

WIMPs can potentially scatter with nuclei through both spin-independent and spin-dependent interactions. The experimental sensitivity to spin-independent couplings benefits from coherent scattering, which leads to cross sections (and rates) proportional to the square of the atomic mass of the target nuclei. The cross sections for spin-dependent scattering, in contrast, are proportional to $J(J+1)$, where $J$ is the spin of the target nucleus, and thus do not benefit from large target nuclei. As a result, the current experimental sensitivity to spin-dependent scattering is far below that of spin-independent interactions. For this reason, we consider first the case of spin-independent scattering of WIMPs with nuclei (we will return to spin-dependent scattering in Sec.~\ref{neutrinotelescopes}).

The spin-independent WIMP-nucleus elastic scattering cross section is given by
\begin{equation}
\label{sig}
\sigma \approx \frac{4 m^2_{\chi^0} m^2_{\rm nucleus}}{\pi (m_{\chi^0}+m_{\rm nucleus})^2} [Z f_p + (A-Z) f_n]^2,
\end{equation}
where $Z$ and $A$ are the atomic number and atomic mass of the nucleus.  $f_p$ and $f_n$ are the WIMP's couplings to protons and neutrons, given by \cite{jungman}
\begin{equation}
f_{p,n}=\sum_{q=u,d,s} f^{(p,n)}_{T_q} a_q \frac{m_{p,n}}{m_q} + \frac{2}{27} f^{(p,n)}_{TG} \sum_{q=c,b,t} a_q  \frac{m_{p,n}}{m_q},
\label{feqn}
\end{equation}
where $a_q$ are the WIMP-quark couplings and $f^{(p)}_{T_u} \approx 0.020\pm0.004$,  
$f^{(p)}_{T_d} \approx 0.026\pm0.005$,  $f^{(p)}_{T_s} \approx 0.118\pm0.062$,  
$f^{(n)}_{T_u} \approx 0.014\pm0.003$,  $f^{(n)}_{T_d} \approx 0.036\pm0.008$, 
$f^{(n)}_{T_s} \approx 0.118\pm0.062$ are quantities measured in nuclear physics experiments~\cite{nuc}. The first term in Eq.~\ref{feqn} 
corresponds to interactions with the quarks in the target nuclei. The second term corresponds to interactions with the gluons in the target through a colored loop diagram. $f^{(p)}_{TG}$ is given by $1 -f^{(p)}_{T_u}-f^{(p)}_{T_d}-f^{(p)}_{T_s} 
\approx 0.84$, and analogously, $f^{(n)}_{TG} \approx 0.83$.

\subsection{Direct Detection of Neutralino Dark Matter}

Neutralinos can elastically scatter with quarks through either $t$-channel CP-even Higgs exchange, or $s$-channel squark exchange:
\vspace{-0.2cm}

\begin{feynartspicture}(222,254)(3,4.3)
\FADiagram{ }
\FAProp(0.,20.)(15.,10.)(0.,){/Straight}{0}
\FALabel(-5.,18.0)[b]{$\chi^0$}
\FAProp(15.,-5.0)(15.,10.0)(0.,){/ScalarDash}{0}
\FAProp(0.,-15.0)(15.,-5.0)(0.,){/Straight}{+1}
\FALabel(-5.,-18.0)[b]{$q$}
\FALabel(20.,0.0)[b]{$H, h$}
\FAProp(15.,10.)(30.,20.)(0.,){/Straight}{0}
\FAProp(15.,-5.)(30.,-15.0)(0.,){/Straight}{+1}
\FALabel(35.,18.0)[b]{$\chi^0$}
\FALabel(35.,-18.)[b]{$q$}

\FAProp(50.,14.)(65.,3.)(0.,){/Straight}{0}
\FALabel(45.,12.0)[b]{$\chi^0$}

\FAProp(50.,-7.0)(65.,3.0)(0.,){/Straight}{+1}
\FALabel(45.,-9.0)[b]{$q$}

\FAProp(65.,3.)(80.,3.0)(0.,){/ScalarDash}{0}
\FALabel(72.5,9.)[t]{$\tilde{q}$}

\FAProp(95.,14.)(80.,3.0)(0.,){/Straight}{0}
\FALabel(100.,12.)[b]{$\chi^0$}

\FAProp(95.,-7.)(80.,3.0)(0.,){/Straight}{+1}
\FALabel(100.,-9.)[b]{$q$}
\end{feynartspicture}

\vspace{-3.5cm}
\noindent

In addition to these diagrams, we can write analogous processes in which the WIMP couples to gluons in the target through a quark/squark loop. By calculating the WIMP-quark couplings, $a_q$, we can also implicitly include the interactions of neutralinos with gluons in the target nuclei as well (see Eq.~\ref{feqn}).

The neutralino-quark coupling, in which all of the supersymmetry model-dependent information is contained, is given by~\cite{scatteraq}
\begin{eqnarray}
\label{aq}
a_q & = & - \frac{1}{2(m^{2}_{1i} - m^{2}_{\chi})} Re \left[
\left( X_{i} \right) \left( Y_{i} \right)^{\ast} \right] 
- \frac{1}{2(m^{2}_{2i} - m^{2}_{\chi})} Re \left[ 
\left( W_{i} \right) \left( V_{i} \right)^{\ast} \right] \nonumber \\
& & \mbox{} - \frac{g_2 m_{q}}{4 m_{W} B} \left[ Re \left( 
\delta_{1} [g_2 N_{12} - g_1 N_{11}] \right) D C \left( - \frac{1}{m^{2}_{H}} + 
\frac{1}{m^{2}_{h}} \right) \right. \nonumber \\
& & \mbox{} +  Re \left. \left( \delta_{2} [g_2 N_{12} - g_1 N_{11}] \right) \left( 
\frac{D^{2}}{m^{2}_{h}}+ \frac{C^{2}}{m^{2}_{H}} 
\right) \right],
\end{eqnarray}
where
\begin{eqnarray}
X_{i}& \equiv& \eta^{\ast}_{11} 
        \frac{g_2 m_{q}N_{1, 5-i}^{\ast}}{2 m_{W} B} - 
        \eta_{12}^{\ast} e_{i} g_1 N_{11}^{\ast}, \nonumber \\
Y_{i}& \equiv& \eta^{\ast}_{11} \left( \frac{y_{i}}{2} g_1 N_{11} + 
        g_2 T_{3i} N_{12} \right) + \eta^{\ast}_{12} 
        \frac{g_2 m_{q} N_{1, 5-i}}{2 m_{W} B}, \nonumber \\
W_{i}& \equiv& \eta_{21}^{\ast}
        \frac{g_2 m_{q}N_{1, 5-i}^{\ast}}{2 m_{W} B} -
        \eta_{22}^{\ast} e_{i} g_1 N_{11}^{\ast}, \nonumber \\
V_{i}& \equiv& \eta_{22}^{\ast} \frac{g_2 m_{q} N_{1, 5-i}}{2 m_{W} B}
        + \eta_{21}^{\ast}\left( \frac{y_{i}}{2} g_1 N_{11},
        + g_2 T_{3i} N_{12} \right),
\label{xywz}
\end{eqnarray}
where throughout $i=1$ for up-type quarks and $i=2$ for down type quarks. $m_{1i}, m_{2i}$ denote the squark mass eigenvalues and $\eta$ is the matrix which diagonalizes the squark mass matrices, $diag(m^2_1, m^2_2)=\eta M^2 \eta^{-1}$. $y_i$, $T_{3i}$ and $e_i$ denote hypercharge, isospin and electric charge of the quarks. For scattering off of up-type quarks
\begin{eqnarray}
\delta_{1} = N_{13},\,\,\,\, \delta_{2} = N_{14}, \,\,\,\, B = \sin{\beta},\,\,\,\, C = \sin{\alpha}, \,\,\,\, D = \cos{\alpha},
\end{eqnarray}
whereas for down-type quarks
\begin{eqnarray}
\delta_{1} = N_{14},\,\,\,\, \delta_{2} = -N_{13}, \,\,\,\, B = \cos{\beta},\,\,\,\, C = \cos{\alpha}, \,\,\,\, D = -\sin{\alpha}.
\end{eqnarray}
The quantity $\alpha$ is the angle that diagonalizes the CP-even Higgs mass matrix. 

The first two terms in Eq.~\ref{aq} correspond to interactions through the exchange of a squark, while the final term is generated through Higgs exchange. To help develop some intuition for what size neutralino-nucleon cross sections we might expect, lets consider a few simple limits. First, consider the case in which the scattering is dominated by heavy Higgs ($H$) exchange through its couplings to strange and bottom quarks. This behavior is often found when the squarks are heavy and $\cos \alpha \approx 1$ (which implies moderate to large $\tan \beta$ and $m_A \sim m_H \sim m_{H^{\pm}}$). In this case, the leading contribution to the neutralino-nucleon cross section is
\begin{equation}
\sigma_{\chi N} \sim \frac{g^2_1 g^2_2 |N_{11}|^2 |N_{13}|^2 \,m^4_N}{4\pi m^2_W \cos^2 \beta \, m^4_H} \bigg(f_{T_s}+\frac{2}{27}f_{TG}\bigg)^2, \,\,\,\, (m_{\tilde{q}}\, \rm{large}, \cos \alpha \approx 1).
\label{case1}
\end{equation}
Here, the cross section scales with $m_H^{-4}$ and with $\tan^2\beta$, a leads to the possibility of very large rates. For a $\sim$100 GeV neutralino, for example, a 200 GeV heavy Higgs mass leads to cross sections with nucleons on the order of $10^{-5}$ to $10^{-7}$ pb for $|\mu| \sim 200$ GeV, or $10^{-7}$ to $10^{-9}$ pb for $|\mu| \sim 1$ TeV. 

Second, we can consider the case in which the cross section is dominated by light Higgs boson ($h$) exchange through its couplings to up-type quarks. This is often found in the case of heavy squarks and heavy to moderate $H$. In this limiting case
\begin{equation}
\sigma_{\chi N} \sim \frac{g^2_1 g^2_2 |N_{11}|^2 |N_{14}|^2 \, m^4_N}{4\pi m^2_W \, m^4_h} \bigg(f_{T_u}+\frac{4}{27}f_{TG}\bigg)^2, \,\,\,\, (m_{\tilde{q}}, m_H\, \rm{large}, \cos \alpha \approx 1).
\label{case2}
\end{equation}
If the heavy Higgs ($H$) is heavier than about $\sim$500 GeV, exchange of the light Higgs generally dominates, leading to cross sections of around $10^{-8}$ to $10^{-10}$ pb for $|\mu|$ in the range of 200 GeV to 1 TeV.

Third, consider the case in which the elastic scattering cross section is dominated by the exchange of squarks through their couplings to strange and bottom quarks. This is found for large to moderate $\tan \beta$ and squarks with masses well below 1 TeV. In this limiting case, and with approximately diagonal squark mass matrices,
\begin{equation}
\sigma_{\chi N} \sim \frac{g^2_1 g^2_2 |N_{11}|^2 |N_{13}|^2 \, m^4_N}{4\pi m^2_W \cos^2 \beta \, m^4_{\tilde{q}}} \bigg(f_{T_s}+\frac{2}{27}f_{TG}\bigg)^2, \,\,\,\, (\tilde{q}\,\, \rm{dominated}, \tan \beta \gg 1).
\label{case3}
\end{equation}
For squarks lighter than $\sim$1 TeV, squark exchange can potentially provide the dominant contribution to neutralino-nuclei elastic scattering.

Using Eq.~\ref{directrate}, we can crudely estimate the minimum target mass required to potentially detect neutralino dark matter.  A detector made up of Germanium targets (such as CDMS or Edelweiss, for example) would expect a WIMP with a nucleon-level cross section of $10^{-6}$ pb ($10^{-42}$ cm$^2$) to yield approximately 1 elastic scattering event per kilogram-day of exposure. Such a target mass could thus be potentially sensitive to strongly mixed gaugino-higgsino neutralinos with light $m_H$ and large $\tan \beta$. The strongest current limits on spin-independent scattering have been obtained using $\sim$$10^2$ kilogram-days of exposure. In contrast, reaching sensitivities near the $10^{-10}$ pb level will require ton-scale detectors capable of operating for weeks, months or longer with very low backgrounds.

  \begin{figure}
    \centering
\includegraphics[width=3.9in,angle=0]{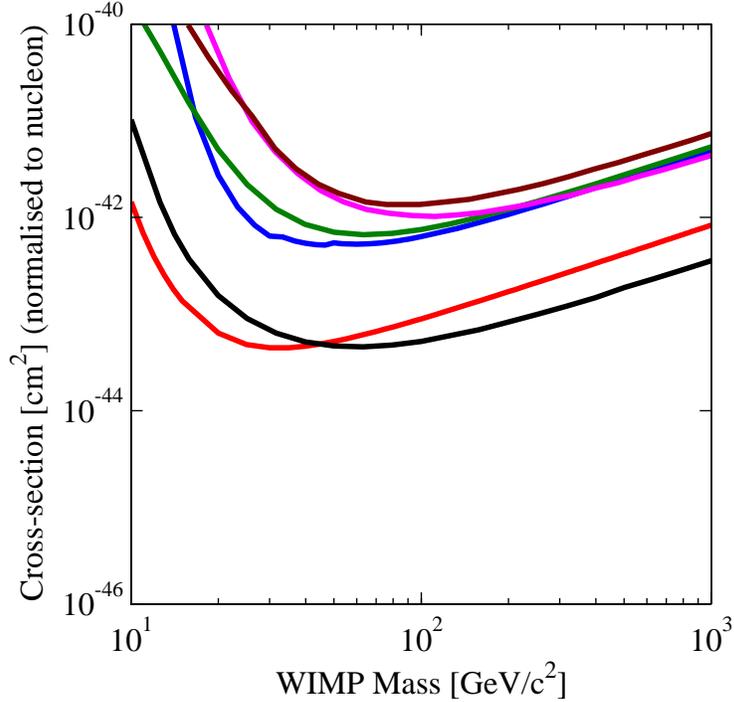}
    \vspace{0cm}
    \caption{Current constraints on the WIMP-nucleon, spin-independent elastic scattering cross section. From bottom-to-top on the right side of the figure, the lines correspond to the limits from the CDMS~\cite{cdms}, XENON-10~\cite{xenon}, WARP~\cite{warp}, CRESST~\cite{cresst}, ZEPLIN~\cite{zeplin}, and Edelweiss~\cite{edelweiss} experiments. This figure was generated using the Dark Matter Limit Plotter~\cite{limitplotter}.}
    \label{directlimits}
  \end{figure}

In Fig.~\ref{directlimits}, we show the current constraints on the WIMP-nucleon spin-independent elastic scattering cross section, as a function of the WIMP's mass. The most stringent constraints currently come from the CDMS and XENON-10 collaborations, which each have obtained limits at around the $10^{-7}$ pb level. The CDMS and XENON-100 collaborations are each anticipated to place limits a factor of several times more stringent within the next year (unless a positive detection is made, that is). Although the more distant future is more difficult to project, it is generally expected that experiments approaching the ton-scale will reach sensitivies near $10^{-9}$ or $10^{-10}$ pb within in the next few to several years.

\subsection{Direct Detection of Kaluza-Klein Dark Matter}

In the case of Kaluza-Klein dark matter (as described in Sec.~\ref{kkdm}), the WIMP-quark coupling $a_q$ receives contributions from the s-channel exchange of KK quarks and the t-channel Higgs boson exchange~\cite{Cheng:2002ej,Servant:2002hb}. This leads to a contribution to the cross section from Higgs exchange which is proportional to $1/(m^2_{B^{(1)}}\,m^4_h)$ and a contribution from KK-quark exchange which is approximately proportional to $1/(m^6_{B^{(1)}}\, \Delta^4)$, where $\Delta=(m_{q^{(1)}}-m_{B^{(1)}})/m_{B^{(1)}}$ is the fractional mass splitting of the KK quarks and the $B^{(1)}$.
The WIMP-quark coupling in this case is given by
\begin{eqnarray}
a_q  =  \frac{m_q \, g^2_1 \, (Y^2_{q_R}+Y^2_{q_L}) \,  ( m^{2}_{B^{(1)}} + m^{2}_{q^{(1)}})}{4 m_{B^{(1)}}(m^{2}_{B^{(1)}} - m^{2}_{q^{(1)}})^2}  + \frac{m_q \, g^2_1}{8 m_{B^{(1)}}\, m^2_h}.
\end{eqnarray}
Numerically, the $B^{(1)}$-nucleon cross section is approximately given by
\begin{equation}
\sigma_{B^{(1)}n,\rm{SI}}  \approx  1.2 \times 10^{-10} \, {\rm pb}\,\bigg(\frac{1\,\rm{TeV}}{m_{B^{(1)}}}\bigg)^2 \, \bigg[\bigg(\frac{100\, \rm{GeV}}{m_h}\bigg)^2 + 0.09 \, \bigg(\frac{1\,\rm{TeV}}{m_{B^{(1)}}}\bigg)^2 \bigg(\frac{0.1}{\Delta}\bigg)^2\bigg]^2.
\end{equation}

With such small cross sections, we will most likely have to wait for ton-scale detectors before this model will be tested by direct detection experiments.

\subsection{Some Model Independent Comments Regarding Direct Detection}

In the cases of the two dark matter candidates discussed thus far in this section, we are lead to expect a rather small elastic scattering cross sections between WIMPs and nuclei -- typically below or well below current experimental constraints. This is {\it not} a universal prediction for a generic WIMP, however. In fact, both neutralinos and Kaluza-Klein dark matter represent somewhat special cases in which direct detection rates are found to be particularly low.

To illustrate this point, consider a Dirac fermion or a scalar WIMP which annihilates in the early universe to fermions with roughly equal couplings to each species -- a heavy 4th generation neutrino or sneutrino, for example. We can take the Feynman diagram for the process of this WIMP annihilating to quarks and turn it on its side, and then calculate the resulting elastic scattering cross section. What we find is that, if the interaction is of scalar or vector form, such a WIMP will scatter with nuclei several orders of magnitude more often than is allowed by the limits of CDMS, XENON and other direct detection experiments. Similar conclusions are reached for many otherwise acceptable WIMP candidates~\cite{modelindependent}. A warning well worth keeping in mind for any WIMP model builder is, ``Beware the crossing symmetry!''.

So what is it about neutralinos or Kaluza-Klein dark matter than enable them to evade these constraints?  In the case of neutralinos, the single most important feature is the suppression of its couplings to light fermions.  Being a Majorana fermion, a neutralino's annihilation cross section to fermion pairs (at low velocity) scales with $\sigma v \propto m^2_f/m^2_{\chi^0}$. As a result, neutralinos annihilate preferentially to heavy fermions (top quarks, bottom quarks, and taus) or gauge/Higgs bosons. As heavy fermions (and gauge/Higgs bosons) are largely absent from nuclei, the potentially dangerous crossing symmetry does not apply. More generally speaking, current direct detection constraints can be fairly easily evaded for any WIMP which interacts with quarks through Higgs exchange, as the Yukawa couplings scale with the fermion's mass.

Alternatively, if the WIMP's couplings are simply very small, direct detection constraints can also be evaded. Small couplings, however, leave us in need of a mechanism for efficiently depleting the WIMP in the early universe. But even with very small couplings, a WIMP might efficiently coannihilate in the early universe, or annihilate through a resonance, leading to an acceptable relic abundance. In this way, coannihilations and resonances can considerably suppress the rates expected in direct detection experiments.

\section{Indirect Detection}
\label{indirect}

Direct detection experiments are not the only technique being pursued in the hope of identifying the particle nature of dark matter. Another major class of dark matter searches are those which attempt to detect the products of WIMP annihilations, including gamma rays, neutrinos, positrons, electrons, and antiprotons. These methods are collectively known as indirect detection and are the topic of this section.

\subsection{Gamma Rays From WIMP Annihilations}

Searches for the photons generated in dark matter annihilations have a key
advantage over other indirect detection techniques in that these particles travel
essentially unimpeded.  In particular, unlike charged particles, gamma rays are not
deflected by magnetic fields, and thus can potentially provide valuable angular information. For example, point-like sources of
dark matter annihilation radiation might appear from high density regions such as the
Galactic Center or dwarf spheroidal galaxies.  Furthermore, over
galactic distance scales, gamma rays are not attenuated, and thus retain their
spectral information. In other words, the spectrum that is measured is the
same as the spectrum generated in the dark matter annihilations.

\begin{figure}

\resizebox{5.65cm}{!}{\includegraphics{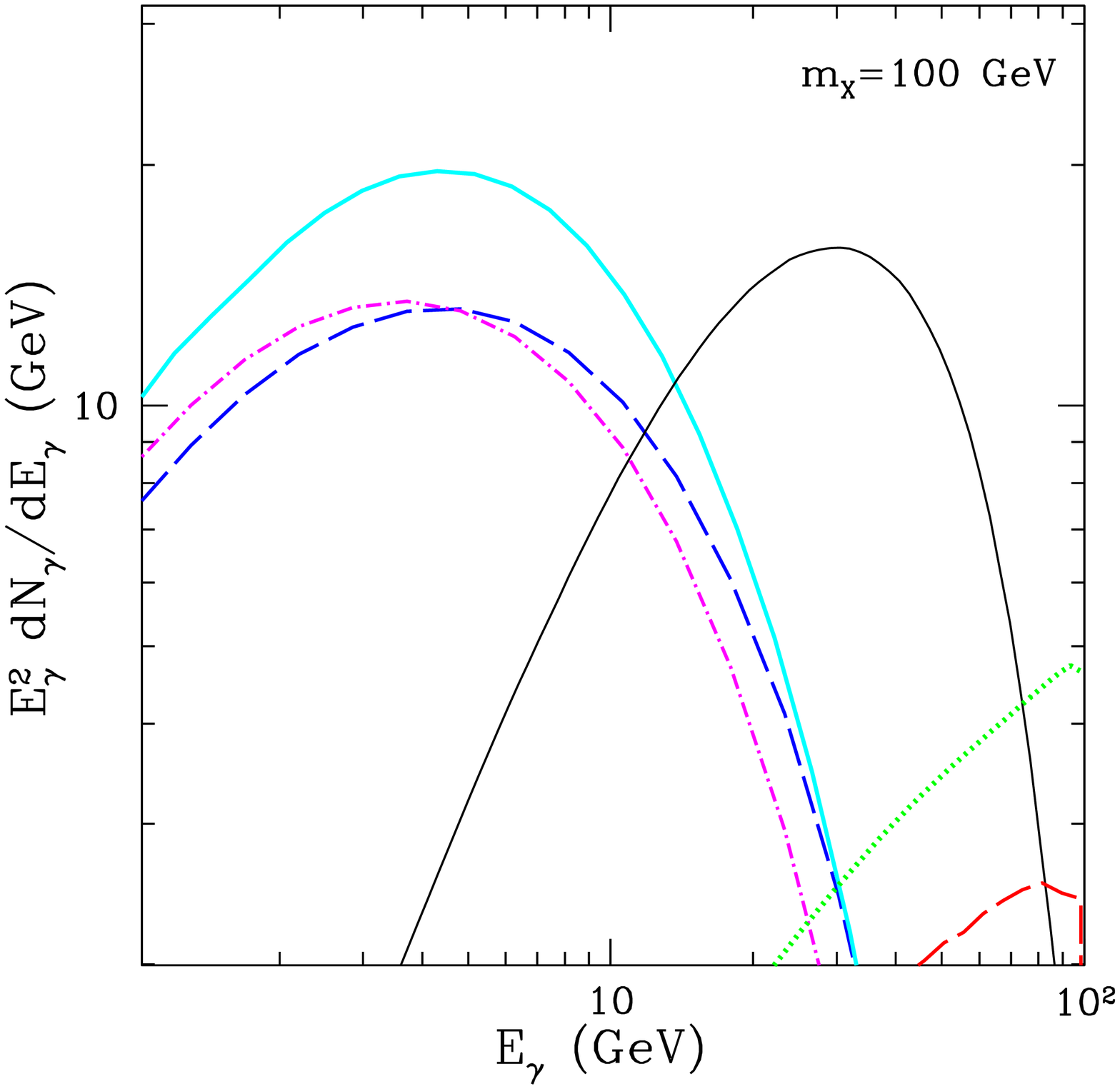}}
\resizebox{5.65cm}{!}{\includegraphics{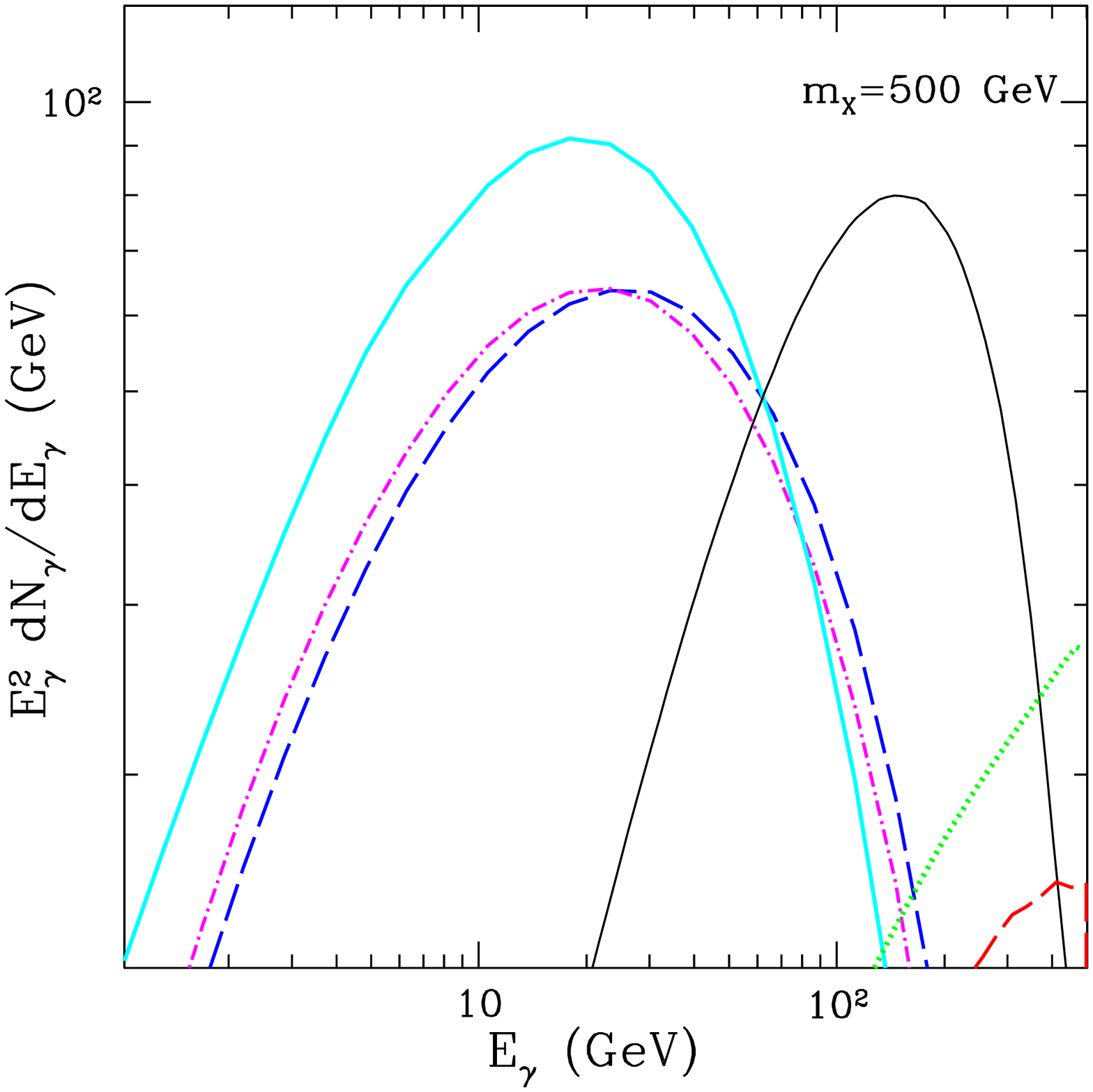}}
\caption{The gamma ray spectrum per WIMP annihilation for a 100 GeV (left) and 500 GeV (right) WIMP. Each curve denotes a different choice of the dominant annihilation mode: $b \bar{b}$ (solid cyan), $ZZ$ (magenta dot-dashed), $W^+ W^-$ (blue dashed), $\tau^+ \tau^-$ (black solid), $e^+ e^-$ (green dotted) and $\mu^+ \mu^-$ (red dashed).}
\label{spectra}
\end{figure}

The spectrum of photons produced in dark matter annihilations depends on the
details of the WIMP being considered. Neutralinos, for example,
typically annihilate to final states consisting of heavy fermions ($b \bar{b}$, $t \bar{t}$, $\tau^+ \tau^-$) or gauge and/or Higgs bosons ($ZZ$, $W^+ W^-$, $HA$, $hA$, $ZH$, $Zh$, $ZA$, $W^{\pm} H^{\pm}$, where $H$, $h$, $A$ and $H^{\pm}$ are the Higgs bosons of the minimal supersymmetric standard model)~\cite{jungman}. With the exception of the $\tau^+ \tau^-$ channel, each of these annihilation modes result in a very similar spectrum of gamma rays. The gamma ray
spectrum from a WIMP which annihilates to leptons can be quite different,
however. This can be particularly important in the case of Kaluza-Klein dark matter in models with one universal extra dimension, for example, in which dark matter particles annihilate significantly to $e^+ e^-$, $\mu^+ \mu^-$ and $\tau^+ \tau^-$~\cite{Servant:2002aq,Cheng:2002ej}. In Fig.~\ref{spectra}, we show the predicted gamma ray spectrum, per annihilation, for several possible WIMP annihilation modes.

In addition to generating continuum gamma rays through the decays of quarks, leptons, Higgs bosons or gauge bosons, dark matter particles can produce gamma rays directly, leading to monoenergetic spectral signatures. If a gamma ray line could be identified, it would constitute a ``smoking gun'' for dark matter annihilations. By definition, however, WIMPs do not annihilate through tree level processes to final states containing photons (if they did, they would be EMIMPs rather than WIMPs). On the other hand, they may be able to produce final states such as $\gamma \gamma$, $\gamma Z$ or $\gamma h$ through loop diagrams. Neutralinos, for example, can annihilate directly to $\gamma \gamma$ \cite{gg} or $\gamma Z$ \cite{gz} through a variety of charged loops. These final states lead to gamma ray lines with energies of $E_{\gamma}=m_{\rm{dm}}$ and $E_{\gamma}=m_{\rm{dm}}(1-m^2_Z/4m^2_{\rm{dm}})$, respectively. Such photons are produced in only a very small fraction of neutralino annihilations, however. The largest neutralino annihilation cross sections to $\gamma \gamma$ and $\gamma Z$ are about $10^{-28}$ cm$^3$/s, and even smaller values are more typical~\cite{buckley}.

The Galactic Center has long been considered to be one of the most promising
regions of the sky in which to search for gamma rays from dark matter
annihilations~\cite{buckley,gchist}. The prospects for this depend, however, on a
number of factors including the nature of the WIMP, the distribution of dark
matter in the region around the Galactic Center, and our ability to understand the astrophysical backgrounds present.

The gamma ray flux from dark matter annihilations is given by
\begin{equation}
\Phi_{\gamma}(E_{\gamma},\psi) = \frac{1}{2}<\sigma_{XX} |v|> \frac{dN_{\gamma}}{dE_{\gamma}} \frac{1}{4\pi m^2_X} \int_{\rm{los}} \rho^2(r) dl(\psi) d\psi.
\label{flux1}
\end{equation}
Here, $<\sigma_{XX} |v|>$ is the WIMP's annihilation cross section (times relative velocity), $\psi$ is the angle observed relative to the direction of the Galactic Center, $\rho(r)$ is the dark matter density as a function of distance to the Galactic Center, and
the integral is performed over the line-of-sight. $dN_{\gamma}/dE_{\gamma}$ is
the gamma ray spectrum generated per WIMP annihilation. Averaging over a solid angle centered around a direction, $\psi$, we arrive
at
\begin{equation}
\Phi_{\gamma}(E_{\gamma}) \approx 2.8 \times 10^{-12} \, \rm{cm}^{-2} \, \rm{s}^{-1} \, \frac{dN_{\gamma}}{dE_{\gamma}} \bigg(\frac{<\sigma_{XX} |v|>}{3 \times 10^{-26} \,\rm{cm}^3/\rm{s}}\bigg)  \bigg(\frac{1 \, \rm{TeV}}{m_{\rm{X}}}\bigg)^2 J(\Delta \Omega, \psi) \Delta \Omega,
\label{flux2}
\end{equation}
where $\Delta \Omega$ is the solid angle observed. The quantity $J(\Delta \Omega, \psi)$ depends only on the dark matter distribution, and is the average over the observed solid angle of the quantity
\begin{equation}
J(\psi) = \frac{1}{8.5 \, \rm{kpc}} \bigg(\frac{1}{0.3 \, \rm{GeV}/\rm{cm}^3}\bigg)^2 \, \int_{\rm{los}} \rho^2(r(l,\psi)) dl.
\label{jpsi}
\end{equation}
$J(\psi)$ is normalized such that a completely flat halo profile, with a density equal to the value at the solar circle, integrated along the line-of-sight to the Galactic Center would yield a value of one. In dark matter distributions favored by N-body simulations, however, this value is much larger. A commonly used parameterization of halo profiles is given by 
\begin{equation}
\rho(r) = \frac{\rho_0}{(r/R)^{\gamma} [1 + (r/R)^{\alpha}]^{(\beta - \gamma)/\alpha}} \,,
\label{profile}
\end{equation}
where $R \sim 20$ kpc is the scale radius and $\rho_0$ is fixed by imposing that the dark matter density at the distance of the Sun from the Galactic Center is equal to the value inferred by observations ($\sim\,$0.3 GeV/cm$^3$). Among the most frequently used parameterizations is the Navarro-Frenk-White (NFW) profile, which is described by $\alpha = 1$, $\beta=3$ and $\gamma =1 $~\cite{nfw}. When considering the region of the Galactic Center, the most important feature of the halo profile is the inner slope, $\gamma$. In some halo profiles, this slope can be considerably steeper than the value used in the NFW parameterization. For example, the Moore {\it et al.} profile is described by $\alpha=1.5$, $\beta=3$, $\gamma=1.5$~\cite{moore}. Note that for  $\gamma\geq 1.5$,  the integral in Eq.~(\ref{jpsi}) diverges. To avoid this behavior, one must truncate the profile within very small galactic radii.

The Narvarro-Frenk-White and Moore {\it et al.} profiles lead to values of  $J(\Delta \Omega=10^{-5} \, {\rm sr}, \psi=0) \sim 10^5$ and $\sim 10^8$, respectively. And although these profiles serve as useful benchmarks, they certainly do not exhaust the range of possibilities. For a number of reasons, it is very difficult to predict the dark matter distribution in the highly important inner parsecs of the Galaxy. Firstly, the resolution of N-body simulations is limited to scales of approximately $\sim 10^{2}$ parsecs or so. Furthermore, the gravitational potential in the inner region of the Milky Way is dominated not by dark matter, but by baryons, whose effects are not generally included in such simulations. The precise impact of baryons on the dark matter distribution is difficult to predict, although an enhancement in the dark matter annihilation rate due to adiabatic compression is generally expected~\cite{ac}. The adiabatic accretion of dark matter onto the central supermassive black hole may also lead to the formation of a density spike in the dark matter distribution. If present, such a spike would result in a very high dark matter annihilation rate~\cite{spike}.


The recent observation of a bright, very high energy gamma ray source in the
region of the Galactic Center by HESS and other ground based Atmospheric Cerenkov Telescopes~\cite{acts} has made efforts to identify gamma rays from dark matter annihilations more
difficult. This source appears to be coincident with the dynamical center of
the Milky Way (Sgr A$^*$) and has no detectable angular extension (less than
1.2 arcminutes). Its spectrum is well described by a power-law,
$dN_{\gamma}/dE_{\gamma} \propto E_{\gamma}^{-\alpha}$, where $\alpha \approx 2.25$ over the range of 160 GeV to 20 TeV, and thus appears to be inconsistent with a dark matter interpretation. Although this gamma ray source represents a formidable background for experiments searching for dark matter
annihilation radiation~\cite{gabi}, it may be possible to reduce the impact of this and other backgrounds by studying the angular distribution of gamma rays from this region of the sky~\cite{Dodelson:2007gd}.

Telescopes potentially capable of detecting gamma rays from dark matter
annihilations include the satellite based Fermi gamma ray space telescope (formerly known as GLAST), and a number ground based Atmospheric Cerenkov Telescopes,
including HESS, MAGIC and VERITAS. These two classes of experiments play complementary roles in the search for dark matter. On one hand, Fermi
will continuously observe a large fraction of the sky, but with an effective
area far smaller than possessed by ground based telescopes. Ground based
telescopes, in contrast, study the emission from a small angular field, but
with far greater exposure. Furthermore, while ground based telescopes can only
study gamma rays with energy greater than $\sim$100 GeV, Fermi will be able to
directly study gamma rays with energies over the range of 100 MeV to 300 GeV.

\begin{figure}[t]
\centerline{\includegraphics[width=0.7\hsize]{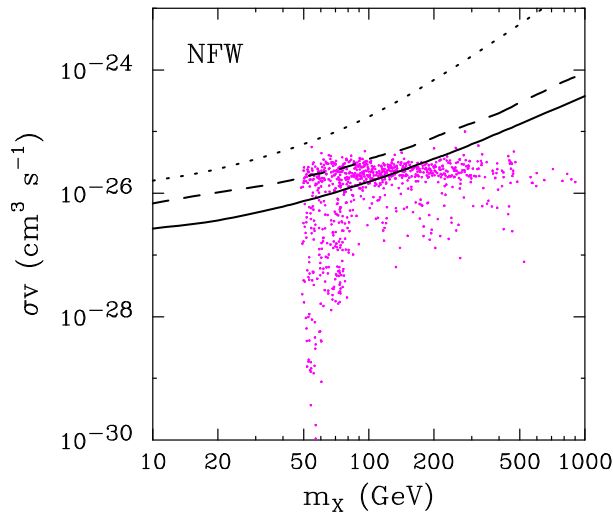}}
\caption{The projected exclusion limits at the 95\% confidence level from five years of observation with the Fermi gamma ray space telescope (formerly known as GLAST) on the WIMP annihilation cross section  as a function of its mass, for the case of an NFW halo profile. The region above the dotted line is already excluded by EGRET~\cite{dingus}. The dashed and solid lines show the projections
for Fermi for an assumed isotropic diffuse background and in the limiting case in which the astrophysical background has the same angular distribution as the dark matter signal, respectively.  Also shown are points representing a random scan of supersymmetric models. Figure from Ref.~\cite{Dodelson:2007gd}.}
\label{glast}
\end{figure}

In Fig.~\ref{glast}, the sensitivity of Fermi to dark matter annihilations in the Galactic Center region is shown for the case of an NFW halo profile and a WIMP annihilating to $W^+W^-$. For this halo profile, WIMPs with an annihilation cross sections of $<\sigma_{XX} |v|> \sim 3 \times 10^{-26}$ cm$^3$/s are near the threshold for detection by Fermi. For WIMPs with approximately this cross section, halo profiles more cuspy than NFW are thus likely to be observable by Fermi, whereas less dense profiles are unlikely to lead to an identifiable signal.

If the dark matter density in the inner parsecs of the Milky Way is not particularly high, or if the astrophysical backgrounds turn out to be particularly foreboding, the prospects for identifying dark matter annihilation radiation from the Galactic Center may be quite unfavorable. In this case, regions of the sky away from the Galactic Center may be more advantageous for dark matter searches. In particular, dwarf spheroidal galaxies within and near the Milky Way provide an opportunity to search for dark matter annihilation radiation with considerably less contamination from astrophysical backgrounds. The flux of gamma rays from dark matter annihilations in such objects, however, is also expected to be lower than from a cusp in the center of the Milky Way~\cite{dwarfs}. As a result, planned experiments are likely to observe dark matter annihilation radiation from dwarf galaxies only in the most favorable range of particle physics models. Alternatively, Fermi may also be sensitive to dark matter annihilations taking place throughout the halo of the Milky Way, or throughout the cosmological distribution of dark matter~\cite{cosmological}.

\subsection{Charged Cosmic Rays From WIMP Annihilations}
\label{charged}

In addition to gamma rays, WIMP annihilations throughout the galactic halo are expected to create charged cosmic rays, including electrons, positrons, protons and antiprotons. Unlike gamma rays, which travel
along straight lines, charged particles move under the influence of the
Galactic Magnetic Field, diffusing and steadily losing energy, resulting in a diffuse
spectrum at Earth. By studying the spectrum of these particles, it may be possible to identify signatures of dark matter annihilations. In fact, multiple experiments have recently announced results which have been interpreted as possible products of WIMPs. 


\begin{figure}
\begin{center}
\hspace{-1.0cm}
\resizebox{12.0cm}{!}{\includegraphics{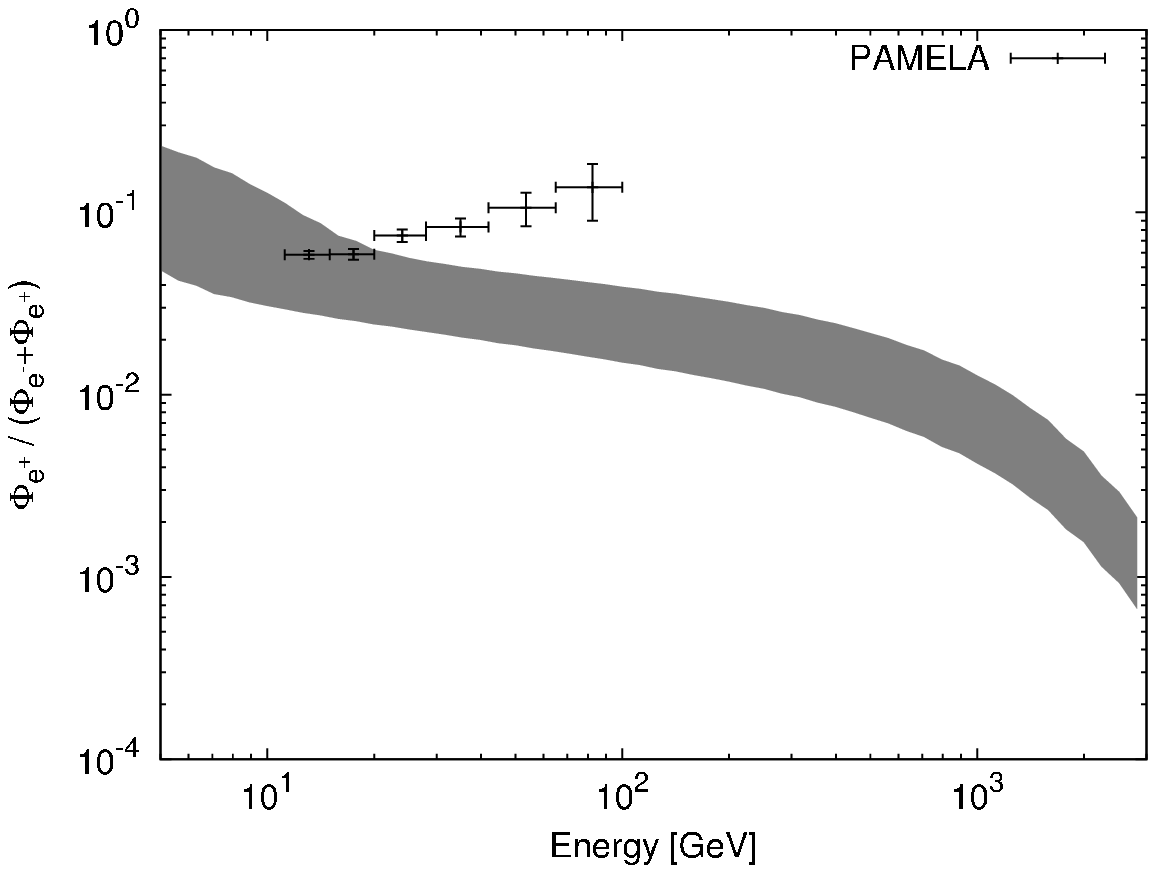}}\\
\vspace{-1.0cm}
\hspace{-1.0cm}
\resizebox{12.0cm}{!}{\includegraphics{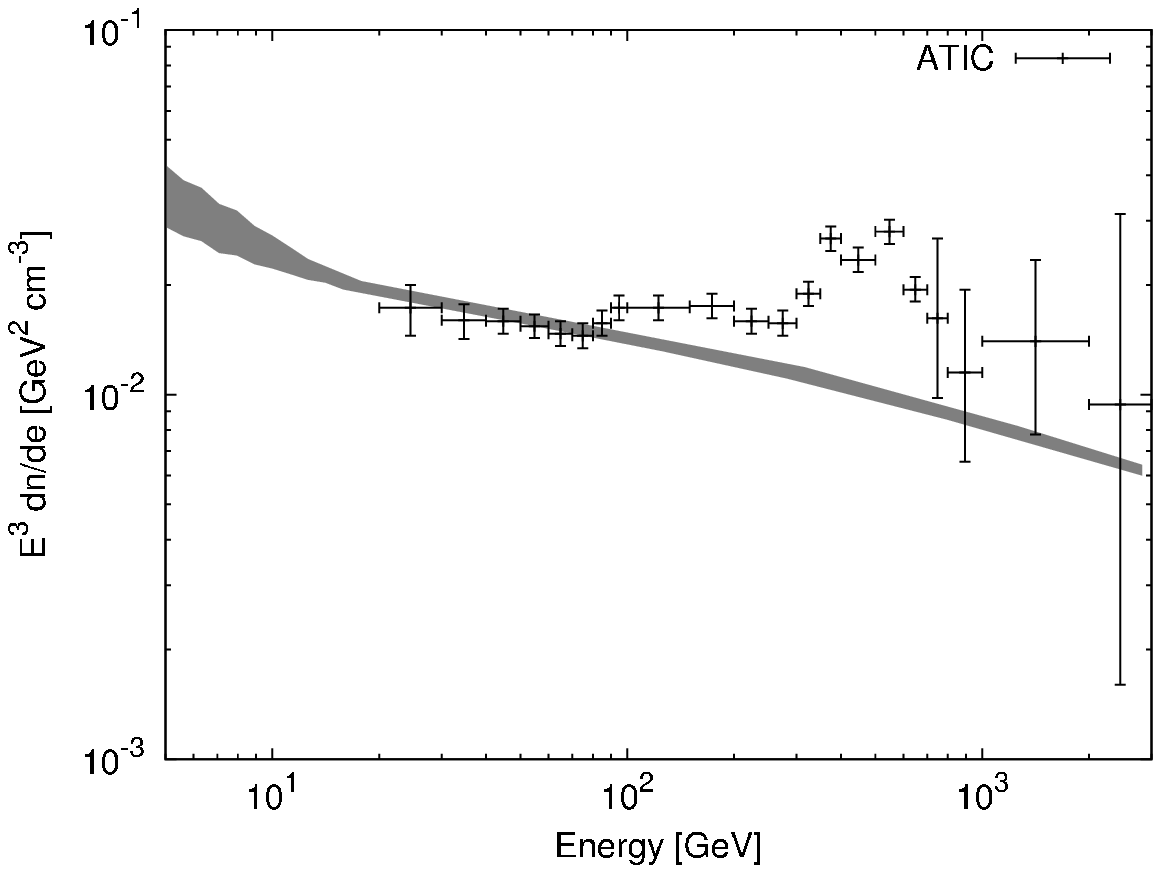}}
\vspace{-1.0cm}
\caption{The positron fraction above 10 GeV as measured by the PAMELA experiment (top)~\cite{pamela} and the electron-plus-positron spectrum as measured by ATIC (bottom)~\cite{atic}. In each frame, the range of standard astrophysical expectations is shown for comparison. These observations clearly require an additional source of energetic electrons and positrons. Figures generated by Melanie Simet.}\label{posdata}
\end{center}
\end{figure}

The PAMELA experiment, which began its three-year satellite mission in June of 2006, recently reported an anomalous rise in the cosmic ray
positron fraction (the positron to positron-plus-electron ratio) above 10 GeV~\cite{pamela}, confirming earlier indications from HEAT~\cite{heat} and AMS-01~\cite{ams01}.  Additionally, the ATIC balloon experiment has recently published data revealing a feature in the cosmic ray electron (plus positron) spectrum between approximately 300 and 800 GeV, peaking at around 600 GeV~\cite{atic}. These measurements are each shown in Fig.~\ref{posdata}, compared to the standard astrophysical predictions. These observations suggest the presence of a relatively local (within $\sim$1 kpc) source or sources of energetic cosmic ray electrons and positrons. Furthermore, in addition to the observations of PAMELA and ATIC, the WMAP experiment has revealed an excess of microwave emission from the central region of the Milky Way which has been interpreted as synchrotron emission from a population of electrons/positrons with a hard spectral index~\cite{haze,hazedark}. Taken together, these observations suggest that energetic electrons and positrons are surprisingly ubiquitous throughout our galaxy.

Although the origin of these electrons and positrons is not currently known, interpretations of the observations have focused two possibilities: emission from pulsars~\cite{pulsars}, and dark matter annihilations~\cite{darkmatter,darkmatter2}. In order for dark matter annihilations throughout in the Milky Way halo to produce a spectrum with a shape similar to that observed by PAMELA and ATIC, however, a large fraction of the annihilations must proceed to electron-positron pairs, or possibly to $\mu^+ \mu^-$ or $\tau^+ \tau^-$~\cite{darkmatter}. Furthermore, WIMPs annihilating to other final states typically exceed the observed flux of cosmic ray antiprotons if normalized to generate the PAMELA and ATIC signals~\cite{antiprotonsexceed}. 

Dark matter particles which annihilate directly to $e^+ e^-$ are predicted to generate a distinctive feature in the cosmic ray electron spectrum: an edge that drops off suddenly at $E_e = m_{X}$. In contrast, pulsars and other astrophysical sources of cosmic ray electrons are expected to produce spectra which fall off more gradually. Although the current data from ATIC is not detailed enough to discriminate between a feature with a sudden edge (dark matter-like) or graduate cutoff (pulsar-like), such a discrimination could become possible if the electron spectrum were measured with greater precision. Interestingly, such a measurement should be possible for ground based gamma ray telescopes such as HESS or VERITAS~\cite{hooperhall}.







Once electrons and positrons are injected into the local halo through dark matter annihilations (or from pulsars), they propagate under the influence of the Galactic Magnetic Field, gradually losing energy through synchrotron emission and through inverse Compton scattering with radiation fields. At energies of a few GeV and higher, the resulting spectrum at Earth can be calculated by solving the diffusion-loss equation~\cite{diffusion}:
\begin{eqnarray}
\frac{\partial}{\partial t}\frac{dn_{e}}{dE_{e}} = \vec{\bigtriangledown} \cdot \bigg[K(E_{e},\vec{x})  \vec{\bigtriangledown} \frac{dn_{e}}{dE_{e}} \bigg] + \frac{\partial}{\partial E_{e}} \bigg[b(E_{e},\vec{x})\frac{dn_{e}}{dE_{e}}  \bigg] + Q(E_{e},\vec{x}),
\label{dif}
\end{eqnarray}
where $dn_{e}/dE_{e}$ is the number density of positrons/electrons per unit
energy, $K(E_{e},\vec{x})$ is the diffusion constant,
$b(E_{e},\vec{x})$ is the rate of energy loss, and $Q(E_{e},\vec{x})$ is
the source term, which contains all of the information regarding the dark matter
annihilation modes, cross section, and spatial distribution. As we expect the cosmic ray distribution from dark matter annihilations to be in or near the steady state limit, the left hand side of the equation is generally set to zero. The energy dependance of the diffusion constant is typically parameterized by $K(E_e) =  K_0 E_e^{\alpha}$.

In the relativistic limit, the energy loss rate resulting from inverse Compton scattering and synchrotron emission is given by
\begin{eqnarray}
b(E_e) &=& \frac{4}{3}\sigma_T \rho_{\rm rad} \bigg(\frac{E_e}{m_e}\bigg)^2    + \frac{4}{3} \sigma_T \rho_{\rm mag} \bigg(\frac{E_e}{m_e}\bigg)^2 \nonumber \\
&\approx& 1.02 \times 10^{-16}\,{\rm GeV/s}\,\bigg(\frac{\rho_{\rm rad}+\rho_{\rm mag}}{{\rm eV}/{\rm cm}^3}\bigg) \times \bigg(\frac{E_e}{{\rm GeV}}\bigg)^2,
\end{eqnarray}
where $\sigma_T$ is the Thompson cross section and $\rho_{\rm rad}$ and $\rho_{\rm mag}$ are the energy densities of radiation (including starlight, cosmic microwave background and emission from dust) and the galactic magnetic field, respectively.

To find a solution to the
diffusion-loss equation, a set of boundary conditions must also be adopted. For example, under the assumption of cylindrical symmetry, the distance from the galactic plane at which electrons and positrons freely escape, $L$, must be selected. 

The diffusion parameters $K_0$, $\alpha$ and $L$ can be constrained by studying the spectra of various species of cosmic ray nuclei. In particular, by studying the ratios of cosmic ray secondaries-to-primaries (most importantly boron-to-carbon) and unstable primaries-to-stable secondaries (Be$^{10}$-to-Be$^{9}$), information can be inferred regarding the size of the diffusion zone and the length of time that cosmic rays are confined to the galaxy. Current cosmic ray data favor values of approximately $K_0\sim 10^{28}$ cm$^2$/s, $\alpha \approx 0.4-0.5$ and $L\sim 1-10$ kpc. For an excellent review of this subject, see Ref.~\cite{StrongMoskPtusk}.

The spectral shape of the source term, $Q$, depends on the leading annihilation modes of the WIMP in the low
velocity limit. Neutralinos and other WIMPs which annihilate primarily to combinations of heavy fermions and Higgs or gauge
bosons generally produce somewhat soft spectra. In contrast, Kaluza-Klein dark matter and other WIMPs which annihilate significantly to electrons and muons are predicted to generate a considerably harder spectrum~\cite{kkpos}.

The normalization of the source term is determined by the annihilation cross section of the WIMP and the spatial distribution of the dark matter in the galactic halo. In particular, small scale inhomogeneities in the dark matter distribution can boost the average annihilation rate.  The factor by which inhomogeneities enhance the resulting cosmic ray flux is called the ``boost factor'', and is defined as
\begin{equation}
{\rm boost}\,\, {\rm factor} = \int_V \frac{<\rho^2>}{<\rho>^2}\frac{dV}{V},
\end{equation}
where the integral is performed over the volume that contributes to the observed spectrum. The results of N-body simulations lead us to expect that this quantity could be as large as 5 to 10. Although considerably larger boost factors are not impossible, they would be somewhat surprising.

In order to generate the signals observed by PAMELA and ATIC with a WIMP annihilating with an annihilation cross section of $\sigma v \approx 3\times 10^{-26}$ cm$^{3}$/s (as required to thermally produce the observed dark matter abundance with a fixed value of $\sigma v$) an annihilation rate many hundred times larger than is naively expected would be required. This could potentially be accommodated in several ways. Firstly, the boost factor may be larger than expected.  Secondly, the dark matter particles may have been generated through a non-thermal mechanism, allowing for considerably larger annihilation cross sections.  Thirdly, WIMPs interacting through the exchange of very light particles can annihilate through non-perturbative processes such that $\sigma v \propto v^{-1}$.  In this case, the annihilation cross section in the present galaxy (with $v\sim 10^{-3} \, c$) would be much larger than the corresponding cross section at the time of thermal freeze-out (when $v\sim 10^{-1} \, c$)~\cite{sommerfeld}.

\subsection{Neutrinos From WIMP Annihilations in the Sun}
\label{neutrinotelescopes}

As the Solar System moves through the halo of the Milky Way, WIMPs become swept up by the Sun. Although dark matter particles interact only weakly, they occasionally scatter elastically with nuclei in the Sun and lose enough momentum to become gravitationally bound. Over the lifetime of the Sun, a sufficient density of WIMPs can accumulate in its center so that an equilibrium is established between their capture and annihilation rates. The annihilation products of these WIMPs include neutrinos, which escape the Sun with minimal absorption, and thus potentially constitute an indirect signature of dark matter. Such neutrinos can be generated through the decays of heavy quarks, gauge bosons, and other products of WIMP annihilation, and then proceed to travel to Earth where can be efficiently identified using large volume neutrino detectors.

Beginning with a simple estimate, we expect WIMPs to be captured in the Sun at a rate approximately given by:
\begin{equation}
C^{\odot} \sim \phi_{X} (M_{\odot}/m_p) \, \sigma_{X p},
\end{equation}
 where $\phi_{X}$ is the flux of WIMPs in the Solar System, $M_{\odot}$ is the mass of the Sun, and $\sigma_{X p}$ is the WIMP-proton elastic scattering cross section. Reasonable estimates of the local distribution of WIMPs leads to a capture rate of $C^{\odot} \sim 10^{20} \, {\rm sec}^{-1} \times (100 \, {\rm GeV}/m_{X})\, (\sigma_{X p}/10^{-6}\, {\rm pb})$. This neglects, however, a number of potentially important factors, including the gravitational focusing of the WIMP flux toward the Sun, and the fact that not every scattered WIMP will ultimately be captured. Taking these and other effects into account leads us to a solar capture rate of~\cite{Gould:1991hx}: 
\begin{eqnarray}
C^{\odot} \approx 1.3 \times 10^{21} \, \mathrm{sec}^{-1}  
\left( \frac{\rho_{\mathrm{local}}}{0.3\, \mathrm{GeV}/\mathrm{cm}^3} \right) 
\left( \frac{270\, \mathrm{km/s}}{\bar{v}_{\mathrm{local}}} \right)  \nonumber \\
\times \left( \frac{100 \, \mathrm{GeV}}{m_{X}} \right) \sum_i \left( \frac{A_i \, (\sigma_{\mathrm{X i, SD}} +\, \sigma_{\mathrm{X i, SI}}) \,S(\frac{m_{X}}{m_{i}})} {10^{-6}\, \mathrm{pb}} \right) ,
\label{capture}
\end{eqnarray}
where $\rho_{\mathrm{local}}$ is the local dark matter density and $\bar{v}_{\mathrm{local}}$ 
is the local rms velocity of halo dark matter particles. $\sigma_{\mathrm{X i, SD}}$ and $\sigma_{\mathrm{X i, SI}}$ are the spin-dependent and spin-independent elastic scattering cross sections of the WIMP with nuclei species $i$, and $A_i$ is a factor denoting the relative abundance and form factor for each species. In the case of the Sun, $A_{\rm H} \approx 1.0$, $A_{\rm He} \approx 0.07$, and $A_{\rm O} \approx 0.0005$. The quantity $S$ contains dynamical information and is given by
\begin{equation}
S(x)=\bigg[\frac{F(x)^{3/2}}{1+F(x)^{3/2}}\bigg]^{2/3}
\end{equation}
where
\begin{equation}
F(x)=\frac{3}{2}\frac{x}{(x-1)^2}\bigg(\frac{v_{\rm esc}}{\bar{v}_{\mathrm{local}}}\bigg)^2,
\end{equation}
and $v_{\rm esc} \approx 1156 \, {\rm km/s}$ is the escape velocity of the Sun. Notice that for WIMPs much heavier than the target nuclei $S \propto 1/m_{X}$, leading the capture rate to be suppressed by two factors of the WIMP mass. In this case ($m_{X} \gsim 30$ GeV), we can write the capture rate as:
\begin{eqnarray}
C^{\odot} \approx 3.35 \times 10^{20} \, \mathrm{sec}^{-1} 
\left( \frac{\rho_{\mathrm{local}}}{0.3\, \mathrm{GeV}/\mathrm{cm}^3} \right) 
\left( \frac{270\, \mathrm{km/s}}{\bar{v}_{\mathrm{local}}} \right)^3 \, \left( \frac{100 \, \mathrm{GeV}}{m_{X}} \right)^2 \nonumber \\ 
\times \left( \frac{\sigma_{X \mathrm{H, SD}} +\, \sigma_{X \mathrm{H, SI}}
+ 0.07 \, \sigma_{X \mathrm{He, SI}}+ 0.0005 \,S(\frac{m_{X}}{m_{\mathrm{O}}})\, \sigma_{X \mathrm{O, SI}}     } {10^{-6}\, \mathrm{pb}} \right).
\label{SIeff}
\end{eqnarray}

If the capture rate and annihilation cross sections are sufficiently large, equilibrium will be reached between these processes.  
For $N(t)$ WIMPs in the Sun, the rate of change of this
quantity is given by
\begin{equation}
\dot{N}(t) = C^{\odot} - A^{\odot} N(t)^2 - E^{\odot} N,
\end{equation}
where $C^{\odot}$ is the capture rate described above and $A^{\odot}$ is the 
annihilation cross section times the relative WIMP velocity per volume. $E^{\odot}$ is the inverse time for a WIMP to escape the Sun via evaporation. Evaporation is highly suppressed for WIMPs heavier than a few GeV~\cite{Gould:1987ir,Griest:1986yu}. $A^{\odot}$ can be approximated by
\begin{equation}
A^{\odot} = \frac{\langle \sigma v \rangle}{V_{\mathrm{eff}}}, 
\end{equation}
where $V_{\mathrm{eff}}$ is the effective volume of the core
of the Sun determined roughly by matching the core temperature with 
the gravitational potential energy of a single WIMP at the core
radius and is given by~\cite{Gould:1987ir,Griest:1986yu}
\begin{equation}
V_{\rm eff} = 5.7 \times 10^{27} \, \mathrm{cm}^3 
\left( \frac{100 \, \mathrm{GeV}}{m_{X}} \right)^{3/2} \;.
\end{equation}
Neglecting evaporation, the present WIMP annihilation rate is given by
\begin{equation} 
\Gamma = \frac{1}{2} A^{\odot} N(t_{\odot})^2 = \frac{1}{2} \, C^{\odot} \, 
\tanh^2 \left( \sqrt{C^{\odot} A^{\odot}} \, t_{\odot} \right) \;, 
\end{equation}
where $t_{\odot} \approx 4.5$ billion years is the age of the solar system.
The annihilation rate is maximized when it reaches equilibrium with
the capture rate.  This occurs when 
\begin{equation}
\sqrt{C^{\odot} A^{\odot}} t_{\odot} \gg 1 \; .
\end{equation}
If this condition is met, the final annihilation rate is determined entirely by the capture rate and has no further dependence on the dark matter particle's annihilation cross section.

Through their annihilations, WIMPs can generate neutrinos through a variety of channels. Annihilations to heavy quarks, tau leptons, gauge bosons and/or Higgs bosons can each generate energetic neutrinos in their subsequent decays~\cite{Jungman:1994jr}. In some models, WIMPs can also annihilate directly to neutrino-antineutrino pairs. Annihilations to light quarks or muons, however, do not contribute to the high energy neutrino spectrum, as these particles come to rest in the solar medium before decaying.

Neglecting the effects of oscillations and interactions with the solar medium, the spectrum of neutrinos from WIMP annihilations to a final state, $Y\bar{Y}$, is given by:
\begin{equation}
\frac{dN_{\nu}}{dE_{\nu}} = \frac{1}{2} \int^{E_{\nu}/\gamma (1-\beta)}_{E_{\nu}/\gamma (1+\beta)}   \frac{1}{\gamma \beta}\frac{dE^{\prime}}{E^{\prime}} \bigg(\frac{dN_{\nu}}{dE_{\nu}}\bigg)^{\rm rest}_{YY},
\end{equation}
where $\gamma=m_{X}/m_Y$, $\beta=\sqrt{1-\gamma^{-2}}$, and $(dN_{\nu}/dE_{\nu})^{\rm rest}_{YY}$ is the spectrum of neutrinos produced in the decay of an $Y$ at rest.


Gauge bosons produced in WIMP annihilations produce the most energetic neutrinos through their decays, $W \rightarrow l \nu$, $Z\rightarrow \nu \bar{\nu}$, but also produce neutrinos through the subsequent decays of muons, taus and other particles. Tau leptons produce neutrinos through a variety of channels, including through the semi-leptonic decays $\tau \rightarrow \mu \nu \nu$, $e \nu \nu$, and the hadronic decays $\tau \rightarrow \pi \nu$, $K \nu$, $\pi \pi \nu$, and $\pi \pi \pi \nu$. Top quarks decay to a $W^{\pm}$ and a bottom quark essentially 100\% of the time, each of which can generate neutrinos in their subsequent decay. For bottom and charm quarks, only the semi-leptonic decays contribute to the neutrino spectrum (with the exception of the neutrinos resulting from taus and $c$-quarks produced in decays of $b$-quarks). In the case of $b$ and $c$ quark decays, the process of hadronization reduces the fraction of energy that is transferred to the resulting neutrinos and other decay products.

Once produced in the Sun's core, neutrinos then propagate through the solar medium and to the Earth. Over this journey, they can potentially be absorbed, lose energy, and/or change flavor. In particular, charged current interactions of electron and muon neutrinos in the Sun lead to their absorption. The probability of absorption taking place can be estimated by $1-\exp(-E_{\nu}/E_{\rm abs})$, where $E_{\rm abs}$ is approximately 130 GeV for electron or muon neutrinos and 200 GeV for electron or muon antineutrinos. Absorption, therefore, only plays an important role for relatively heavy WIMPs.

The effect of charged current interactions on tau neutrinos in the Sun is somewhat more complicated. The tau leptons produced in such interactions quickly decay and thus regenerate the absorbed tau neutrino, albeit with a reduced energy. Neutral current interactions of all three neutrino flavors similarly reduce the neutrinos' energy without depleting their number. 

Vacuum oscillations lead to the full mixing of muon and tau neutrinos over their propagation to the Earth, making the observed muon neutrino spectrum (which is the flux relevant for detection in neutrino telescopes) effectively the average of the muon and tau flavors prior to mixing. Electron neutrinos can also oscillate into muon flavor through matter effects in the Sun (the MSW effect). Electron antineutrinos can generally be neglected, as their oscillations to muon or tau flavors are highly suppressed~\cite{Lehnert:2007fv}. 

Program such as DarkSUSY~\cite{darksusy}, which include effects such as hadronization, absorption, regeneration, and oscillations, are very useful in making detailed predictions for the neutrino spectrum resulting from WIMP annihilations in the Sun.

Once they reach Earth, neutrinos can potentially be detected in large volume neutrino telescopes.  Neglecting oscillations and solar absorption, the muon neutrino spectrum at the Earth resulting from WIMP annihilations in the Sun is given by
\begin{equation}
\frac{dN_{\nu_{\mu}}}{dE_{\nu_{\mu}}} = \frac{ C_{\odot} F_{\rm{Eq}}}{4 \pi D_{\rm{ES}}^2}   \bigg(\frac{dN_{\nu_{\mu}}}{dE_{\nu_{\mu}}}\bigg)^{\rm{Inj}},
\label{wimpflux}
\end{equation}
where $C_{\odot}$ is the WIMP capture rate in the Sun, $F_{\rm{Eq}}$ is the non-equilibrium suppression factor ($F_{\rm Eq}=1$ for capture-annihilation equilibrium), $D_{\rm{ES}}$ is the Earth-Sun distance and $(\frac{dN_{\nu_{\mu}}}{dE_{\nu_{\mu}}})^{\rm{Inj}}$ is the muon neutrino spectrum from the Sun per WIMP annihilating.

Muon neutrinos can produce muons through charged current interactions with ice or water nuclei inside or near the detector volume of a high energy neutrino telescope. The rate of neutrino-induced muons observed in a high energy neutrino telescope is estimated by
\begin{eqnarray}
N_{\rm{events}} &\approx& \int \int \frac{dN_{\nu_{\mu}}}{dE_{\nu_{\mu}}}\, \frac{d\sigma_{\nu}}{dy}(E_{\nu_{\mu}},y) \,[R_{\mu}(E_{\mu})+L]\, A_{\rm{eff}} \, dE_{\nu_{\mu}} \, dy  \nonumber \\
 &+& \int \int \frac{dN_{\bar{\nu}_{\mu}}}{dE_{\bar{\nu}_{\mu}}}\, \frac{d\sigma_{\bar{\nu}}}{dy}(E_{\bar{\nu}_{\mu}},y) \,[R_{\mu}(E_{\mu})+L]\, A_{\rm{eff}} \, dE_{\bar{\nu}_{\mu}} \, dy,
\end{eqnarray}
where $\sigma_{\nu}$ ($\sigma_{\bar{\nu}}$) is the neutrino-nucleon (antineutrino-nucleon) charged current interaction cross section, $(1-y)$ is the fraction of neutrino/antineutrino energy which goes into the muon, $A_{\rm{eff}}$ is the effective area of the detector, $R_{\mu}(E_{\mu})$ is the distance a muon of energy, $(1-y)\,E_{\nu}$, travels before falling below the energy threshold of the experiment (ranging from approximately 1 to 100 GeV), called the muon range, and $L$ is the depth of the detector volume. The muon range in water/ice is approximately given by
\begin{equation}
R_{\mu}(E_{\mu}) \approx 2.4 \, {\rm km} \,\times \, \ln\bigg[\frac{2.0+0.0042\,E_{\mu} ({\rm GeV})}{2.0+0.0042\,E^{\rm thr}_{\mu} ({\rm GeV})}\bigg],
\end{equation}
where $,E^{\rm thr}_{\mu}$ is the threshold of the experiment.

When completed, the IceCube experiment will possess a full square kilometer of effective area and kilometer depth, and will be sensitive to muons above approximately 50 GeV~\cite{icecube}. The Deep Core extension of Icecube will be sensitive down to 10 GeV. The Super-Kamiokande detector, in contrast, has $10^{-3}$ times the effective area of IceCube and a depth of only 36.2 meters~\cite{Desai:2004pq}. For low mass WIMPs, however, Super-Kamiokande benefits over large volume detector such as IceCube by being sensitive to muons with as little energy as $\sim$1 GeV.


  \begin{figure}
    \centering
\includegraphics[width=1.9in,angle=90]{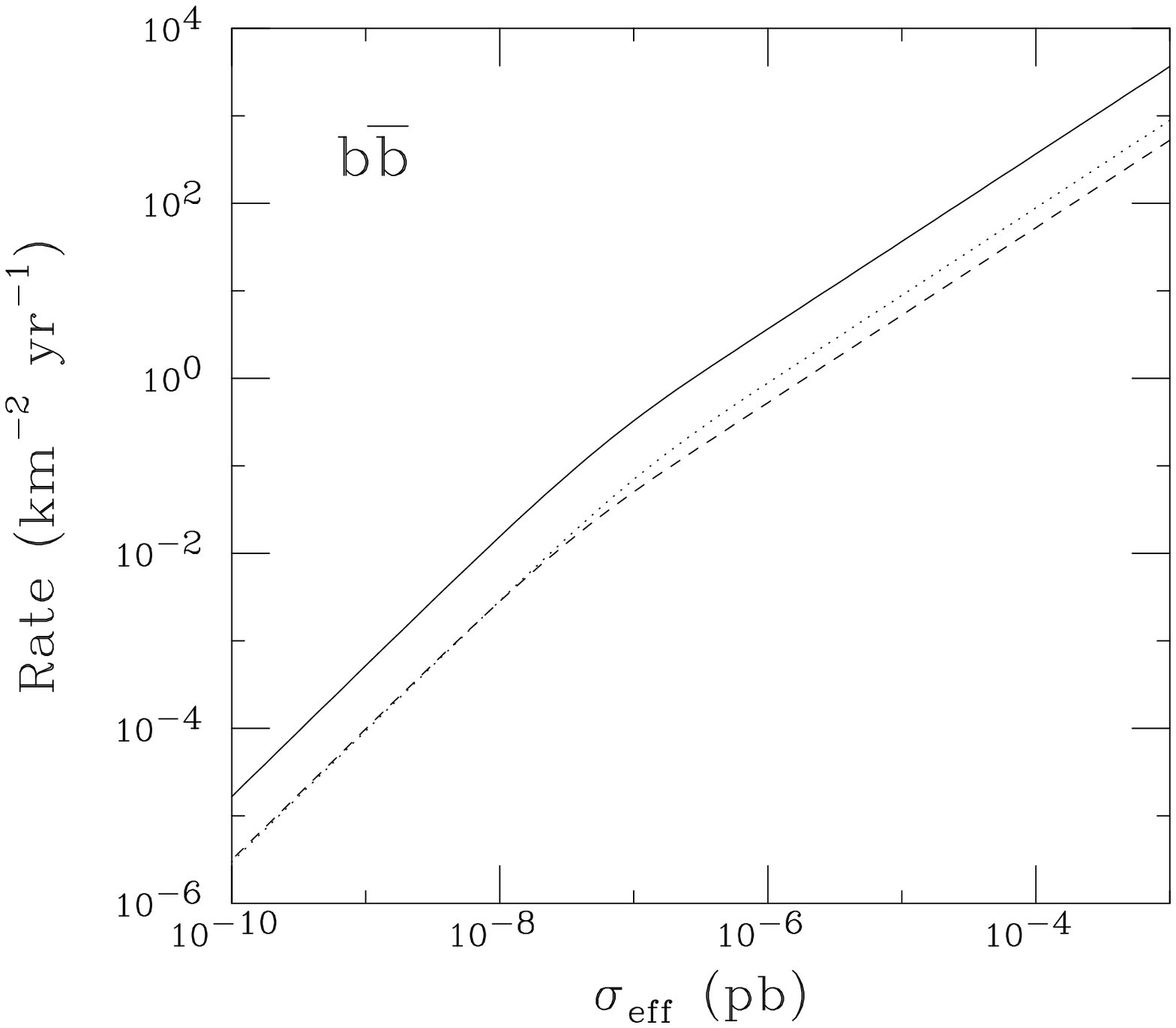}
\includegraphics[width=1.9in,angle=90]{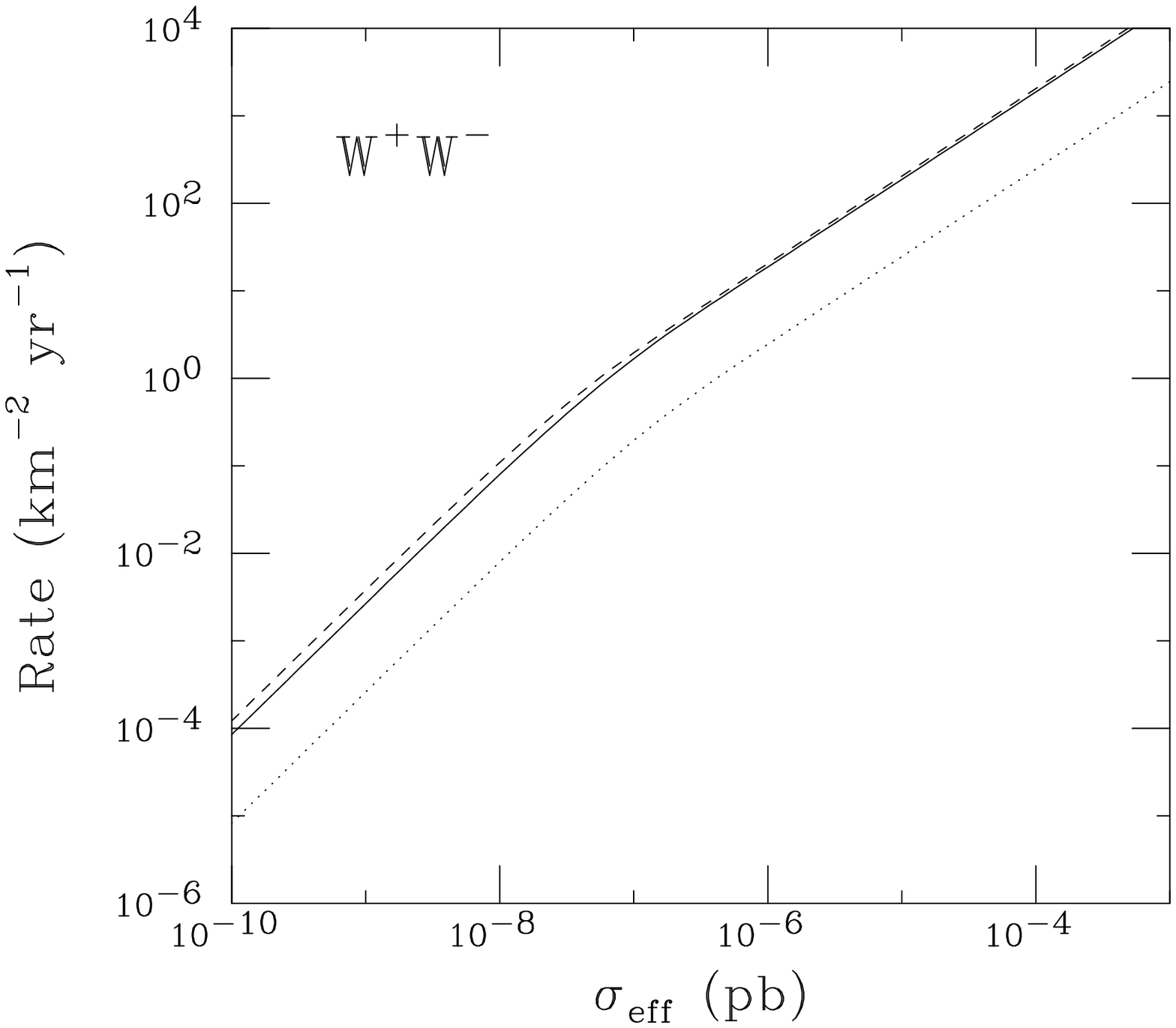}
 \\
\includegraphics[width=1.9in,angle=90]{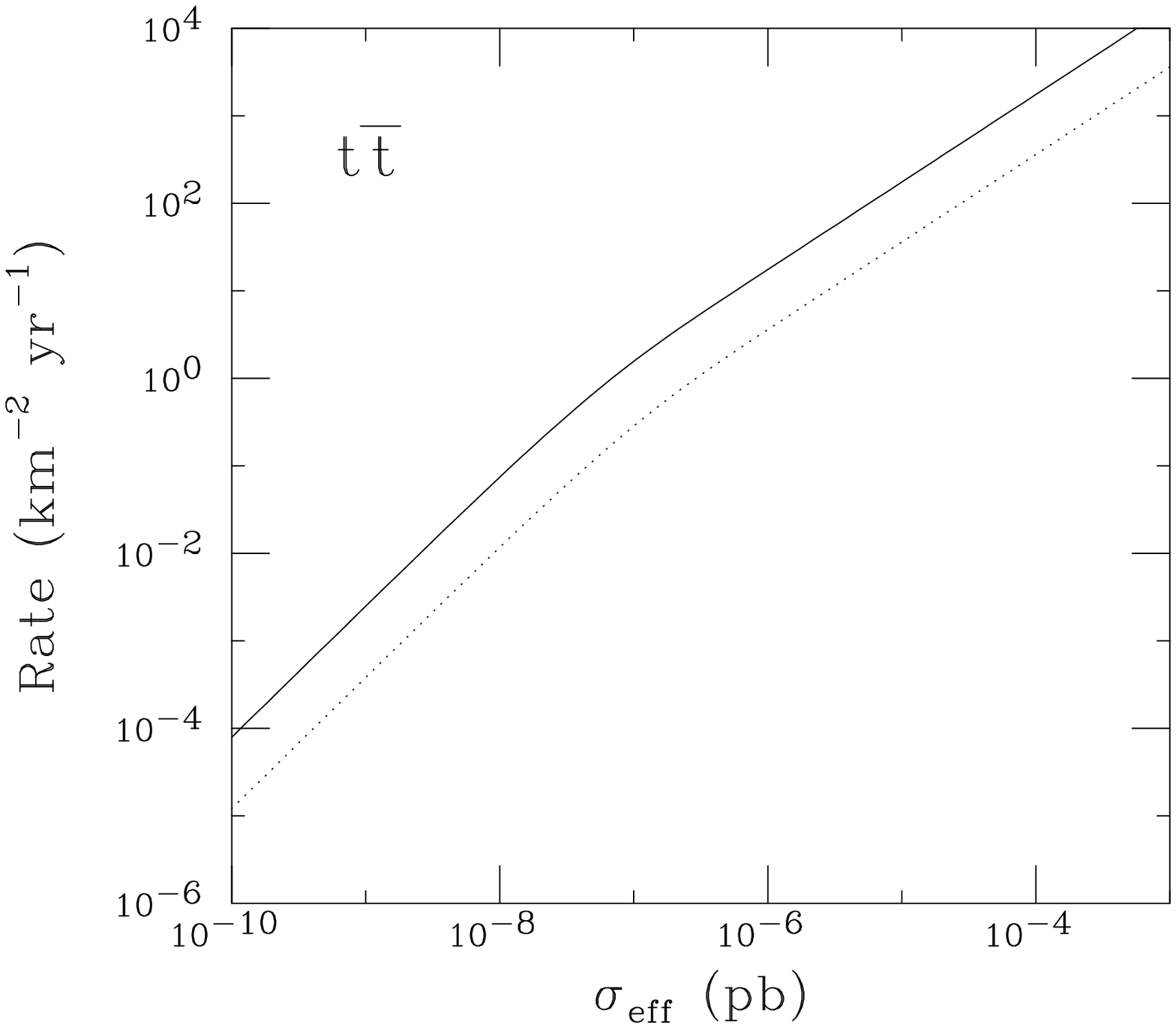}
\includegraphics[width=1.9in,angle=90]{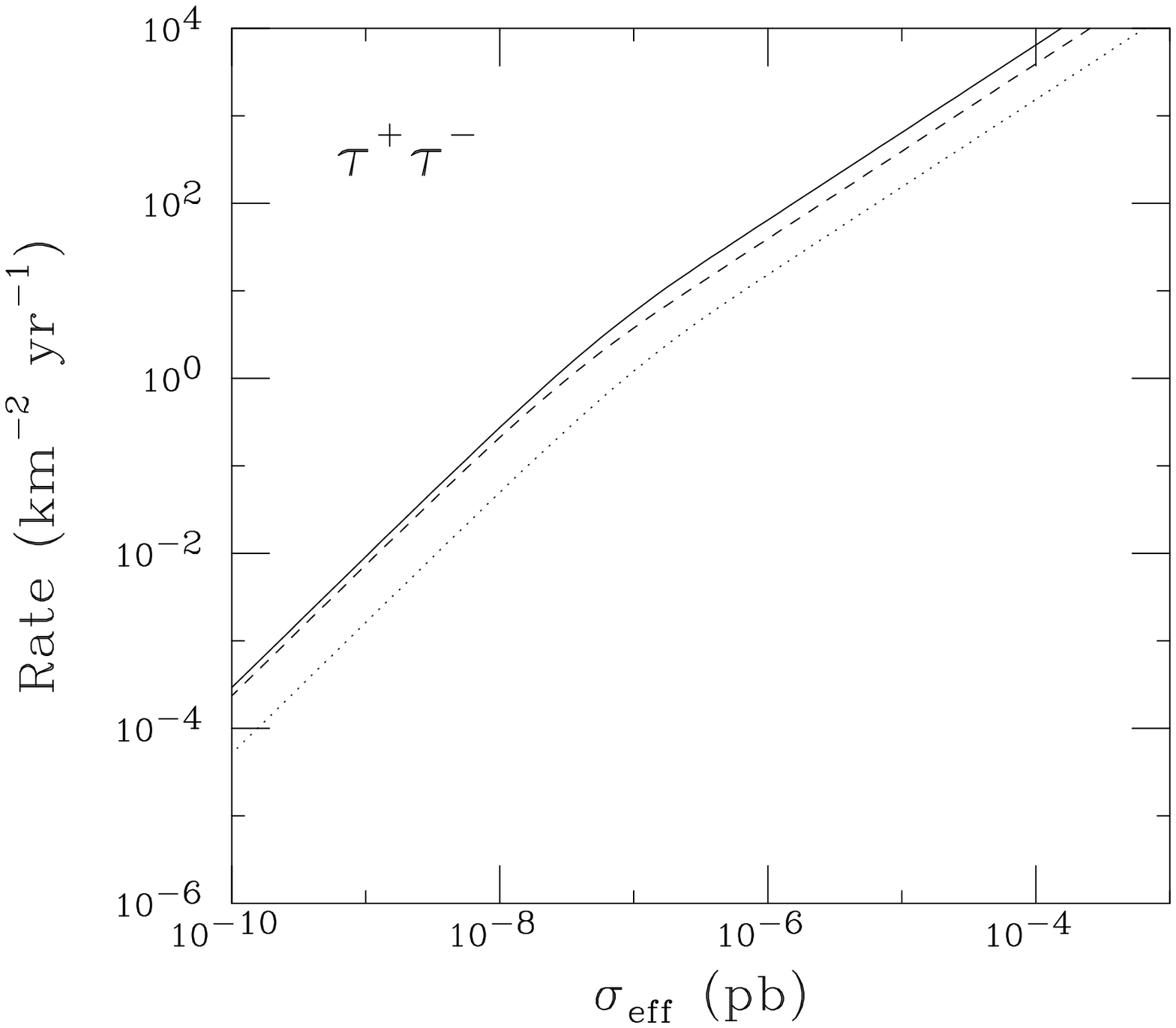}
    \vspace{0cm}
    \caption{The event rate in a kilometer-scale neutrino telescope such as IceCube as a function of the WIMP's effective elastic scattering cross section in the Sun for a variety of annihilation modes. The effective elastic scattering cross section is defined as $\sigma_{\rm{eff}} = \sigma_{\mathrm{X H, SD}} +\, \sigma_{\mathrm{X H, SI}}
+ 0.07 \, \sigma_{\mathrm{X He, SI}} + 0.0005 \, S(m_{X}/m_0) \, \sigma_{\mathrm{X 0, SI}}$ (see Eq.~\ref{SIeff}). The dashed, solid and dotted lines correspond to WIMPs of mass 100, 300 and 1000 GeV, respectively. A 50 GeV muon energy threshold and an annihilation cross section of $3 \times 10^{-26}$ cm$^{3}$ s$^{-1}$ have been adopted.}
    \label{ratecompare}
  \end{figure}

The spectrum and flux of neutrinos generated in WIMP annihilations depends on the annihilation modes which dominate, and thus are model dependent. As long as the majority of annihilations proceed to final states such as $b\bar{b}$, $t\bar{t}$, $\tau^+ \tau^-$, $W^+W^-$, $ZZ$, or some combination of Higgs and gauge bosons, however, the variation between different final states is not dramatic. In Fig.~\ref{ratecompare}, we plot the event rate in a kilometer-scale neutrino telescope such as IceCube as a function of the WIMP's effective elastic scattering cross section for four possible annihilation modes. The effective elastic scattering cross section used here is defined as $\sigma_{\rm{eff}} = \sigma_{\mathrm{X H, SD}} +\, \sigma_{\mathrm{X H, SI}}
+ 0.07 \, \sigma_{\mathrm{X He, SI}} + 0.0005 \, S(m_{X}/m_0) \, \sigma_{\mathrm{X 0, SI}}$ (see Eq.~\ref{SIeff}). 


The elastic scattering cross section of a WIMP is constrained by the absence of a positive signal in direct detection experiments. As described in Sec.~\ref{direct}, the strongest limits on the WIMP-nucleon spin-independent elastic scattering cross section have been made by the CDMS~\cite{cdms} and XENON~\cite{xenon} experiments. These results exclude spin-independent cross sections larger than approximately $5 \times 10^{-8}$ pb for a 25-100 GeV WIMP or $2 \times 10^{-7}$ pb $\times (m_{X}$/500 GeV) for a heavier WIMP.

With these results in mind, consider as an example a 300 GeV WIMP with an elastic scattering cross section with nucleons which is largely spin-independent. With a cross section near the CDMS bound, say $1 \times 10^{-7}$ pb, we can determine from Fig.~\ref{ratecompare} the corresponding rates in a kilometer-scale neutrino telescope, such as IceCube. Sadly, we find that this cross section yields less than 1 event per year for annihilations to $b\bar{b}$, about 2 events per year for annihilations to $W^+W^-$ or $t\bar{t}$ and about 8 per year for annihilations to $\tau^+ \tau^-$, none of which are likely to be sufficient to be identified over the background of atmospheric neutrinos by IceCube or other planned experiments. Clearly, WIMPs that scatter with nucleons mostly through spin-independent interactions are not likely to be detected with IceCube or other planned neutrino telescopes.

The same conclusion is not reached for the case of spin-dependent scattering, however. The strongest bounds on the WIMP-proton spin-dependent cross section have been placed by the COUPP~\cite{coupp} and KIMS~\cite{kims} collaborations. These constraints are approximately 7 orders of magnitude less stringent than those corresponding to spin-independent couplings, however. As a result, a WIMP with a largely spin-dependent scattering cross section with protons may be capable of generating large event rates in high energy neutrino telescopes. Again considering a 300 GeV WIMP with a cross section near the experimental limit, Fig.~\ref{ratecompare} suggests that rates as high as $\sim$$10^6$ per year could be generated if purely spin-dependent scattering contributes to the capture rate of WIMPs in the Sun.

Spin-dependent, axial-vector, scattering of neutralinos with the quarks or gluons within a nucleon is made possible through the t-channel exchange of a $Z$-boson, or the s-channel exchange of a squark. Although the cross sections for these processes can vary dramatically depending on the details of the supersymmetric model under consideration, it is often the case that the neutralino's spin-dependent cross section is considerably larger than its corresponding spin-independent interaction. In particular, very large spin-dependent cross sections ($\sigma_{\rm{SD}} \gsim 10^{-3}$pb) are possible even in models with very small spin-independent scattering rates. Such a model would go easily undetected in all planned direct detection experiments, while still generating on the order of $\sim 1000$ events per year at IceCube.

\begin{figure}[t]
\centerline{\includegraphics[width=0.6\hsize,angle=90]{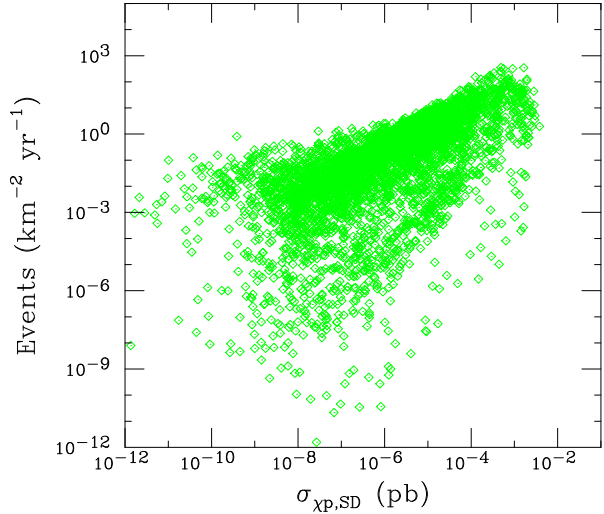}}
\caption{The rate of events at a kilometer-scale neutrino telescope such as IceCube from neutralino annihilations in the Sun, as a function of the neutralino's spin-dependent elastic scattering cross section with protons. Each point shown is beyond the reach of current direct detection experiments.}
\label{ratesd}
\end{figure}

In Fig.~\ref{ratesd}, we demonstrate this by plotting the rate in a kilometer-scale neutrino telescope from WIMP annihilation in the Sun verses the WIMP's spin-dependent cross section with protons. In this figure, each point shown is beyond the reach of present direct detection experiments. We thus conclude that neutralinos may be observable by IceCube while remaining unobserved by current and near future direct detection experiments.

A neutralino which has a large spin-dependent cross section generally has a sizable coupling to the $Z$, and thus has a large higgsino component. In particular, the spin-dependent scattering cross section through the exchange of a $Z$ is proportional to the square of the quantity $|N_{13}|^2 - |N_{14}|^2$. As a result, neutralinos with a few percent higgsino fraction or more are likely to be within the reach of IceCube~\cite{neutrinosun,Halzen:2005ar}. This makes the focus point region of supersymmetric parameter space especially promising. In this region, the lightest neutralino is typically a strong mixture of bino and higgsino components, often leading to the prediction of hundreds of events per year at IceCube.

Considering Kaluza-Klein dark matter, the range of elastic scattering cross sections predicted are quite challenging to reach with direct detection experiments, but are more favorable for detection using neutrino telescopes. The spin-independent $B^{(1)}$-nucleon cross section, which is generated through the exchange of KK quarks and the Higgs boson, is rather small and well beyond the sensitivity of current or upcoming direct detection experiments. The spin-dependent scattering cross section for the $B^{(1)}$ with a proton, however, is considerably larger and is given by \cite{Servant:2002hb}
\begin{eqnarray}
\sigma_{H, SD} &=& \frac{g_1^{4}\, m^2_p}{648 \pi m^4_{B^{(1)}} r^2_{q}} (4 S_u^p + S_d^p + S_s^p)^2 \nonumber \\
&\approx& 4.4 \times 10^{-6} \,\rm{pb}\, \bigg(\frac{800 \,\rm{GeV}}{m_{B^{(1)}}}\bigg)^4 \, \bigg(\frac{0.1}{\Delta_q}\bigg)^2,
\label{kkelastic}
\end{eqnarray}
where $\Delta_q \equiv (m_{q^{(1)}}-m_{B^{(1)}})/m_{B^{(1)}}$ is fractional shift of the KK quark masses over the $B^{(1)}$ mass, which is expected to be on the order of 10\%. The $S$'s parameterize the fraction of spin carried by each variety of quark within the proton.

In addition to this somewhat large spin-dependent scattering cross section, the annihilation products of the $B^{(1)}$ are very favorable for the purposes of generating observable neutrinos. In sharp contrast to neutralinos, approximately 60\% of $B^{(1)}$ annihilations generate a pair of charged leptons (20\% to each type). Although most the remaining 40\% of annihilations produce up-type quarks, about 4\% generate neutrino pairs directly. The neutrino and tau lepton final states each contribute substantially to the event rate in a neutrino telescope.


Taken together, this leads to the prediction of a fairly high neutrino event rate from Kaluza-Klein dark matter annihilating in the Sun. In particular, for masses in the 500-1000 GeV range and a 10\% mass degeneracy with the Kaluza-Klein quarks, we expect $\sim$10-1000 events per year in a kilometer-scale neutrino telescope such as IceCube~\cite{Hooper:2002gs}.

Currently, the strongest constraints on the neutrino flux from WIMPs annihilating in the Sun have been placed by the IceCube (for $m_X \gsim 200$ GeV) and Super-Kamiokande ($m_{X} \lsim 200$ GeV) collaborations. The current IceCube limit constrains the neutrino-induced muon rate from dark matter to be less than $\sim$400-500 per square kilometer per year. Ultimately, IceCube is expected to reach a sensitivity about an order of magnitude below this current level.

\section{Signals, Hints, and... Otherwise}

Over the past several years, a number of observations have been interpreted as possible products of dark matter annihilations. In this section, I take the opportunity to summarize and discuss some of these observations.

\subsection{The PAMELA and ATIC Excesses}
\label{sec:2} 

As I have already discussed the recent observations of the ATIC~\cite{atic} and PAMELA~\cite{pamela} experiments in Sec.~\ref{charged}, in this section I will only briefly summarize some of the arguments in favor of and against these signals being likely products of dark matter annihilations. 

Although the observations of ATIC and PAMELA support a very compelling case that a powerful source of energetic positrons and electrons is present within $\sim$1 kpc of the Solar System, the nature of this source or sources is not yet clear. In order to produce the PAMELA and ATIC signals through the annihilations of dark matter distributed throughout the galactic halo, the WIMPs must annihilate dominantly to charged leptons (to produce a sufficiently hard spectrum, and to avoid the overproduction of cosmic ray antiprotons)~\cite{darkmatter,darkmatter2,antiprotonsexceed}. Furthermore, a very large annihilation rate is also required -- hundreds or thousands of times higher than is naively expected for a thermal relic. To accommodate such a high annihilation rate, we must require that either the WIMPs possess a very large annihilation cross section (which, in turn, requires a non-thermal mechanism for their production in the early universe), or that local inhomogeneities in the dark matter distribution boost the annihilation rate far more efficiently than N-body simulations would lead us to expect. Large annihilation cross sections might also result from non-perturbative processes in some models~\cite{sommerfeld}. Alternatively, in the relatively unlikely event that a large and dense clump of dark matter happened to reside within $\sim$1 kpc of the Solar System, a sufficiently high annihilation rate, hard electron/positron spectrum, and low antiproton flux could plausibly be generated~\cite{clump}.

Taken together, these considerations lead me to conclude that although the ATIC and PAMELA signals could potentially be explained by dark matter annihilations, such a scenario would require WIMPs which possess rather special properties, or that are distributed in a somewhat unlikely way. The leading astrophysical alternative for the origin of these signals is a nearby pulsar (or pulsars)~\cite{pulsars}. Although pulsars are known to be sites of electron-positron pair production, it is not possible to reliably predict the spectrum of or total power injected from these objects. To accommodate the PAMELA and ATIC observations, a nearby (within $\sim$1 kpc) and somewhat young ($10^5$-$10^6$ years) pulsar or pulsars must have deposited a few percent or more of their total energy output in the form of a very hard spectrum of electron-positron pairs. While this appears to be a larger fraction of the total energy than was generally expected, it is certainly not implausible. 

In the relatively near future, measurements by the Fermi gamma ray space telescopes, as well as ground based telescopes~\cite{hooperhall}, should clarify this situation considerably. The origin of the PAMELA and ATIC signals will likely not remain a mystery for long.

\subsection{The WMAP Haze}
\label{sec:6} 

In addition to its measurements of the cosmic microwave background, data from the Wilkinson Microwave Anisotropy Probe (WMAP) has been used to provide the best measurements to date of the standard interstellar medium emission mechanisms, including thermal dust, spinning dust, ionized gas, and synchrotron. In addition to these expected foregrounds, the observations have revealed an excess of microwave emission in the inner $20^{\circ}$ around the center of the Milky Way, distributed with approximate radial symmetry.  This excess is known as the ``WMAP Haze''~\cite{haze}.

Although the WMAP Haze was initially thought likely to be thermal bremsstrahlung (free-free emission) from hot gas ($10^4 \, \rm{K}\gg T \gg 10^6\, \rm{K}$), this interpretation can be ruled out by the relative absence of the H$\alpha$ recombination line and X-ray emission. Other possible origins for this signal, such as thermal dust, spinning dust, and Galactic synchrotron as traced by low-frequency surveys, also seem unlikely. Alternatively, it has been suggested that
the WMAP Haze could be a product of dark matter annihilations~\cite{hazedark}. In particular, annihilating dark matter particles produce relativistic electrons and positrons which travel under the influence of the Galactic magnetic field. As they do, they will emit synchrotron photons, which naturally fall within the frequency range measured by WMAP.

The angular distribution of the Haze can be used to constrain the shape of the required dark matter halo profile. In particular, the morphology of the Haze is consistent with originating from dark matter distributed as $\rho(r) \propto r^{-1.2}$ within the inner kiloparsecs of our galaxy~\cite{hazedark}. This slope falls between those predicted by the NFW, $\rho(r) \propto r^{-1}$, and Moore {\it et al.}, $\rho(r) \propto r^{-1.5}$, halo profiles. The annihilation cross section required of a $\sim$100-1000 GeV WIMP to produce the observed intensity of the WMAP Haze is of the same order magnitude as the value predicted for a simple thermal relic.  No large boost factors are needed to generate this signal.

It is also interesting to note that the dark matter halo profile and annihilation cross section required to generate the WMAP Haze with dark matter imply a flux of prompt gamma rays from the Galactic Center region that is within the reach of the upcoming Fermi gamma ray space telescope~\cite{Hooper:2007gi}. Additionally, the upcoming Planck satellite will provide substantially improved measurements of the spectrum and angular distribution of the Haze.

\subsection{DAMA's Annual Modulation}

The direct detection experiment DAMA has reported evidence for an annual modulation in its rate of nuclear recoil events~\cite{dama}. It has been claimed that this signal is the result WIMP interactions, which result from variations in the relative velocity of the Earth with respect to the dark matter halo as the Earth orbits the Sun. This effect is predicted to lead to a variation in the flux of dark matter particles and their velocity distribution, with expected extrema occurring at June 2 and December 2.  The DAMA experiment 
observes a maximum rate at low nuclear recoil energies on May 24, plus or minus 8 days, and have accumulated enough data to put the significance of the observed modulation at approximately $8\sigma$.  The collaboration has not been able to identify any other systematic effects capable of producing this signal.  The claim that this signal is the result of dark matter interactions has been 
controversial, in part because a number of other experiments appear to be in direct conflict with the DAMA result.  

Several studies have attempted to reconcile the DAMA modulation signal with the null results of other direct-detection experiments~\cite{damadark}.  In particular, an elastically scattering WIMP with a mass in the several GeV range can satisfy the results of DAMA while remaining marginally consistent with the null results of CDMS~\cite{cdms}, CRESST~\cite{cresst}, CoGeNT~\cite{cogent}, and XENON~\cite{xenon}.  The allowed parameter region depends crucially on the occurrence of an effect known as channeling in the NaI crystals of the DAMA apparatus~\cite{Drobyshevski:2007zj}.

Another possibility to have been proposed is that the DAMA signal might arise from a WIMP which does not scatter with nuclei elastically, but instead scatters inelastically, leaving the interaction in the form of a slightly heavier state (with a mass splitting on the order of 100 keV)~\cite{inelastic}. For kinematic reasons, this scenario allows for the efficient scattering of the WIMPs with iodine nuclei in DAMA, while suppressing the scattering rate off of germanium and other comparatively light nuclei used in other experiments.

\subsection{The INTEGRAL 511 keV Line}
\label{sec:3} 

In 2003, the SPI spectrometer onboard the INTEGRAL satellite confirmed the very bright emission of 511 keV photons from the region of the Galactic Bulge, corresponding to an injection rate of approximately $3 \times 10^{42}$ positrons per second in the inner Galaxy~\cite{integral}. This is orders of magnitude larger than the expected rate from pair creation via cosmic ray interactions with the interstellar medium. The signal appears to be approximately (but not perfectly~\cite{asymmetric}) spherically symmetric (with a full-width-half-maximum of approximately 6$^{\circ}$), with little of the emission tracing the Galactic Disk. A stellar origin of this signal, such as type Ia supernovae, hypernovae or gamma ray bursts, would therefore also require a mechanism (such as substantial coherent magnetic fields) by which the positrons could be transported from the disk to throughout the volume of the Bulge~\cite{Prantzos:2005pz}. Furthermore, type Ia supernovae do not inject enough positrons to generate the observed intensity of this signal~\cite{Kalemci:2006bz}. It is possible, however, that hypernovae~\cite{Casse:2003fh} or gamma ray bursts~\cite{Casse:2003fh,Parizot:2004ph} might be capable of injecting positrons at a sufficient rate. A large population of several thousand X-ray binaries has also been proposed as a possible source of these photons~\cite{Bandyopadhyay:2008ts}.

Given the challenges involved with generating the observed 511 keV emission with astrophysical sources, it was suggested that this signal could potentially be the product of dark matter annihilations~\cite{511dark}. In order for dark matter particles to generate the observed spectral line width of this signal, however, their annihilations must inject positrons with energies below a few MeV~\cite{beacom}. This, in turn, implies that the dark matter's mass be near the 1-3 MeV range -- much lighter than annihilating dark matter particles in most theoretically attractive models (for an interesting exception, see Ref.~\cite{zurekmodel}).

Although weakly interacting particles with masses smaller than a few GeV (but larger than $\sim$1 MeV) tend to be overproduced in the early universe relative to the measured dark matter abundance~\cite{leeweinberg}, this can be avoided if a new light mediator is introduced which makes dark matter annihilations more efficient~\cite{lightok}. For example, although neutralinos within the MSSM are required by relic abundance considerations to be heavier than $\sim$20 GeV~\cite{susycase}, they can be much lighter in extended supersymmetric models in which light Higgs bosons can mediate neutralino annihilations~\cite{nmssm}.

For dark matter particles with MeV-scale masses to generate the measured dark matter abundance, they must annihilate during the freeze-out epoch with a cross section of $\sigma v \sim 3 \times 10^{-26}$ cm$^3$/s. To inject the flux of positrons needed to generate the signal observed by SPI/INTEGRAL, however, an annihilation cross section four to five orders of magnitude smaller is required. Together, these requirements force us to consider dark matter particles with annihilation cross sections of the form, $\sigma v \propto v^2$. Such behavior can be found, for example, in the case of fermionic or scalar dark matter particles annihilating through a vector mediator. For such a dark matter particle to not be overabundant today, the mediating boson also must be quite light~\cite{scalar}.

As an alternative explanation for the 511 keV line, it has been proposed that $\sim \,$500 GeV dark matter particles could be collisionally excited to states which are $\sim$1 MeV heavier, which produce electron-positron pairs in their subsequent de-excitations to the ground state~\cite{exciting}.

\subsection{EGRET's Diffuse Galactic Spectrum}
\label{sec:4}

The satellite-based gamma ray detector, EGRET, has measured the diffuse spectrum of gamma rays over the entire sky. When compared to conventional galactic models, these measurements appear to contain an excess at energies above approximately 1 GeV. This has been interpreted as evidence for dark matter annihilations in the halo of the Milky Way~\cite{deboer}. 

Among the most intriguing features of the observed EGRET excess is its similar spectral shape over all regions of the sky. Furthermore, this shape is consistent with that predicted from the annihilations of a 50-100 GeV WIMP. There are, however, some substantial challenges involved with interpreting the EGRET excess as a product of dark matter annihilations. In particular, to accommodate the required normalization for the annihilation rate in various regions of the Galaxy, the distribution of dark matter has to depart substantially from the predictions of standard dark matter halo profiles. In particular, Ref.~\cite{deboer} adopts a distribution which includes two very massive ($\sim 10^{10}\, M_{\odot}$) toroidal rings of dark matter near or within the Galactic Plane, at distances of approximately 4 and 14 kiloparsecs from the Galactic Center. The authors motivate the presence of these rings by observed features in the Galactic rotation curve, and suggest that they may be remnants of very massive dwarf galaxies which have been tidally disrupted.

The other difficulty involved with the interpretation of the EGRET excess as dark matter annihilation radiation is the large flux of antiprotons which is expected to be generated in such a scenario~\cite{bergstromdeboer}. In particular, the flux of cosmic antiprotons produced is expected to exceed the measured flux by more than an order of magnitude. To avoid this conclusion, one is forced to consider significant departures from standard galactic diffusion models. In particular, an anisotropic diffusion model featuring strong convection away from the Galactic Disk and a large degree of inhomogeneities in the local environment could reduce the cosmic antiproton flux to acceptable levels~\cite{deBoer:2006ck}.

The dark matter interpretation of EGRET's measurement of the galactic diffuse spectrum has also been challenged on the grounds that the data could plausibly be explained without the addition of an exotic component, from dark matter or otherwise. In particular, uncertainties in the cosmic ray propagation and diffusion model lead to considerable variations in the predicted diffuse gamma ray backgrounds~\cite{Moskalenko:2006zy}. More recently, it has also been suggested that the observed excess could also be the result of systematic errors in EGRET's calibration~\cite{Stecker:2007xp}.

The Fermi gamma ray space telescope should be able to quickly clarify the origin of the EGRET diffuse emission.


\subsection{EGRET's Diffuse Extragalactic Spectrum}
\label{sec:5}

In addition to galactic emission, it has also been proposed that dark matter annihilation radiation might constitute a significant fraction of the extragalactic (isotropic) diffuse gamma ray flux as measured by EGRET~\cite{Elsaesser:2004ap}.

The origin of EGRET's diffuse extragalactic gamma ray background is currently unknown. Although all or most of the observed spectrum could be the product of astrophysical source such as blazars, much of the flux observed in the 1-20 GeV range could also plausibly be the result of dark matter annihilations taking place throughout the universe~\cite{cosmological}. In particular, the observed spectrum fits reasonably well the predictions for a WIMP with a mass of roughly 500 GeV.

The dark matter annihilation rate needed to normalize to the diffuse flux measured by EGRET is, however, quite high and requires either a very large dark matter annihilation cross section or dark matter halos which are very cuspy. In particular, if an NFW profile~\cite{nfw} is adopted for all halos throughout the universe, then an annihilation cross section $10^2$ to $10^3$ times larger than is predicted for a thermal relic is required to generate the observed gamma ray flux. If the dark matter distribution in the Milky Way is similar to that found in halos throughout the universe, however, then the gamma ray flux from the center of our galaxy would far exceed that which is observed~\cite{Ando:2005hr}. To generate the isotropic diffuse flux observed by EGRET without conflicting with observations of the Galactic Center, therefore, requires extremely cusped halo profiles in most or at least many of the galaxies throughout the universe, and a far less dense cusp in our own Milky Way. 

\section{Summary and Outlook}

The next few years will be a very exciting time for research in particle dark matter. As direct detection experiments move closer to the ton-scale, many of the most attractive models of dark matter will become within their reach.  The lack of a detection in these experiments would be capable of severely constraining the nature of supersymmetry or other TeV-scale physics containing a WIMP candidate. A confirmed detection of a WIMP would open the door to the age of precision dark matter studies, in which measurements of the WIMP's mass, interactions, and distribution would begin to be made.

Indirect detection efforts are currently developing very rapidly. In particular, much of what I have written here about electron, positron and gamma ray signals for dark matter will be hopelessly out of date even a year from now. But rapid progress makes for exciting times!  With new data from the Fermi gamma ray space telescope on the horizon, the near future will likely hold many exciting results for indirect detection.

The single most remarkable aspect of these lecture notes is that I have managed to write almost 50 pages about TeV-scale dark matter without mentioning the Large Hadron Collider. This should not be taken as an indication that the LHC is not important in the hunt for dark matter's identity. In contrast, I fully expect the next twenty years of particle physics (including particle-astrophysics) to be largely defined by what this incredible machine reveals to us. And while I remain largely agnostic regarding what the LHC is likely to discover, it will almost certainly give us insights into the nature of dark matter (even if that insight is that dark matter is not made up of WIMPs). 

Colliders and astrophysical experiments each provide very different and complementary types of information regarding dark matter. It is very unlikely that any single experiment or class of experiments will be sufficient to
conclusively identify the particle nature of dark matter. The direct or indirect
detection of the dark matter particles making up our galaxy's halo will probably not be able to provide enough information to reveal the underlying physics (supersymmetry, etc.) behind these particles.  In contrast, collider experiments may identify a
long-lived, weakly interacting particle, but will not be able to test its
cosmological stability or abundance. Only by combining the information provided
by many different experimental approaches is the mystery of dark
matter's particle nature likely to be solved. Although a confirmed detection of dark matter in any one search channel would constitute a discovery of the utmost importance, it would almost certainly leave many important questions unanswered as well.


\section*{Acknowledgments}

I would like to thank Tao Han and Robin Erbacher for organizing TASI 2008, and K.~T. Mahanthappa for his hospitality. This work was supported in part by the Fermi Research Alliance, LLC under Contract No.~DE-AC02-07CH11359 with the US Department of Energy and by NASA grant NNX08AH34G.



\begin{thebibliography}{999}



\bibitem{rotationcurves}
A.~Borriello and P.~Salucci, 
Mon.\ Not.\ Roy.\ Astron.\ Soc.\  {\bf 323}, 285 (2001) 
[arXiv:astro-ph/0001082].  



\bibitem{weaklensing}
H.~Hoekstra, H.~Yee and M.~Gladders,
New Astron.\ Rev.\  {\bf 46}, 767 (2002)
[arXiv:astro-ph/0205205].


\bibitem{strong}
R.~B.~Metcalf, L.~A.~Moustakas, A.~J.~Bunker and I.~R.~Parry,
arXiv:astro-ph/0309738;
L.~A.~Moustakas and R.~B.~Metcalf,
Mon.\ Not.\ Roy.\ Astron.\ Soc.\  {\bf 339}, 607 (2003)
[arXiv:astro-ph/0206176].



\bibitem{clusters}


N. Bahcall and X. Fan, Astrophys. J. {\bf 504} (1998) 1;
A. Kashlinsky, Phys. Rep. {\bf 307} (1998) 67;
R.~G.~Carlberg et al., 
Astrophys. J. {\bf 516} (1999) 552;
J.~A.~Tyson, G.~P.~Kochanski and I.~P.~Dell'Antonio,  
 Astrophys.\ J.\  {\bf 498}, L107 (1998)
[arXiv:astro-ph/9801193].
H.~Dahle, 
arXiv:astro-ph/0701598.




\bibitem{zwicky}
F. Zwicky, Helv. Phys. Acta 6 (1933) 110.

\bibitem{wmap}
  E.~Komatsu {\it et al.}  [WMAP Collaboration],
  arXiv:0803.0547 [astro-ph].

\bibitem{bbn}
K.~A.~Olive, G.~Steigman and T.~P.~Walker,
Phys.\ Rept.\  {\bf 333}, 389 (2000)
[arXiv:astro-ph/9905320].

\bibitem{lss}
 M.~Tegmark {\it et al.} 
[SDSS Collaboration],
Astrophys.\ J.\  {\bf 606}, 702 (2004)
[arXiv:astro-ph/0310725].


\bibitem{mond}
  M.~Milgrom,
  Astrophys.\ J.\  {\bf 270}, 365 (1983).

\bibitem{Bekenstein:2004ne}
  J.~D.~Bekenstein,
  Phys.\ Rev.\  D {\bf 70}, 083509 (2004)
  [Erratum-ibid.\  D {\bf 71}, 069901 (2005)]
  [arXiv:astro-ph/0403694].

\bibitem{Skordis:2005xk}
  C.~Skordis, D.~F.~Mota, P.~G.~Ferreira and C.~Boehm,
  Phys.\ Rev.\ Lett.\  {\bf 96}, 011301 (2006)
  [arXiv:astro-ph/0505519].

\bibitem{bullet}
D.~Clowe, M.~Bradac, A.~H.~Gonzalez, M.~Markevitch, S.~W.~Randall, C.~Jones and D.~Zaritsky,
arXiv:astro-ph/0608407.


\bibitem{Srednicki:1988ce}
  M.~Srednicki, R.~Watkins and K.~A.~Olive,
  Nucl.\ Phys.\  B {\bf 310}, 693 (1988).


\bibitem{Gondolo:1990dk}
P.~Gondolo and G.~Gelmini,
Nucl.\ Phys.\ B {\bf 360} (1991) 145.

\bibitem{kolbturner}
E.~W.~Kolb and M.~S.~Turner, ``The Early Universe''. 


\bibitem{leeweinberg}
B.~W.~Lee and S.~Weinberg,
Phys.\ Rev.\ Lett.\  {\bf 39}, 165 (1977).


\bibitem{pdg}
C.~Amsler {\it et al}. (Particle Data Group), 
Phys.\ Lett.\ B  {\bf 667}, 1 (2008).




\bibitem{Griest:1990kh}
K.~Griest and D.~Seckel,
Phys.\ Rev.\ D {\bf 43} (1991) 3191.


\bibitem{Feng:2003xh}
J.~L.~Feng, A.~Rajaraman and F.~Takayama,
Phys.\ Rev.\ Lett.\  {\bf 91} (2003) 011302
[arXiv:hep-ph/0302215].




\bibitem{susyreview}
For a review of supersymmetry phenomenology, see:  S.~P.~Martin,
  arXiv:hep-ph/9709356.

\bibitem{gut}
  J.~R.~Ellis, S.~Kelley and D.~V.~Nanopoulos,
  Phys.\ Lett.\ B {\bf 260}, 131 (1991).


\bibitem{neutralinodm}
  H.~Goldberg,
  Phys.\ Rev.\ Lett.\  {\bf 50}, 1419 (1983);
  J.~R.~Ellis, J.~S.~Hagelin, D.~V.~Nanopoulos, K.~A.~Olive and M.~Srednicki,
  Nucl.\ Phys.\ B {\bf 238}, 453 (1984).



\bibitem{falksneutrino}
T.~Falk, K.~A.~Olive and M.~Srednicki,
Phys.\ Lett.\ B {\bf 339}, 248 (1994)
[arXiv:hep-ph/9409270].







\bibitem{jungman}
  G.~Jungman, M.~Kamionkowski and K.~Griest,
  Phys.\ Rept.\  {\bf 267}, 195 (1996).





\bibitem{Edsjo:1997bg}
J.~Edsjo and P.~Gondolo,
Phys.\ Rev.\ D {\bf 56} (1997) 1879
[arXiv:hep-ph/9704361].

\bibitem{Ellis:1999mm}
  J.~R.~Ellis, T.~Falk, K.~A.~Olive and M.~Srednicki,
  Astropart.\ Phys.\  {\bf 13}, 181 (2000)
  [Erratum-ibid.\  {\bf 15}, 413 (2001)]
  [arXiv:hep-ph/9905481].

\bibitem{Ellis:2001nx}
  J.~R.~Ellis, K.~A.~Olive and Y.~Santoso,
  Astropart.\ Phys.\  {\bf 18}, 395 (2003)
  [arXiv:hep-ph/0112113].

\bibitem{Edsjo:2003us}
J.~Edsjo, M.~Schelke, P.~Ullio and P.~Gondolo,
JCAP {\bf 0304} (2003) 001
[arXiv:hep-ph/0301106].






\bibitem{darksusy}
  P.~Gondolo, J.~Edsjo, P.~Ullio, L.~Bergstrom, M.~Schelke and E.~A.~Baltz,
  JCAP {\bf 0407}, 008 (2004)
  [arXiv:astro-ph/0406204].




\bibitem{micromegas}
  G.~Belanger, F.~Boudjema, A.~Pukhov and A.~Semenov,
  Comput.\ Phys.\ Commun.\  {\bf 177}, 894 (2007).






\bibitem{gm2SM}
K.~Hagiwara, A.D.~Martin,  D.~Nomura, and T.~Teubner, 
(2006) [arXiv:hep-ph/0611102].




\bibitem{Arkani-Hamed:1998rs}
  N.~Arkani-Hamed, S.~Dimopoulos and G.~R.~Dvali,
  Phys.\ Lett.\ B {\bf 429}, 263 (1998)
  [arXiv:hep-ph/9803315].

\bibitem{Arkani-Hamed:1998nn}
  N.~Arkani-Hamed, S.~Dimopoulos and G.~R.~Dvali,
  Phys.\ Rev.\ D {\bf 59}, 086004 (1999)
  [arXiv:hep-ph/9807344].


\bibitem{Randall:1999ee}
  L.~Randall and R.~Sundrum,
  Phys.\ Rev.\ Lett.\  {\bf 83}, 3370 (1999)
  [arXiv:hep-ph/9905221].

\bibitem{Appelquist:2000nn}
  T.~Appelquist, H.~C.~Cheng and B.~A.~Dobrescu,
  Phys.\ Rev.\ D {\bf 64}, 035002 (2001)
  [arXiv:hep-ph/0012100].

\bibitem{uedreview}
  D.~Hooper and S.~Profumo,
  Phys.~Rept., in press, arXiv:hep-ph/0701197.

\bibitem{Servant:2002aq}
G.~Servant and T.~M.~Tait,
Nucl.\ Phys.\ B {\bf 650} (2003) 391
[arXiv:hep-ph/0206071].

\bibitem{Cheng:2002ej}
H.~C.~Cheng, J.~L.~Feng and K.~T.~Matchev,
Phys.\ Rev.\ Lett.\  {\bf 89}, 211301 (2002)
[arXiv:hep-ph/0207125].


\bibitem{Burnell:2005hm}
  F.~Burnell and G.~D.~Kribs,
  Phys.\ Rev.\  D {\bf 73}, 015001 (2006)
  [arXiv:hep-ph/0509118].

\bibitem{Kong:2005hn}
  K.~Kong and K.~T.~Matchev,
  JHEP {\bf 0601}, 038 (2006)
  [arXiv:hep-ph/0509119].



\bibitem{tparity}
H.~C.~Cheng and I.~Low,
JHEP {\bf 0309}, 051 (2003)
[arXiv:hep-ph/0308199].



\bibitem{cdms}
  Z.~Ahmed {\it et al.}  [CDMS Collaboration],
  arXiv:0802.3530 [astro-ph].

\bibitem{xenon}
  J.~Angle {\it et al.}  [XENON Collaboration],
  Phys.\ Rev.\ Lett.\  {\bf 100}, 021303 (2008)
  [arXiv:0706.0039 [astro-ph]].



\bibitem{zeplin}
  G.~J.~Alner {\it et al.},
  Astropart.\ Phys.\  {\bf 28}, 287 (2007)
  [arXiv:astro-ph/0701858];
  G.~J.~Alner {\it et al.}  [UK Dark Matter Collaboration],
  Astropart.\ Phys.\  {\bf 23}, 444 (2005).

\bibitem{edelweiss}
  V.~Sanglard {\it et al.}  [The EDELWEISS Collaboration],
  Phys.\ Rev.\ D {\bf 71}, 122002 (2005)
  [arXiv:astro-ph/0503265].


\bibitem{cresst}
  G.~Angloher {\it et al.},
  Astropart.\ Phys.\  {\bf 23}, 325 (2005)
  [arXiv:astro-ph/0408006].


\bibitem{cogent}
  C.~E.~Aalseth {\it et al.},
  arXiv:0807.0879 [astro-ph].


\bibitem{dama}
  R.~Bernabei {\it et al.}  [DAMA Collaboration],
  arXiv:0804.2741 [astro-ph];
  R.~Bernabei {\it et al.}  [DAMA Collaboration],
  arXiv:0804.2741 [astro-ph].


\bibitem{coupp}
  E.~Behnke {\it et al.}  [COUPP Collaboration],
  Science {\bf 319}, 933 (2008)
  [arXiv:0804.2886 [astro-ph]].

\bibitem{warp}
  P.~Benetti {\it et al.},
  arXiv:astro-ph/0701286;
  R.~Brunetti {\it et al.},
  New Astron.\ Rev.\  {\bf 49}, 265 (2005)
  [arXiv:astro-ph/0405342].

\bibitem{kims}
  H.~S.~Lee. {\it et al.}  [KIMS Collaboration],
  Phys.\ Rev.\ Lett.\  {\bf 99}, 091301 (2007)
  [arXiv:0704.0423 [astro-ph]].



\bibitem{nuc}
  A.~Bottino, F.~Donato, N.~Fornengo and S.~Scopel,
  Astropart.\ Phys.\  {\bf 18}, 205 (2002)
  [arXiv:hep-ph/0111229];
 Astropart.\ Phys.\  {\bf 13}, 215 (2000)
  [arXiv:hep-ph/9909228];
  J.~R.~Ellis, K.~A.~Olive, Y.~Santoso and V.~C.~Spanos,
  Phys.\ Rev.\ D {\bf 71}, 095007 (2005)
  [arXiv:hep-ph/0502001].


\bibitem{scatteraq}
G.~B.~Gelmini, P.~Gondolo and E.~Roulet,
Nucl.\ Phys.\ B {\bf 351}, 623 (1991);
M.~Srednicki and R.~Watkins,
Phys.\ Lett.\ B {\bf 225}, 140 (1989);
M.~Drees and M.~Nojiri,
Phys.\ Rev.\ D {\bf 48}, 3483 (1993)
[arXiv:hep-ph/9307208];
M.~Drees and M.~M.~Nojiri,
Phys.\ Rev.\ D {\bf 47}, 4226 (1993)
[arXiv:hep-ph/9210272];
J.~R.~Ellis, A.~Ferstl and K.~A.~Olive, 
Phys.~Lett.~B  481, (2000) 304,
[arXiv:hep-ph/0001005].











\bibitem{Servant:2002hb}
G.~Servant and T.~M.~Tait,
New J.\ Phys.\  {\bf 4}, 99 (2002)
[arXiv:hep-ph/0209262].


\bibitem{limitplotter}
http://dendera.berkeley.edu/plotter/entryform.html


\bibitem{modelindependent}
  M.~Beltran, D.~Hooper, E.~W.~Kolb and Z.~C.~Krusberg,
  arXiv:0808.3384 [hep-ph].






\bibitem{gg}
  L.~Bergstrom and P.~Ullio,
  Nucl.\ Phys.\ B {\bf 504}, 27 (1997)
  [arXiv:hep-ph/9706232].

\bibitem{gz}
  P.~Ullio and L.~Bergstrom,
  Phys.\ Rev.\ D {\bf 57}, 1962 (1998)
  [arXiv:hep-ph/9707333].

\bibitem{buckley}
  L.~Bergstrom, P.~Ullio and J.~H.~Buckley,
  Astropart.\ Phys.\  {\bf 9}, 137 (1998)
  [arXiv:astro-ph/9712318].



\bibitem{gchist}
  V.~Berezinsky, A.~Bottino and G.~Mignola,
  Phys.\ Lett.\ B {\bf 325}, 136 (1994)
  [arXiv:hep-ph/9402215].




\bibitem{nfw}
 J.~F.~Navarro, C.~S.~Frenk and S.~D.~M.~White,
  Astrophys.\ J.\  {\bf 462}, 563 (1996)
  [arXiv:astro-ph/9508025];
  J.~F.~Navarro, C.~S.~Frenk and S.~D.~M.~White,
  Astrophys.\ J.\  {\bf 490}, 493 (1997).


\bibitem{moore}
  B.~Moore, S.~Ghigna, F.~Governato, G.~Lake, T.~Quinn, J.~Stadel and P.~Tozzi,
  Astrophys.\ J.\  {\bf 524}, L19 (1999).

\bibitem{ac}
  F.~Prada, A.~Klypin, J.~Flix, M.~Martinez and E.~Simonneau,
  arXiv:astro-ph/0401512;
  G.~Bertone and D.~Merritt,
  Mod.\ Phys.\ Lett.\ A {\bf 20}, 1021 (2005)
  [arXiv:astro-ph/0504422];
  G.~Bertone and D.~Merritt,
  Phys.\ Rev.\ D {\bf 72}, 103502 (2005)
  [arXiv:astro-ph/0501555].

\bibitem{spike}
  P.~Gondolo and J.~Silk,
  Phys.\ Rev.\ Lett.\  {\bf 83}, 1719 (1999)
  [arXiv:astro-ph/9906391];
  P.~Ullio, H.~Zhao and M.~Kamionkowski,
  Phys.\ Rev.\ D {\bf 64}, 043504 (2001)
  [arXiv:astro-ph/0101481];
  G.~Bertone, G.~Sigl and J.~Silk,
  Mon.\ Not.\ Roy.\ Astron.\ Soc.\  {\bf 337}, 98 (2002)
  [arXiv:astro-ph/0203488].

\bibitem{acts}
F.~Aharonian {\it et al.}  [The HESS Collaboration],
  arXiv:astro-ph/0408145;
  J.~Albert {\it et al.}  [MAGIC Collaboration],
  Astrophys.\ J.\  {\bf 638}, L101 (2006)
  [arXiv:astro-ph/0512469].





\bibitem{gabi}
  G.~Zaharijas and D.~Hooper,
  Phys.\ Rev.\  D {\bf 73}, 103501 (2006).
  [arXiv:astro-ph/0603540].


\bibitem{Dodelson:2007gd}
  S.~Dodelson, D.~Hooper and P.~D.~Serpico,
  arXiv:0711.4621 [astro-ph].



\bibitem{dingus}
  D.~Hooper and B.~L.~Dingus,
  Phys.\ Rev.\ D {\bf 70}, 113007 (2004)
  [arXiv:astro-ph/0210617].


\bibitem{dwarfs}
  N.~W.~Evans, F.~Ferrer and S.~Sarkar,
  Phys.\ Rev.\  D {\bf 69}, 123501 (2004)
  [arXiv:astro-ph/0311145];
  L.~Bergstrom and D.~Hooper,
  Phys.\ Rev.\  D {\bf 73}, 063510 (2006)
  [arXiv:hep-ph/0512317];
  L.~E.~Strigari, S.~M.~Koushiappas, J.~S.~Bullock, M.~Kaplinghat, J.~D.~Simon, M.~Geha and B.~Willman,
  arXiv:0709.1510 [astro-ph].





\bibitem{cosmological}
  P.~Ullio, L.~Bergstrom, J.~Edsjo and C.~G.~Lacey,
  Phys.\ Rev.\ D {\bf 66}, 123502 (2002)
  [arXiv:astro-ph/0207125];
  D.~Elsaesser and K.~Mannheim,
  Astropart.\ Phys.\  {\bf 22}, 65 (2004)
  [arXiv:astro-ph/0405347].





\bibitem{pamela}
  O.~Adriani {\it et al.},
  arXiv:0810.4995 [astro-ph].


\bibitem{heat}
S.~W.~Barwick {\it et al.}  [HEAT Collaboration],
Astrophys.\ J.\  {\bf 482}, L191 (1997)
[arXiv:astro-ph/9703192];
S.~Coutu {\it et al.}  [HEAT-pbar Collaboration],
in Proceedings of 27th ICRC (2001).

\bibitem{ams01}
Olzem, Jan [AMS Collaboration],
Talk given at the 7th UCLA Symposium on Sources and Detection of Dark Matter and Dark Energy in the Universe, Marina del Ray, CA, Feb 22-24, 2006.

\bibitem{atic}
J.~Chang {\it et al.} [ATIC Collaboration],
Nature\ {\bf 456}, 362 (2008).

\bibitem{haze}
  G.~Dobler and D.~P.~Finkbeiner,
  Astrophys.\ J.\  {\bf 680}, 1222 (2008)
  [arXiv:0712.1038 [astro-ph]].

\bibitem{hazedark}
 D.~P.~Finkbeiner,
  arXiv:astro-ph/0409027;
  D.~Hooper, D.~P.~Finkbeiner and G.~Dobler,
  Phys.\ Rev.\  D {\bf 76}, 083012 (2007)
  [arXiv:0705.3655 [astro-ph]].

\bibitem{pulsars}
  D.~Hooper, P.~Blasi and P.~D.~Serpico,
  arXiv:0810.1527 [astro-ph];
  S.~Profumo,
  arXiv:0812.4457 [astro-ph];
  F.~A~Aharonian, A.~M.~Atoyan and H.~J.~Volk,
  Astron.\ Astrophys.\  {\bf 294}, L41-L44 (1995);
  L.~Zhang and K.~S.~Cheng,
  Astron.\ Astrophys.\  {\bf 368}, 1063-1070 (2001);
  I.~Buesching, O.~C.~de Jager, M.~S.~Potgieter and C.~Venter,
  arXiv:0804.0220 [astro-ph].

\bibitem{darkmatter}
  I.~Cholis, L.~Goodenough, D.~Hooper, M.~Simet and N.~Weiner,
  arXiv:0809.1683 [hep-ph].

\bibitem{darkmatter2}
  L.~Bergstrom, T.~Bringmann and J.~Edsjo,
  arXiv:0808.3725 [astro-ph];
  M.~Cirelli and A.~Strumia,
  arXiv:0808.3867 [astro-ph];
  V.~Barger, W.~Y.~Keung, D.~Marfatia and G.~Shaughnessy,
  arXiv:0809.0162 [hep-ph];
  N.~Arkani-Hamed, D.~P.~Finkbeiner, T.~Slatyer and N.~Weiner,
  arXiv:0810.0713 [hep-ph];
  I.~Cholis, G.~Dobler, D.~P.~Finkbeiner, L.~Goodenough and N.~Weiner,
  arXiv:0811.3641 [astro-ph];
  P.~J.~Fox and E.~Poppitz,
  arXiv:0811.0399 [hep-ph];
 K.~M.~Zurek,
  arXiv:0811.4429 [hep-ph].

\bibitem{antiprotonsexceed}
  F.~Donato, D.~Maurin, P.~Brun, T.~Delahaye and P.~Salati,
  arXiv:0810.5292 [astro-ph];
 M.~Cirelli, M.~Kadastik, M.~Raidal and A.~Strumia,
  arXiv:0809.2409 [hep-ph].


\bibitem{hooperhall}
  J.~Hall and D.~Hooper,
  arXiv:0811.3362 [astro-ph];
  E.~A.~Baltz and D.~Hooper,
  JCAP {\bf 0507}, 001 (2005)
  [arXiv:hep-ph/0411053].



\bibitem{diffusion}
W.~R.~Webber, M.~A.~Lee and M.~Gupta,
Astrophys.\ J.{\bf 390} (1992) 96;
I.~V.~Moskalenko, A.~W.~Strong, S.~G.~Mashnik and J.~F.~Ormes,
Astrophys.\ J.\  {\bf 586}, 1050 (2003)
[arXiv:astro-ph/0210480];
I.~V.~Moskalenko and A.~W.~Strong,
Phys.\ Rev.\ D {\bf 60}, 063003 (1999)
[arXiv:astro-ph/9905283];
E.~A.~Baltz and J.~Edsjo,
Phys.\ Rev.\ D {\bf 59} (1999) 023511
[arXiv:astro-ph/9808243];
D.~Hooper and J.~Silk,
Phys.\ Rev.\ D {\bf 71}, 083503 (2005)
[arXiv:hep-ph/0409104].




\bibitem{StrongMoskPtusk}
A.~W.~Strong, I~.V.~Moskalenko and V.~S.~Ptuskin,
ARNPS, 57, 285 (2007).

\bibitem{kkpos}
  D.~Hooper and G.~D.~Kribs,
  Phys.\ Rev.\ D {\bf 70}, 115004 (2004)
  [arXiv:hep-ph/0406026].


\bibitem{sommerfeld}
  M.~Cirelli and A.~Strumia,
  arXiv:0808.3867 [astro-ph];
  I.~Cholis, G.~Dobler, D.~P.~Finkbeiner, L.~Goodenough and N.~Weiner,
  arXiv:0811.3641 [astro-ph];
  N.~Arkani-Hamed, D.~P.~Finkbeiner, T.~Slatyer and N.~Weiner,
  arXiv:0810.0713 [hep-ph].








\bibitem{Gould:1991hx}
A.~Gould, Astrophys.\ J.\ {\bf 388}, 338 (1991).

\bibitem{Gould:1987ir}
A.~Gould,
Astrophys.\ J.\  {\bf 321}, 571 (1987).


\bibitem{Griest:1986yu}
K.~Griest and D.~Seckel,
Nucl.\ Phys.\ B {\bf 283}, 681 (1987)
[Erratum-ibid.\ B {\bf 296}, 1034 (1988)].


\bibitem{Jungman:1994jr}
G.~Jungman and M.~Kamionkowski,
Phys.\ Rev.\ D {\bf 51} (1995) 328
[arXiv:hep-ph/9407351];
For a more recent calculation see:
 M.~Cirelli, N.~Fornengo, T.~Montaruli, I.~Sokalski, A.~Strumia and F.~Vissani,
  arXiv:hep-ph/0506298.

\bibitem{Lehnert:2007fv}
R.~Lehnert and T.~J.~Weiler,
Phys.\ Rev.\ D {\bf 77} (2008) 125004
[arXiv:0708.1035].




\bibitem{icecube}
T.~DeYoung  [IceCube Collaboration],
  Int.\ J.\ Mod.\ Phys.\ A {\bf 20}, 3160 (2005);
J.~Ahrens {\it et al.}  [The IceCube Collaboration],
  Nucl.\ Phys.\ Proc.\ Suppl.\  {\bf 118}, 388 (2003)
  [arXiv:astro-ph/0209556].


\bibitem{Desai:2004pq}
  S.~Desai {\it et al.}  [Super-Kamiokande Collaboration],
  Phys.\ Rev.\  D {\bf 70}, 083523 (2004)
  [Erratum-ibid.\  D {\bf 70}, 109901 (2004)]
  [arXiv:hep-ex/0404025].



\bibitem{neutrinosun}
 L.~Bergstrom, J.~Edsjo and P.~Gondolo,
  Phys.\ Rev.\  D {\bf 55}, 1765 (1997)
  [arXiv:hep-ph/9607237];
  Phys.\ Rev.\  D {\bf 58}, 103519 (1998)
  [arXiv:hep-ph/9806293];
 V.~D.~Barger, F.~Halzen, D.~Hooper and C.~Kao,
  Phys.\ Rev.\  D {\bf 65}, 075022 (2002)
  [arXiv:hep-ph/0105182].


\bibitem{Halzen:2005ar}
  F.~Halzen and D.~Hooper,
  Phys.\ Rev.\  D {\bf 73}, 123507 (2006)
  [arXiv:hep-ph/0510048].



\bibitem{Hooper:2002gs}
D.~Hooper and G.~D.~Kribs,
Phys.\ Rev.\ D {\bf 67}, 055003 (2003)
[arXiv:hep-ph/0208261].






\bibitem{clump}
  D.~Hooper, A.~Stebbins and K.~M.~Zurek,
  arXiv:0812.3202 [hep-ph].




\bibitem{Hooper:2007gi}
  D.~Hooper, G.~Zaharijas, D.~P.~Finkbeiner and G.~Dobler,
  Phys.\ Rev.\  D {\bf 77}, 043511 (2008)
  [arXiv:0709.3114 [astro-ph]].







  \bibitem{damadark}
  A.~Bottino, F.~Donato, N.~Fornengo and S.~Scopel,
  Phys.\ Rev.\  D {\bf 77}, 015002 (2008)
  [arXiv:0710.0553 [hep-ph]];
  R.~Foot,
  arXiv:0804.4518 [hep-ph];
  J.~L.~Feng, J.~Kumar and L.~E.~Strigari,
  arXiv:0806.3746 [hep-ph];
  F.~Petriello and K.~M.~Zurek,
  arXiv:0806.3989 [hep-ph];
  A.~Bottino, F.~Donato, N.~Fornengo and S.~Scopel,
  arXiv:0806.4099 [hep-ph].





  \bibitem{Drobyshevski:2007zj}
  E.~M.~Drobyshevski,
  arXiv:0706.3095 [physics.ins-det];
  R.~Bernabei {\it et al.},
  Eur.\ Phys.\ J.\  C {\bf 53}, 205 (2008)
  [arXiv:0710.0288 [astro-ph]].


\bibitem{inelastic}
  S.~Chang, G.~D.~Kribs, D.~Tucker-Smith and N.~Weiner,
  arXiv:0807.2250 [hep-ph];
D.~R.~Smith and N.~Weiner,
Phys.\ Rev.\ D {\bf 64}, 043502 (2001)
[arXiv:hep-ph/0101138];
  D.~Tucker-Smith and N.~Weiner,
  arXiv:hep-ph/0402065.






\bibitem{integral}
  P.~Jean {\it et al.},
  Astron.\ Astrophys.\  {\bf 407}, L55 (2003)
  [arXiv:astro-ph/0309484].

\bibitem{asymmetric}
  G.~Weidenspointner {\it et al.},
  Nature {\bf 451}, 159 (2008).

\bibitem{Prantzos:2005pz}
  N.~Prantzos,
  Astron.\ Astrophys.\  {\bf 449}, 869 (2006)
  [arXiv:astro-ph/0511190].

\bibitem{Kalemci:2006bz}
  E.~Kalemci, S.~E.~Boggs, P.~A.~Milne and S.~P.~Reynolds,
  Astrophys.\ J.\  {\bf 640}, L55 (2006)
  [arXiv:astro-ph/0602233].

\bibitem{Casse:2003fh}
  M.~Casse, B.~Cordier, J.~Paul and S.~Schanne,
  Astrophys.\ J.\  {\bf 602}, L17 (2004)
  [arXiv:astro-ph/0309824].


\bibitem{Parizot:2004ph}
  G.~Bertone, A.~Kusenko, S.~Palomares-Ruiz, S.~Pascoli and D.~Semikoz,
  Phys.\ Lett.\  B {\bf 636}, 20 (2006)
  [arXiv:astro-ph/0405005];
  E.~Parizot, M.~Casse, R.~Lehoucq and J.~Paul,
  arXiv:astro-ph/0411656.


\bibitem{Bandyopadhyay:2008ts}
  R.~M.~Bandyopadhyay, J.~Silk, J.~E.~Taylor and T.~J.~Maccarone,
  arXiv:0810.3674 [astro-ph].




\bibitem{511dark}
 C.~Boehm, D.~Hooper, J.~Silk, M.~Casse and J.~Paul,
  Phys.\ Rev.\ Lett.\  {\bf 92}, 101301 (2004)
  [arXiv:astro-ph/0309686].

\bibitem{beacom}
  J.~F.~Beacom and H.~Yuksel,
  Phys.\ Rev.\ Lett.\  {\bf 97}, 071102 (2006)
  [arXiv:astro-ph/0512411];
  J.~F.~Beacom, N.~F.~Bell and G.~Bertone,
  Phys.\ Rev.\ Lett.\  {\bf 94}, 171301 (2005)
  [arXiv:astro-ph/0409403].

\bibitem{zurekmodel}
 D.~Hooper and K.~M.~Zurek,
  Phys.\ Rev.\  D {\bf 77}, 087302 (2008)
  [arXiv:0801.3686 [hep-ph]].

\bibitem{lightok}
  C.~Boehm, T.~A.~Ensslin and J.~Silk,
  J.\ Phys.\ G {\bf 30}, 279 (2004)
  [arXiv:astro-ph/0208458].



\bibitem{susycase}
  D.~Hooper and T.~Plehn,
  Phys.\ Lett.\ B {\bf 562}, 18 (2003)
  [arXiv:hep-ph/0212226];
  A.~Bottino, F.~Donato, N.~Fornengo and S.~Scopel,
  Phys.\ Rev.\ D {\bf 68}, 043506 (2003)
  [arXiv:hep-ph/0304080].

\bibitem{nmssm}
  J.~F.~Gunion, D.~Hooper and B.~McElrath,
  Phys.\ Rev.\ D {\bf 73}, 015011 (2006)
  [arXiv:hep-ph/0509024].

\bibitem{scalar}
  C.~Boehm and P.~Fayet,
  Nucl.\ Phys.\ B {\bf 683}, 219 (2004)
  [arXiv:hep-ph/0305261];
  P.~Fayet,
  Phys.\ Rev.\ D {\bf 70}, 023514 (2004)
  [arXiv:hep-ph/0403226].


\bibitem{exciting}
  D.~P.~Finkbeiner and N.~Weiner,
  arXiv:astro-ph/0702587.




\bibitem{deboer}
  W.~de Boer, M.~Herold, C.~Sander, V.~Zhukov, A.~V.~Gladyshev and D.~I.~Kazakov,
  arXiv:astro-ph/0408272;
  W.~de Boer, C.~Sander, V.~Zhukov, A.~V.~Gladyshev and D.~I.~Kazakov,
  Astron.\ Astrophys.\  {\bf 444}, 51 (2005)
  [arXiv:astro-ph/0508617];
  W.~de Boer, C.~Sander, V.~Zhukov, A.~V.~Gladyshev and D.~I.~Kazakov,
  Phys.\ Lett.\  B {\bf 636}, 13 (2006)
  [arXiv:hep-ph/0511154];
  W.~de Boer, C.~Sander, V.~Zhukov, A.~V.~Gladyshev and D.~I.~Kazakov,
  Phys.\ Rev.\ Lett.\  {\bf 95}, 209001 (2005)
  [arXiv:astro-ph/0602325].


\bibitem{bergstromdeboer}
  L.~Bergstrom, J.~Edsjo, M.~Gustafsson and P.~Salati,
  JCAP {\bf 0605}, 006 (2006)
  [arXiv:astro-ph/0602632].

\bibitem{deBoer:2006ck}
  W.~de Boer, I.~Gebauer, C.~Sander, M.~Weber and V.~Zhukov,
  AIP Conf.\ Proc.\  {\bf 903}, 607 (2007)
  [arXiv:astro-ph/0612462];
See also: W.~de Boer and V.~Zhukov,
  arXiv:0709.4576 [astro-ph].




\bibitem{Moskalenko:2006zy}
  I.~V.~Moskalenko, S.~W.~Digel, T.~A.~Porter, O.~Reimer and A.~W.~Strong,
  arXiv:astro-ph/0609768.


\bibitem{Stecker:2007xp}
  F.~W.~Stecker, S.~D.~Hunter and D.~A.~Kniffen,
  arXiv:0705.4311 [astro-ph].






\bibitem{Elsaesser:2004ap}
  D.~Elsaesser and K.~Mannheim,
  Phys.\ Rev.\ Lett.\  {\bf 94}, 171302 (2005)
  [arXiv:astro-ph/0405235].



\bibitem{Ando:2005hr}
  S.~Ando,
  Phys.\ Rev.\ Lett.\  {\bf 94}, 171303 (2005)
  [arXiv:astro-ph/0503006].





\end{thebibliography}
\end{document}